\documentclass[twocolumn]{aastex701}
\usepackage{graphicx} % Required for inserting images
\usepackage{amsmath}
\usepackage{caption}
\usepackage{afterpage}
\usepackage{multirow}
\usepackage{array}
\usepackage{tabularx}
\usepackage{pifont}

\makeatletter
\newcommand{\clearafterfulltwocolumnpage}{%
  \if@twocolumn
    \if@firstcolumn
      \afterpage{\afterpage{\clearpage}}%
    \else
      \afterpage{\clearpage}%
    \fi
  \else
    \afterpage{\clearpage}%
  \fi
}
\makeatother

\definecolor{blazeorange}{rgb}{1.0, 0.6, 0.2}
\definecolor{seagreen}{rgb}{0.18, 0.55, 0.34}
\definecolor{rufous}{rgb}{0.66, 0.11, 0.03}
\definecolor{royalfuchsia}{rgb}{0.79, 0.17, 0.57}
\definecolor{scarlet}{rgb}{1.0, 0.13, 0.0}
\definecolor{royalpurple}{rgb}{0.47, 0.32, 0.66}
\definecolor{darkblue}{rgb}{0, 0, 0.66}
\definecolor{violet}{rgb}{0.5,0.,0.5}
\begin{document}
\def\tjet{\ensuremath{t_{\rm jet}}}
\def\tovertjet{\ensuremath{\tilde{t}}}
\def\PL{\ensuremath{P_{\rm L}}}
\def\PiL{\ensuremath{\Pi_{\rm L}}}
\def\Pimax{\ensuremath{\Pi_{\rm max}}}
\def\thetajet{\ensuremath{\theta_{\rm jet}}}
\def\thetaobs{\ensuremath{\theta_{\rm obs}}}
\def\EKiso{\ensuremath{E_{\rm K,iso}}}
\def\Egammaiso{\ensuremath{E_{\gamma,\rm iso}}}
\newcommand{\cmark}{\ding{51}}
\newcommand{\xmark}{\ding{55}}
\def\thetaB{\ensuremath{\theta_{\rm B}}}
\def\thetac{\ensuremath{\theta_{\rm c}}}
\def\thetavis{\ensuremath{\theta_{\rm vis}}}

\title{Revealing the Structure and Magnetization of GRB Jets with ALMA Polarization Observations}

\newcommand{\UA}{\affiliation{Steward Observatory, University of Arizona, 933 North Cherry Avenue, Tucson, AZ 85721-0065, USA}}

\author[orcid=0000-0003-1792-2338,gname=Tanmoy,sname=Laskar]{Tanmoy~Laskar}
\affiliation{Department of Physics \& Astronomy, University of Utah, Salt Lake City, UT 84112, USA}
\email{tanmoy.laskar@utah.edu}

\author[orcid=0000-0003-0528-202X,gname=Collin,sname=Christy]{Collin~T.~Christy}
\UA
\email{collinchristy@arizona.edu}

\author[orcid=0000-0003-4537-3575,gname=Noah,sname=Franz]{Noah~Franz}
\email{nfranz@arizona.edu}
\UA

\author[orcid=0009-0008-5392-4190,gname=Gavin,sname=Farley]{Gavin~Farley}
\affiliation{Department of Physics \& Astronomy, University of Utah, Salt Lake City, UT 84112, USA}
\email{gavinfarley04@gmail.com}

\author[orcid=0000-0002-8297-2473,gname=Kate,sname=Alexander]{Kate~D.~Alexander}
\UA
\email{kdalexander@arizona.edu}

\author[orcid=0000-0001-8530-8941, gname=Jonathan, sname=Granot]{Jonathan~Granot}
\affiliation{Astrophysics Research Center of the Open University (ARCO), The Open University of Israel, P.O Box 808, Ra'anana 43537, Israel}
\affiliation{Astroparticule et Cosmologie (APC), CNRS – Université Paris Cité, F-75013 Paris, France}
\affiliation{Department of Physics, The George Washington University, 725 21st Street NW., Washington, DC 20052, USA}
\email{granot@openu.ac.il}

\author[orcid=0000-0003-0516-2968,gname=Ramandeep,sname=Gill]{Ramandeep~Gill}
\affiliation{Instituto de Radioastronomía y Astrof\'isica, Universidad Nacional Aut\'onoma de M\'exico, Antigua Carretera a P\'atzcuaro \# 8701,\\ Ex-Hda. San Jos\'e de la, Huerta, Morelia, Michoac\'an, M\'exico C.P. 58089, M\'exico}
\affiliation{Astrophysics Research Center of the Open University (ARCO), The Open University of Israel, P.O Box 808, Ra'anana 43537, Israel}
\email{r.gill@irya.unam.mx}

\author[orcid=0000-0003-0307-9984, gname=Tarraneh, sname=Eftekhari]{Tarraneh Eftekhari}
\affiliation{Center for Interdisciplinary Exploration and Research in Astrophysics (CIERA), Northwestern University, 1800 Sherman Avenue, Evanston, IL 60201, USA }
\email{teftekhari@northwestern.edu}

\author[orcid=0000-0001-7946-4200,gname=Shiho,sname=Kobayashi]{Shiho Kobayashi}
\affiliation{Astrophysics Research Institute, Liverpool John Moores University, 146 Brownlow Hill, Liverpool L3 5RF, UK}
\email{S.Kobayashi@ljmu.ac.uk}

\author[orcid=0000-0002-8680-8718, gname=Hendrik, sname=Van Eerten]{Hendrik van Eerten}
\affiliation{Department of Physics, University of Bath, Claverton Down, Bath BA2 7AY, UK}
\email{hjve20@bath.ac.uk}

\author[orcid=0000-0003-4768-7586, gname=Raffaella, sname=Margutti]{Raffaella Margutti}
\affiliation{Department of Astronomy, University of California, Berkeley, CA 94720-3411, USA}
\affiliation{Berkeley Center for Multi-messenger Research on Astrophysical Transients and Outreach (Multi-RAPTOR), University of California, Berkeley, CA 94720-3411, USA}
\affiliation{Department of Physics, University of California, 366 Physics North MC 7300, Berkeley, CA 94720, USA}
\email{rmargutti@berkeley.edu}

\author[0000-0002-9392-9681]{Edo Berger}
\affiliation{Center for Astrophysics \textbar{} Harvard \& Smithsonian, 60 Garden Street, Cambridge, MA 02138-1516, USA}
\email{eberger@cfa.harvard.edu}

% AUTHOR INFO IN NO PARTICULAR ORDER

\begin{abstract}
We present a systematic study of the currently available ALMA millimeter polarimetric sample of gamma-ray burst (GRB) afterglows. Our sample comprises 24 observations (20 new) of 11 long-duration GRBs spanning $\approx0.1$--87\,days after the burst. We detect significant linear polarization in 8 observations across 6 events, with polarization degrees ranging from $\PiL\approx0.6\%$ to $2.4\%$. For the remaining observations, we place deep upper limits (median $\PiL\lesssim1\%$). Multi-epoch observations reveal diverse polarization evolution. GRB\,190114C yields the best-sampled $\PiL$ for a radio afterglow to date, with early evolution favoring patchy magnetic fields in the reverse shock (RS) and later polarization broadly consistent with forward-shock (FS) models. GRB\,220921A exhibits a rapid rise in polarization from $\PiL\lesssim0.4\%$ to $2.4\%$ over 1.9--6.8\,days, inconsistent with toroidal RS magnetic-field models but broadly consistent with several FS random-field models. GRB\,221009A yields the highest-significance polarization detections in the sample ($\PiL\approx1.4$--$1.6\%$), yet neither existing FS nor RS polarization models reproduce both the observed polarization evolution and the viewing geometry inferred from broadband afterglow modeling, with the exception of patchy fields in the RS. Deep upper limits for GRB\,171205A rule out RS toroidal-field models for the published off-axis geometry, while a strong, single-epoch detection of GRB\,190829A, if RS-dominated, requires a nearly on-axis geometry for toroidal-field configurations. Interpreting the observations within a patchy-field framework implies magnetic-field coherence scales of order $\theta_B\sim10^{-3}$--$10^{-2}$\,rad. These observations demonstrate the diagnostic power of radio/mm polarimetry for probing the magnetic-field structure, emission region, and viewing geometry of relativistic GRB jets.

\end{abstract}

\section{Introduction}
Long-duration gamma-ray bursts (GRBs), resulting from the deaths of massive stars, produce a relativistic jet that interacts with the ambient medium. These jets, forming and subsequently dissipating on human timescales, allow us to investigate key questions around jet physics, such as their formation, level of collimation, and magnetization \citep{mes06,kz15}. Synchrotron polarization studies of their afterglows, in particular, are expected to provide constraints on the magnetic field structure and viewing geometry  \citep{gra03,gk03,rlsg04,gt05,kob17}.

As the relativistic jet interacts with the surrounding medium, two shocks are produced: a forward shock (FS) propagating into the ambient medium, which powers the long-lived multiwavelength afterglow, and a short-lived reverse shock (RS) that propagates into the ejecta \citep{sp95,sp99,zk05,gra12}. Both shocks produce synchrotron radiation that is polarized to some degree; RS linear polarization is dependent on the magnetic field structure of the ejecta, and the FS linear polarization probes the jet angular structure, viewing geometry, and local magnetic field structure (shock-generated or amplified due to magnetohydrodynamic instabilities; \citealt{sal03,rlsg04,mkca+13,aak+24,bgb26}. High linear polarization ($\PiL \gtrsim 50\%$) requires large-scale ordered magnetic fields, while a low polarization signal ($\PiL \lesssim 10\%$) suggests tangled or patchy magnetic fields \citep{gw99,gk03,gra03,no04,gt05}. 

Due to the long-lived nature of RS emission at low frequencies, searching for polarization signals in the radio and millimeter band is an immediately attractive choice. However, searches in this frequency range have historically been limited by instrumental sensitivity \citep{gt05,vdhpdb+14,utc+23}. The advent of ALMA allowed for the first detection of linearly polarized RS millimeter emission at the $5\sigma$ level from the GRB\,190114C \citep{lag+19}. This detection constrained the angular coherence scale of the magnetic fields in this GRB to $\theta_B\approx10^{-3}\,$rad. The first measurement of Faraday rotation in the surrounding environment of GRB\,260310A, which also represented the first polarization detection in the cm-band, constrained both the intrinsic magnetization of the jet and the properties of the progenitor's surrounding HII region \citep{cla+26}. 

Here we present Atacama Large Millimeter/submillimeter Array (ALMA) polarimetry observations in Band 3 (97.5 GHz) and Band 4 (145 GHz) of a sample of 11 GRBs, spanning from $<1$ to 86.97 days post-trigger. Our data represent the most comprehensive GRB polarization sample to date, with significant ($\gtrsim 3 \sigma$) detections for 6 of 11 GRBs (and 8 of 20 new observations): GRB\,190114C (2 epochs), GRB\,190829A (1 epoch), GRB\,191221B (1 epoch), GRB\,220921A (1 epoch), GRB\,221009A (2 epochs), and GRB\,251013C (1 epoch). 

\begin{table*}
    \centering
    \caption{Summary of Observations}
    \begin{tabular}{llcclllr}
        \hline
        Target & Project (P.I.) & Date & Band & Leakage & Bandpass & Gain & $F_{\nu,\rm I}$ ($-\beta$)\tablenotemark{\dag} \\
        &&(yyyy-mm-dd)&&Cal.&Cal.&Cal.&(Jy) \\
        \hline 
        GRB\,171205A & 2017.1.00801.T (Urata) & 2017-12-14\tablenotemark{*} & 3 & J1256-0547 & J1127-1857 & J1130-1449 & --- \\
        & 2017.1.00801.T (Urata)  & 2017-12-16\tablenotemark{\ddag} & 3 & J1256-0547 & J1127-1857 & J1130-1449 & --- \\
        & 2017.A.00021.T (Urata) & 2018-01-10\tablenotemark{\ddag} & 3 & J1256-0547 & J1127-1857 & J1130-1449 & --- \\
        \hline
        GRB\,190114C & 2018.1.01405.T (Laskar) & 2019-01-14\tablenotemark{*} & 3 & J0423-0120 & J0423-012 & J0348-2749 & --- \\
        
        & 2018.1.01405.T (Laskar)& 2019-01-15 & 3 & J0423-0120 & J0522-3627 & J0348-2749 & 4.154(0.33) \\
        & 2018.1.00579.T (Urata) & 2019-01-16 & 3 & J0210-5101 & J0238+1636 & J0348-2749 & 0.425(0.49) \\
        & 2018.1.01405.T (Laskar)& 2019-01-17 & 3 & J0210-5101 & J0238+1636 & J0348-2749 & 0.425(0.49) \\
        & 2018.1.01405.T (Laskar)& 2019-01-24 & 3 & J0210-5101 & J0522-3627 & J0348-2749 & 4.170(0.24) \\
        & 2018.1.01405.T (Laskar)& 2019-04-11\tablenotemark{\ddag} & 3 & J0210-5101 & J0519-4546 & J0348-2749 & --- \\
        \hline
        GRB\,190829A& 2018.1.00579.T (Urata)  & 2019-09-01 & 3 & J0433+0531	& J0423-0120 & J0257-1212 & 3.620(0.68) \\
        \hline 
        GRB\,191221B & 2019.1.01482.T (Laskar) & 2019-12-22 & 3 & J1058+1033 & J1037-2934 & J1036-3744 & 0.616($0.46$) \\
        & 2019.1.01482.T (Laskar)& 2019-12-23 & 3 & J1058+1033 & J1037-2934 & J1036-3744 & 0.620($0.46$) \\
        & 2019.1.01016.T (Urata)& 2019-12-24\tablenotemark{*} & 3 & J1058+1033 & J1037-2934 & J1036-3744 & --- \\
        & 2019.1.01016.T (Urata)& 2019-12-27\tablenotemark{*} & 3 & J1058+1033 & J1037-2934 & J1036-3744 & --- \\
        \hline
        GRB\,210610B& 2019.1.01482.T (Laskar) & 2021-06-11 & 3 & J1773-1304 & J1550+0527 & J1608+1029 & $1.240(0.59)$ \\
        \hline
        GRB\,210702A& 2019.1.01482.T (Laskar) & 2021-07-03 & 3 & J1256-0547 & J1107-4449 & J1126-3828 & $1.106(0.69)$ \\
        \hline
        GRB\,220521A & 2019.1.01482.T (Laskar) & 2022-07-17\tablenotemark{\ddag} & 3 & J1957-3845 & J1550+0527 & J1824+1044 & --- \\
        & 2019.1.01482.T (Laskar) & 2022-08-12\tablenotemark{\ddag} & 3 & J1924-2914 & J1550+0527 & J1824+1044 & --- \\
        \hline
        GRB\,220921A & 2021.1.00657.T (Laskar) &  2022-09-23 & 3 & J0423-0120 & J0334-4008 & J0424-3756 & 0.767($0.47$) \\
        & 2021.1.00657.T (Laskar) & 2022-09-28 & 3 & J0348-2749 & J0334-4008 & J0424-3756 & 0.752($0.44$) \\
        \hline 
        GRB\,221009A & 2022.A.00009.T (Huang) & 2022-10-12 & 4 & J1957-3845 & J1924-2914 & J1914+1636 & 8.884($0.56$) \\
        & 2022.A.00009.T (Huang) & 2022-10-14 & 3 & J1957-3845 & J1924-2914 & J1914+1636 & 11.607($0.56$) \\
        \hline
        GRB\,250327B & 2024.1.01174.T (Laskar) & 2025-03-29 & 3 & J1256-0547 & J1229+0203 & J1159+2914 & $8.342(0.81)$ \\
        \hline
        GRB\,251013C & 2025.A.00003.T (Christy) & 2025-10-21 & 3 & J2158-1501 & J2232+1143 & J2301-0158 & $5.266(0.65)$\\
        \hline
    \end{tabular}
    \tablenotetext{*}{These epochs have previously been published: 171205A, \citealt{uth+19,lhc20}; 190114C, \citealt{lag+19}; and 191221B, \citealt{utc+23}.}
    \tablenotetext{\dag}{Stokes I flux density $F_{\nu, \rm I}$ of the flux calibrator (the same as the bandpass calibrator in each case) at the time of observation (along with the negative of the spectral index, $-\beta$), as queried from the ALMA database using the ALMA flux service\footnote{\url{https://almascience.eso.org/alma-data/calibrator-catalogue}}.}
    \tablenotetext{\ddag}{For these observations we use the ALMA pipeline reduction images rather than re-reducing the data with our custom reduction pipeline. Calibrator information is taken from the ALMA QA0/QA2 report.}
    \label{tab:obs}
\end{table*}

This paper is structured as follows. In section \ref{sec:reduction}, we summarize our data reduction and systematic check process. In section \ref{sec:results}, we describe the polarization properties of our sample and the constraints on non-detections. In section \ref{sec:modelsummary}, we summarize the existing models and in section \ref{sec:individual_targets}, we discuss constraints placed by our observations on the magnetic field structure of the emission region and the viewing geometry. In section \ref{sec:conclusion}, we conclude and summarize our findings. We use the convention $F_\nu\propto t^\alpha\nu^\beta$ throughout, where $t$ is the time since trigger. All times and frequencies are in the observer frame, unless otherwise stated. 

\section{Data Analysis} 
\begin{table*}
\centering
\caption{Radio polarization measurements for the GRBs in our sample. Reported uncertainties correspond to the 68\% credible interval derived from Monte Carlo realizations of the measured Stokes parameters assuming Gaussian errors. Upper limits are quoted at the 95\% confidence level for non-detections.}
\begin{tabular}{l l l c c c c c c c c c}
\hline
GRB & Date & $\nu$ & $t$ & $I$ & $Q$ & $U$ & $V$ & $P_L$ & $\PiL$ & $\chi$ & Z \\
 & UTC & GHz & d & $\rm mJy$ & $\rm \mu Jy$ & $\rm \mu Jy$ & $\rm \mu Jy$ & $\rm \mu Jy$ & \% & ($^\circ$) &  \\

\hline
171205A$^*$ & 2017 Dec 10 & 97.5 & 5.20 & $32.44_{-0.03}^{+0.03}$ & -- & -- & -- & -- & $<0.3$ & -- & -- \\
171205A & 2017 Dec 16 & 97.6 & 11.17 & $15.409_{-0.142}^{+0.142}$ & $-18_{-10}^{+10}$ & $26_{-10}^{+10}$ & -- & $<49$ & $<0.32$ & -- & $2.47$ \\
171205A & 2018 Jan 10 & 97.5 & 36.04 & $21.690_{-0.174}^{+0.174}$ & $8_{-8}^{+8}$ & $1_{-7}^{+7}$ & -- & $<24$ & $<0.11$ & -- & $-0.21$ \\
190114C$^*$ & 2019 Jan 14 & 97.5 & 0.11 & $10.700_{-0.026}^{+0.026}$ & $90_{-15}^{+15}$ & $32_{-15}^{+15}$ & -- & $96_{-15}^{+15}$ & $0.87_{-0.13}^{+0.13}$ & $9.8_{-4.6}^{+4.6}$ & $6.04$ \\
190114C$^*$ & 2019 Jan 14 & 97.5 & 0.15 & $8.600_{-0.030}^{+0.030}$ & $56_{-14}^{+14}$ & $-33_{-15}^{+15}$ & -- & $67_{-14}^{+14}$ & $0.76_{-0.16}^{+0.16}$ & $-15.4_{-6.6}^{+6.6}$ & $3.91$ \\
190114C$^*$ & 2019 Jan 14 & 97.5 & 0.20 & $7.040_{-0.024}^{+0.024}$ & $2_{-14}^{+14}$ & $-41_{-15}^{+15}$ & -- & $44_{-14}^{+14}$ & $0.60_{-0.19}^{+0.19}$ & $-43.7_{-11.7}^{+11.7}$ & $2.08$ \\
190114C & 2019 Jan 15 & 97.5 & 1.12 & $2.692_{-0.007}^{+0.007}$ & $14_{-6}^{+6}$ & $-24_{-6}^{+6}$ & $4_{-7}^{+7}$ & $28_{-6}^{+6}$ & $1.05_{-0.22}^{+0.22}$ & $-29.4_{-6.8}^{+6.4}$ & $4.09$ \\
190114C & 2019 Jan 16 & 97.5 & 2.06 & $1.904_{-0.008}^{+0.008}$ & $21_{-6}^{+6}$ & $-19_{-5}^{+5}$ & $4_{-7}^{+7}$ & $29_{-6}^{+6}$ & $1.53_{-0.30}^{+0.30}$ & $-20.8_{-5.9}^{+5.5}$ & $4.50$ \\
190114C & 2019 Jan 17 & 97.5 & 3.06 & $1.930_{-0.006}^{+0.006}$ & $8_{-6}^{+6}$ & $-1_{-6}^{+6}$ & $5_{-6}^{+6}$ & $<19$ & $<1.01$ & -- & $0.39$ \\
190114C & 2019 Jan 24 & 97.5 & 10.09 & $1.043_{-0.013}^{+0.013}$ & $7_{-12}^{+12}$ & $1_{-12}^{+12}$ & $-2_{-13}^{+13}$ & $<32$ & $<3.03$ & -- & $-1.05$ \\
190114C & 2019 Apr 11 & 97.5 & 86.97 & $0.042_{-0.005}^{+0.005}$ & $7_{-8}^{+8}$ & $2_{-8}^{+8}$ & -- & $<22$ & $<54.61$ & -- & $-0.33$ \\
190829A & 2019 Sep 01 & 97.5 & 2.50 & $5.859_{-0.006}^{+0.006}$ & $-28_{-4}^{+4}$ & $19_{-4}^{+4}$ & $3_{-5}^{+5}$ & $34_{-4}^{+4}$ & $0.58_{-0.07}^{+0.07}$ & $73.4_{-3.7}^{+3.7}$ & $7.34$ \\
191221B & 2019 Dec 22 & 97.5 & 0.48 & $3.563_{-0.006}^{+0.006}$ & $22_{-6}^{+6}$ & $10_{-5}^{+5}$ & $5_{-5}^{+5}$ & $24_{-5}^{+5}$ & $0.68_{-0.15}^{+0.15}$ & $12.2_{-6.1}^{+6.3}$ & $3.77$ \\
191221B & 2019 Dec 23 & 97.5 & 1.46 & $4.753_{-0.008}^{+0.008}$ & $-7_{-6}^{+6}$ & $-7_{-6}^{+6}$ & $-9_{-7}^{+7}$ & $<22$ & $<0.46$ & -- & $0.69$ \\
191221B$^*$ & 2019 Dec 24 & 97.5 & 2.52 & $3.949_{-0.007}^{+0.007}$ & $0_{-6}^{+6}$ & $-23_{-6}^{+6}$ & -- & $<25$ & $<0.64$ & -- & $3.22$ \\
191221B$^*$ & 2019 Dec 27 & 97.5 & 5.48 & $2.371_{-0.090}^{+0.090}$ & $0_{-7}^{+7}$ & $0_{-7}^{+7}$ & -- & $<42$ & $<1.79$ & -- & -- \\
210610B & 2021 Jun 11 & 97.5 & 0.30 & $2.235_{-0.006}^{+0.006}$ & $-1_{-5}^{+5}$ & $15_{-5}^{+5}$ & $-8_{-6}^{+6}$ & $<24$ & $<1.09$ & -- & $2.27$ \\
210702A & 2021 Jul 03 & 97.5 & 1.10 & $2.133_{-0.005}^{+0.005}$ & $-5_{-5}^{+5}$ & $9_{-5}^{+5}$ & $-9_{-5}^{+5}$ & $<20$ & $<0.94$ & -- & $1.17$ \\
220521A & 2022 Jul 17 & 97.5 & 56.12 & $0.064_{-0.006}^{+0.006}$ & $-4_{-5}^{+5}$ & $3_{-6}^{+6}$ & -- & $<15$ & $<24.53$ & -- & $-0.45$ \\
220521A & 2022 Aug 12 & 97.5 & 82.09 & $0.031_{-0.007}^{+0.007}$ & $7_{-6}^{+6}$ & $-6_{-6}^{+6}$ & -- & $<21$ & $<78.31$ & -- & $0.45$ \\
220921A & 2022 Sep 23 & 97.5 & 1.89 & $4.579_{-0.006}^{+0.006}$ & $-1_{-6}^{+6}$ & $9_{-6}^{+6}$ & $3_{-6}^{+6}$ & $<20$ & $<0.44$ & -- & $0.51$ \\
220921A & 2022 Sep 28 & 97.5 & 6.75 & $1.556_{-0.006}^{+0.006}$ & $28_{-6}^{+6}$ & $24_{-6}^{+6}$ & $-2_{-6}^{+6}$ & $38_{-6}^{+6}$ & $2.44_{-0.39}^{+0.39}$ & $20.4_{-4.8}^{+4.8}$ & $5.60$ \\
221009A & 2022 Oct 12 & 145.0 & 3.39 & $6.961_{-0.011}^{+0.011}$ & $10_{-7}^{+7}$ & $-98_{-6}^{+6}$ & $22_{-7}^{+7}$ & $99_{-6}^{+6}$ & $1.42_{-0.09}^{+0.09}$ & $-42.2_{-2.0}^{+2.0}$ & $15.91$ \\
221009A & 2022 Oct 14 & 97.5 & 5.32 & $5.865_{-0.007}^{+0.007}$ & $-19_{-7}^{+7}$ & $-93_{-7}^{+7}$ & $13_{-7}^{+7}$ & $95_{-7}^{+7}$ & $1.62_{-0.12}^{+0.12}$ & $-50.8_{-2.1}^{+2.1}$ & $13.51$ \\
250327B & 2025 Mar 29 & 97.5 & 1.20 & $0.530_{-0.005}^{+0.005}$ & $6_{-6}^{+6}$ & $8_{-5}^{+5}$ & $-13_{-5}^{+5}$ & $<20$ & $<3.72$ & -- & $1.09$ \\
251013C & 2025 Oct 21 & 97.5 & 8.24 & $2.903_{-0.006}^{+0.006}$ & $-24_{-7}^{+7}$ & $1_{-6}^{+6}$ & $7_{-6}^{+6}$ & $25_{-6}^{+6}$ & $0.85_{-0.22}^{+0.22}$ & $77.9_{-9.1}^{+9.1}$ & $2.97$ \\
\hline
\end{tabular}
\tablenotetext{*}{Reported values are from literature: 171205A, \citealt{lhc20}; 190114C, \citealt{lag+19}; and 191221B, \citealt{utc+23}. The first epoch of observations of GRB\,171205A were shown to contain strong instrumental systematics and we use an upper limit of $\Pi<0.30\%$ for this epoch \citep{lhc20}. For GRB\,191221B, the published upper limits do not include $QU$ forced photometry \citep{utc+23} and these are reported here with central values of zero and an uncertainty corresponding to the reported image RMS.}
\label{tab:grb_pol}
\end{table*}

\label{sec:reduction}
We searched the ALMA Science Archive for observations with ``GRB'' in the source name and selected those obtained in Full Stokes mode. After excluding unrelated observations of Sgr~B2, our sample comprises 24 individual observations of 11 GRB afterglows. Details of the observations are given in Table~\ref{tab:obs}. To the best of our knowledge, this represents essentially the complete ALMA polarimetric sample of GRB afterglows currently available.
We reduce the ALMA observations using a custom, semi-interactive calibration and imaging pipeline (Appendix~\ref{sec:obspipe}) built using existing CASA tasks. While the standard ALMA polarization calibration pipeline is highly effective for rapidly producing science-ready products in a non-interactive manner, it does not readily generate several auxiliary products required for the instrumental systematic tests (Appendix~\ref{sec:sys-checks}), and its non-interactive structure can make it difficult to monitor or interrupt processing when issues arise. Our pipeline is designed to facilitate both efficient calibration and imaging and the simultaneous evaluation of instrumental systematics through the automated production of diagnostic plots and auxiliary files. All code used in this pipeline is publicly available on Zenodo \citep{tanmoy_laskar_2026_20836999} and GitHub\footnote{\url{https://github.com/alexander-group/almapol}}.

The calibration steps are described in detail in Appendix~\ref{sec:obspipe}. In summary, our reduction pipeline comprises three steps: pre-calibration, parallel hand calibration, and polarization calibration. In pre-calibration, we convert the raw ASDM into CASA measurement sets, perform flagging, and generate and apply system temperature $T_{\rm sys}$ and atmospheric opacity corrections. This is followed by parallel-hand calibration, where we set the flux-density scale, calibrate the frequency-dependent bandpass and time-dependent complex gain, derive the flux density scaling, and apply these calibrations to the calibrator fields. Finally, in the polarization calibration step, we solve for the complex gain of the leakage calibrator and then derive the cross-hand delays, residual cross-hand phase, and the intrinsic polarization of the leakage and gain calibrators. We then revise the gain calibration using the full polarization model of the gain calibrator, solve for the leakage (D-terms), derive the XY gain amplitude, and apply all calibrations to the target field. Our pipeline produces a fully polarization-calibrated measurement set, along with a series of auxiliary files and plots for quality assurance and error diagnosis. 

We create and deconvolve images of each target field using \texttt{tclean} in CASA, employing Briggs weighting with robust$=2$ (natural weighting) for maximum point-source sensitivity, a cell size corresponding to $\lambda_{\rm ref}/(8B_{\rm max})$, where $\lambda_{\rm ref}$ is the reference wavelength of the dataset and $B_{\rm max}$ is the length of the longest baseline in wavelengths, and using the \texttt{clarkstokes} deconvolver to clean each Stokes plane separately. Our images yield strong detections in Stokes $I$ (total intensity) for all targets. Unless otherwise noted, we perform phase-only self-calibration on each target using a solution interval longer than the minimum solution interval (see Equation \ref{eq:solint_min}) setting \texttt{gaintype=`T'} to avoid introducing polarization-dependent systematics in the data (see Section~\ref{sec:sys-checks}, points 6 and 7). 
We obtain flux densities and associated uncertainties by fitting an elliptical Gaussian fixed to the size of the synthesized beam using the CASA task \verb'imfit'. For Stokes $QUV$ images, we fix the position of the elliptical Gaussian to the source location inferred from the point-source fit to the Stokes $I$ image. In Figure \ref{fig:imgs}, we show the Stokes $I$, $Q$, $U$, and linearly polarized flux density, $P_L=\sqrt{Q^2+U^2}$ images for all 20 individual observations in our sample. Statistically significant ($\gtrsim3\sigma$) detections in any Stokes parameter are labeled by stars inside the beam ellipse in the corresponding sub-image.

\newcommand{\figwidth}{0.97}
\begin{figure*}[p!]
    \centering
    {\includegraphics[width=\figwidth\textwidth]{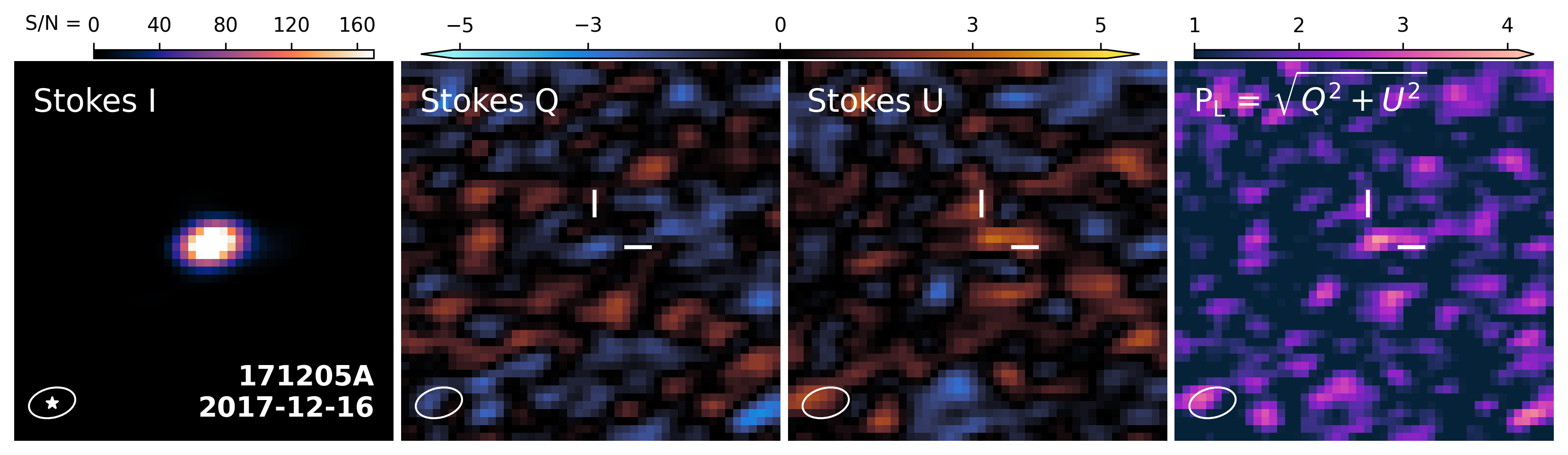}} \vfill
    {\includegraphics[width=\figwidth\textwidth]{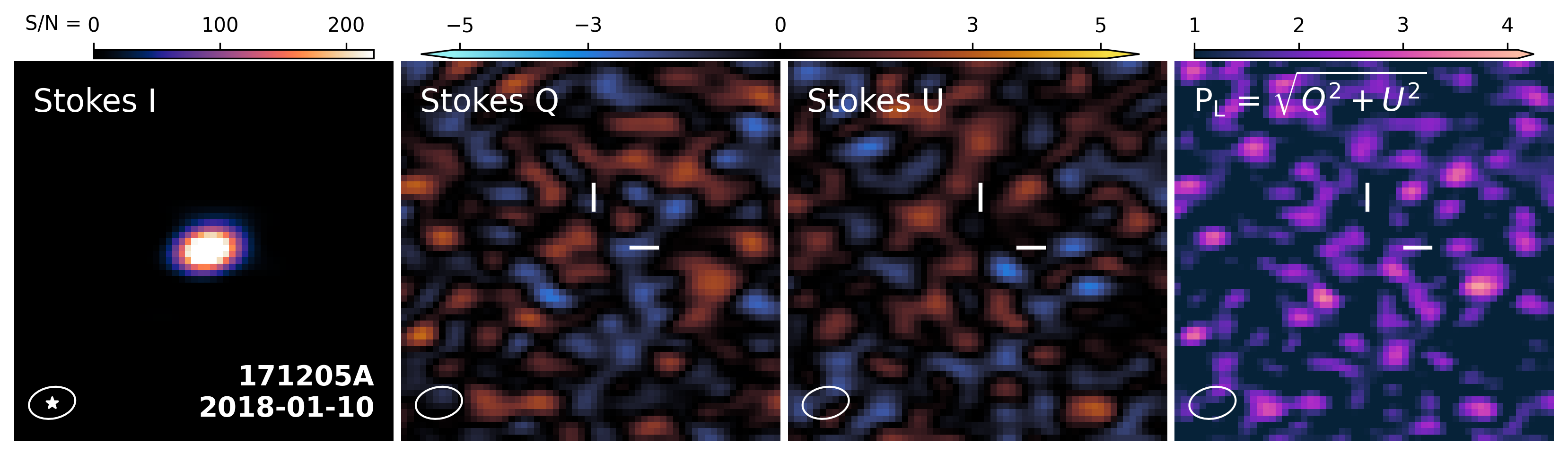}} \vfill
    {\includegraphics[width=\figwidth\textwidth]{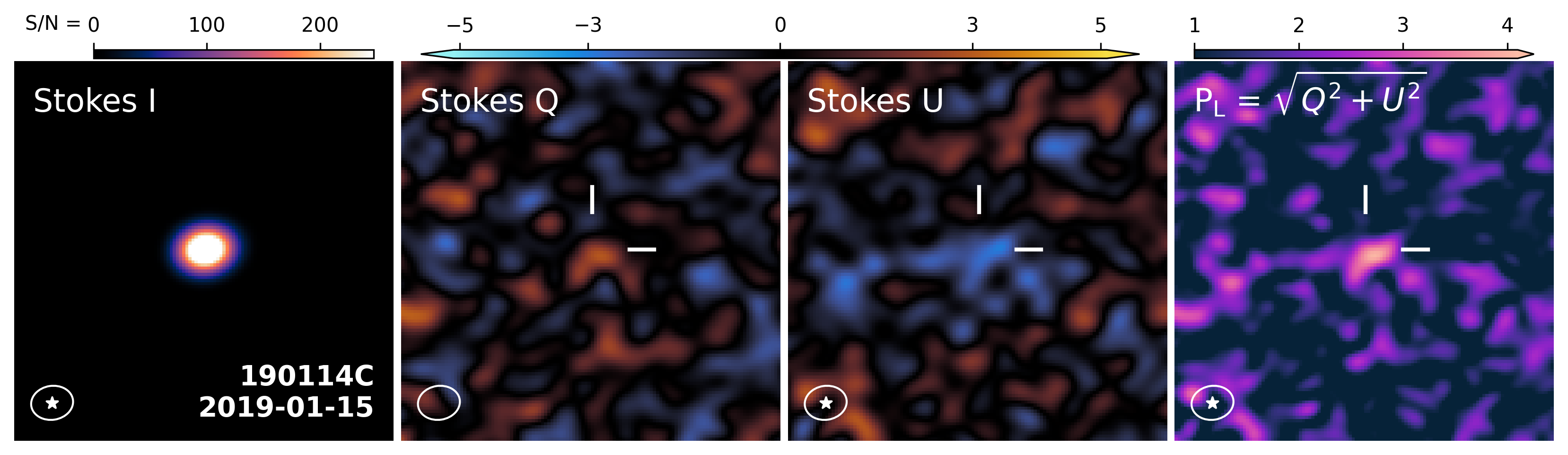}} \vfill
    {\includegraphics[width=\figwidth\textwidth]{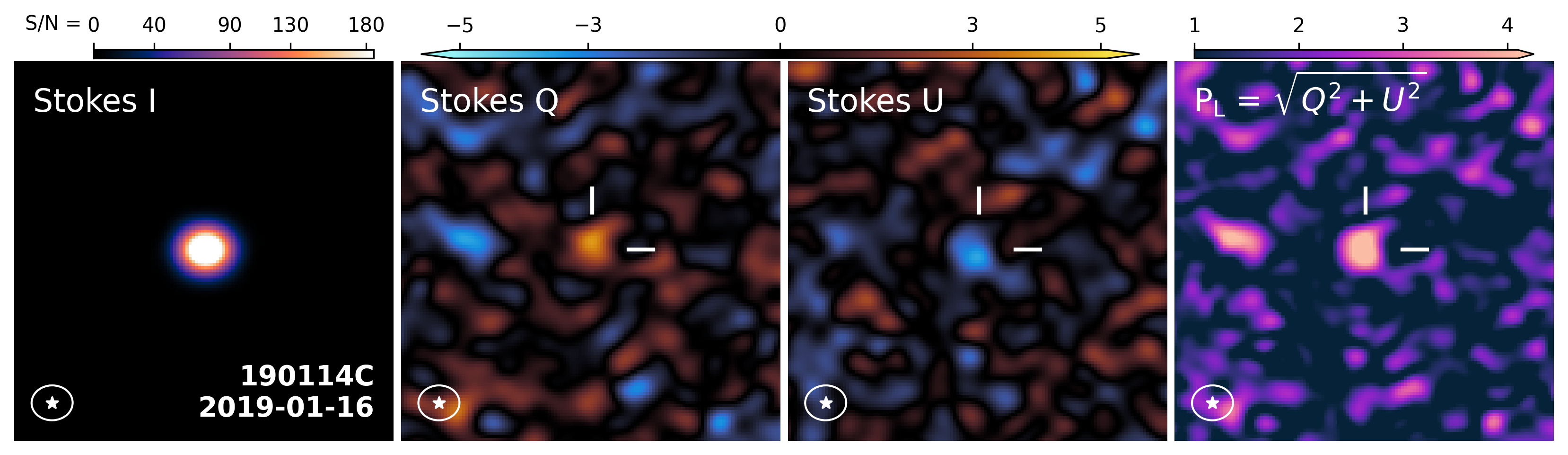}} \vfill
    \caption{Stokes $IQU$, and linear polarization, $P_L=\sqrt{Q^2+U^2}$ images for the GRBs in our sample (see Table~\ref{tab:grb_pol} for the corresponding measurements). Statistically significant ($\gtrsim3\sigma$) detections for a given Stokes parameter or $P_L$ (see Table~\ref{tab:grb_pol}) are denoted by a star in the center of the beam. The color axis indicates the S/N ratio.}
    \label{fig:imgs}
\end{figure*}
\begin{figure*}[p!]
    \ContinuedFloat
    \centering
    {\includegraphics[width=\figwidth\textwidth]{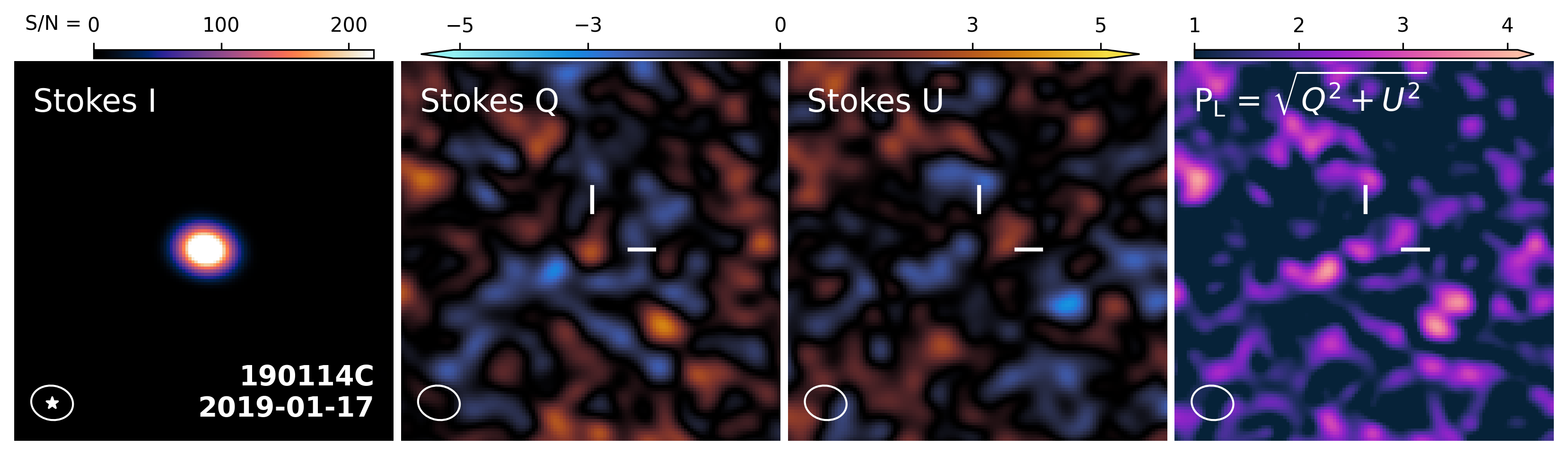}} \vfill
    {\includegraphics[width=\figwidth\textwidth]{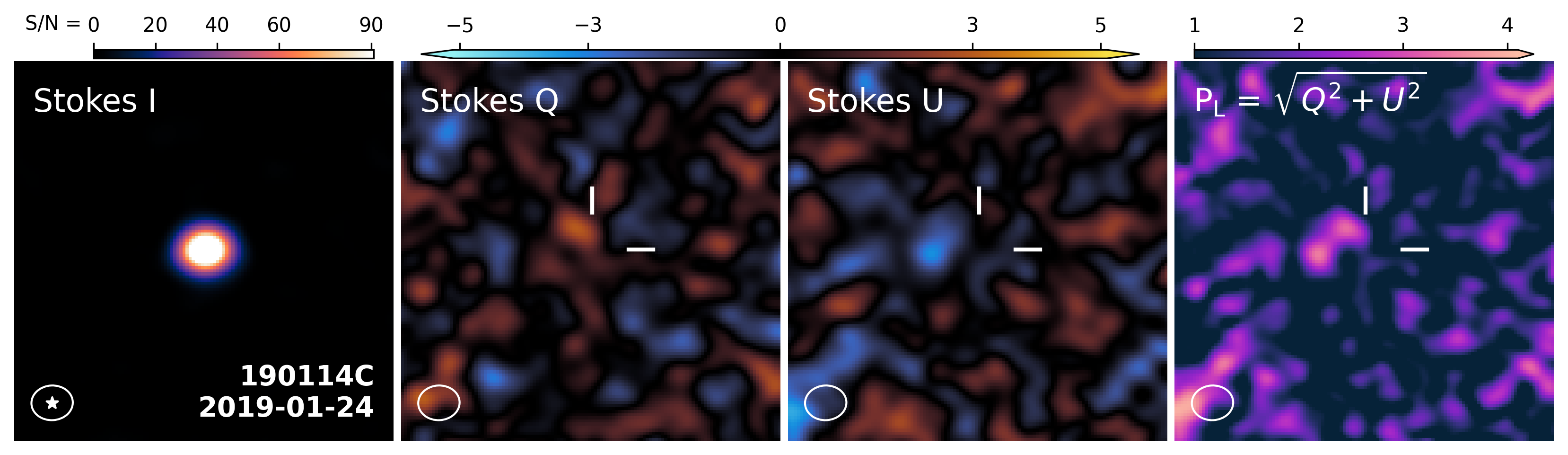}} \vfill
    {\includegraphics[width=\figwidth\textwidth]{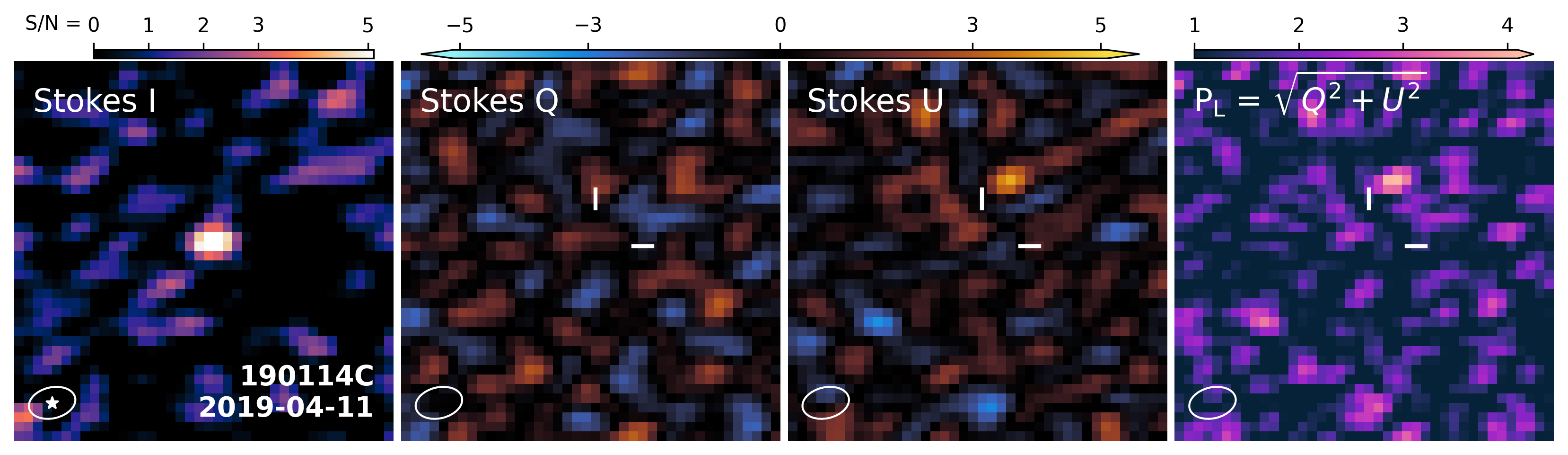}} \vfill
    {\includegraphics[width=\figwidth\textwidth]{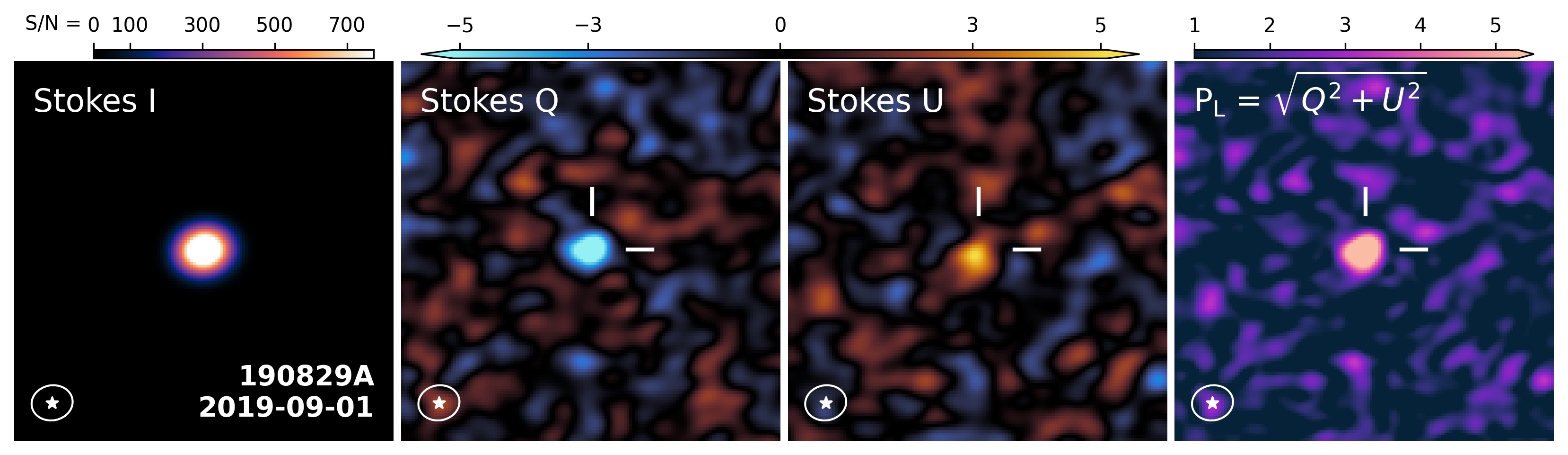}} \vfill
    \caption[]{\emph{Continued.}}
\end{figure*}
\begin{figure*}[p!]
    \ContinuedFloat
    \centering
    {\includegraphics[width=\figwidth\textwidth]{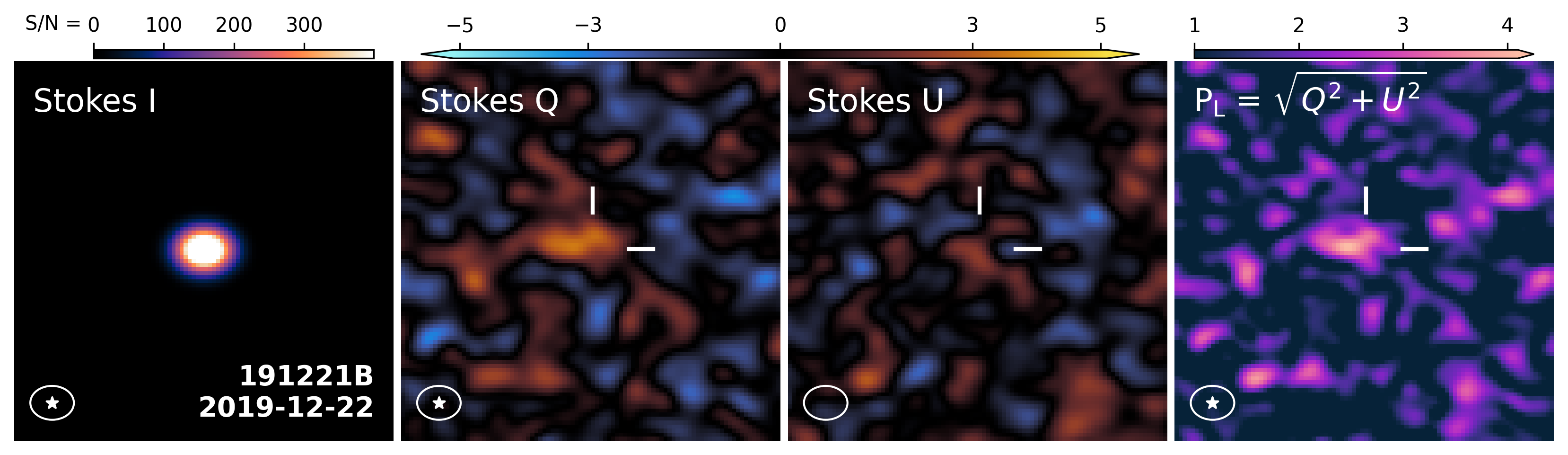}} \vfill
    {\includegraphics[width=\figwidth\textwidth]{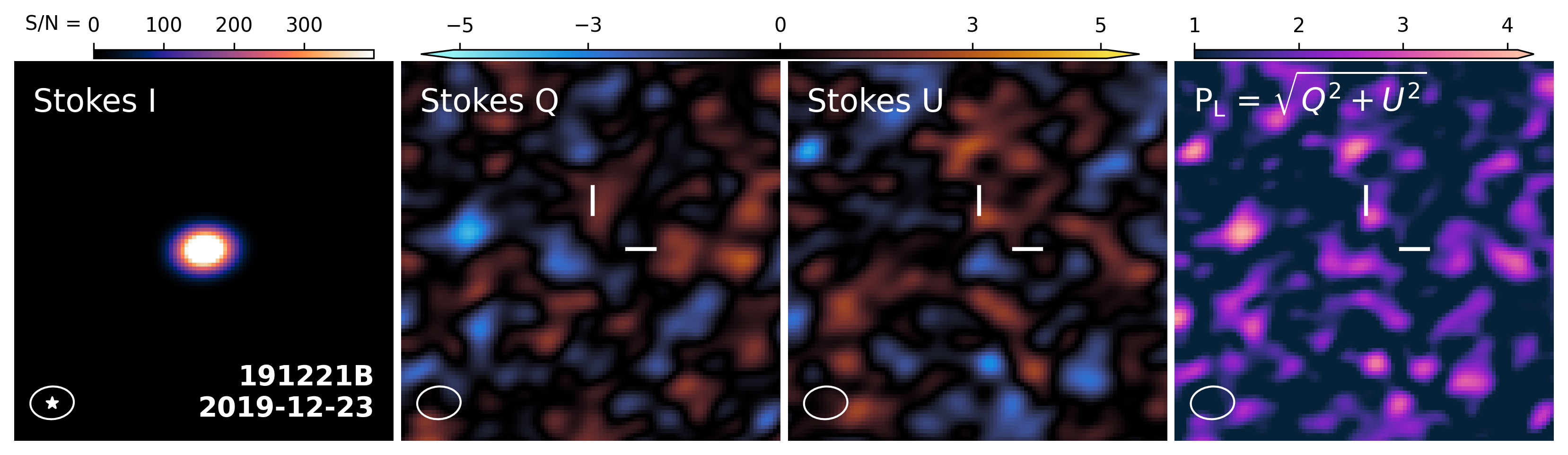}} \vfill  
    {\includegraphics[width=\figwidth\textwidth]{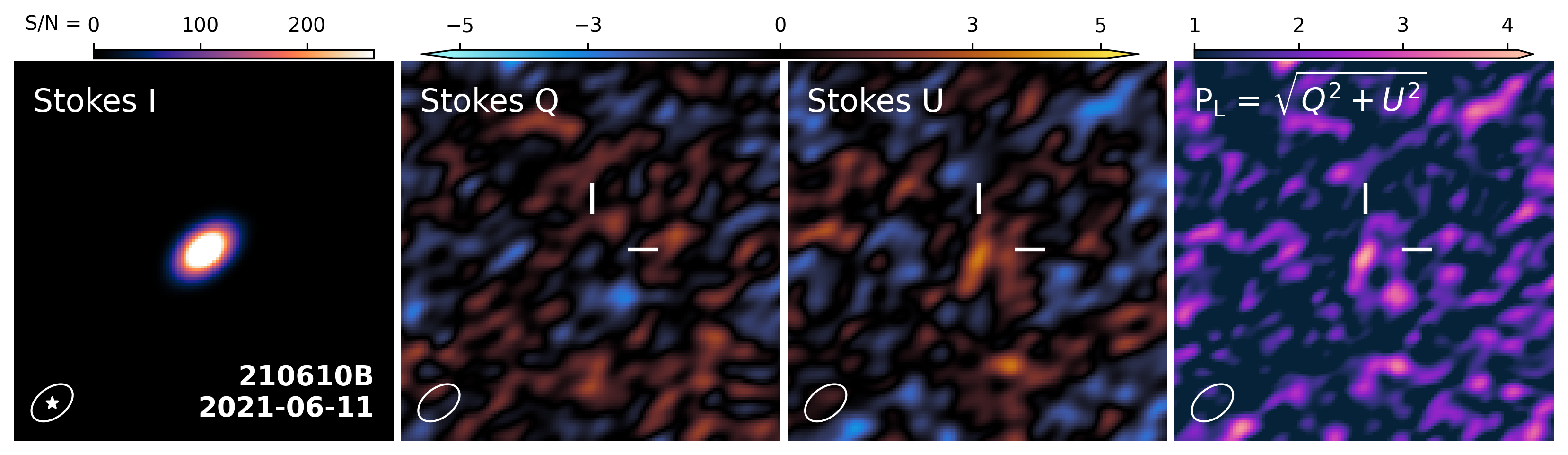}} \vfill
    {\includegraphics[width=\figwidth\textwidth]{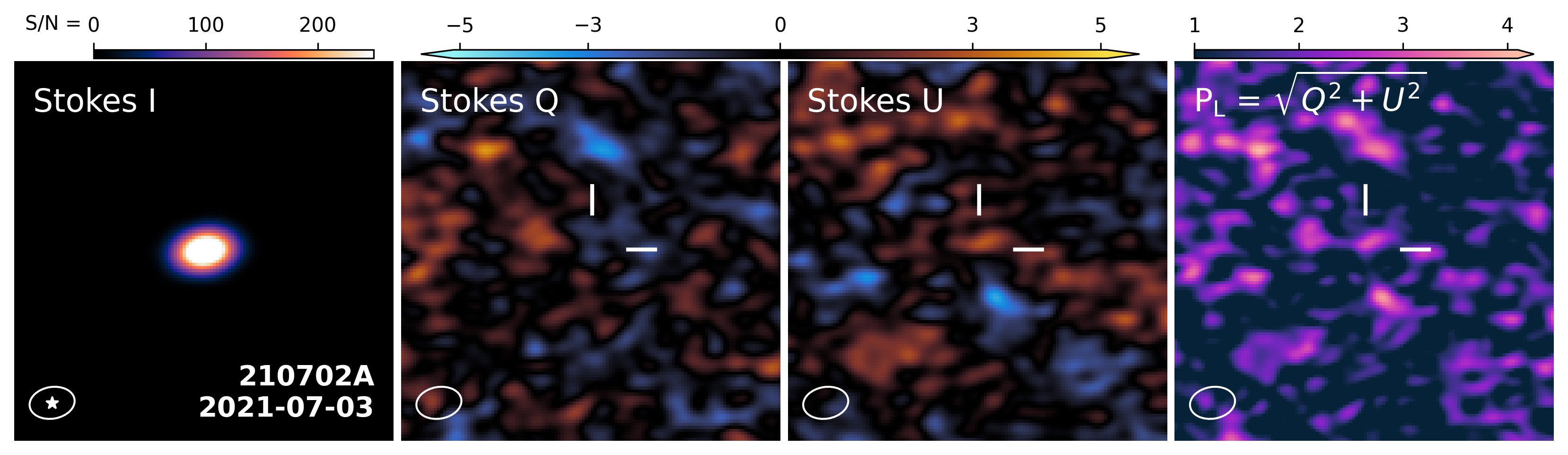}} \vfill
    \caption[]{\emph{Continued.}}
\end{figure*}
\begin{figure*}[p!]
    \ContinuedFloat
    \centering
    {\includegraphics[width=\figwidth\textwidth]{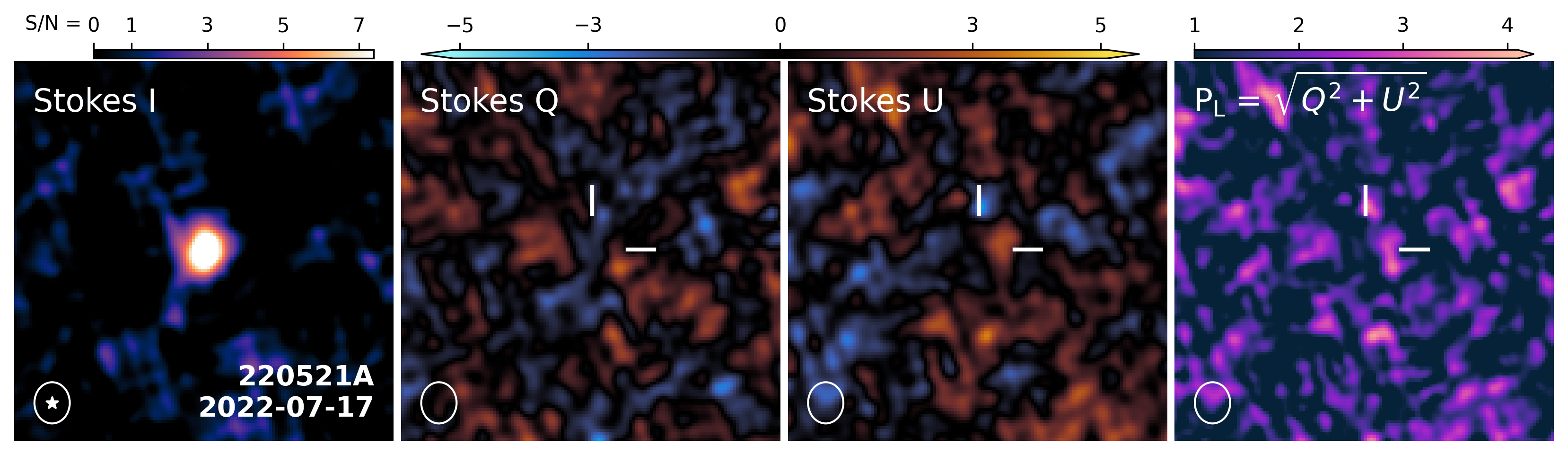}} \vfill
    {\includegraphics[width=\figwidth\textwidth]{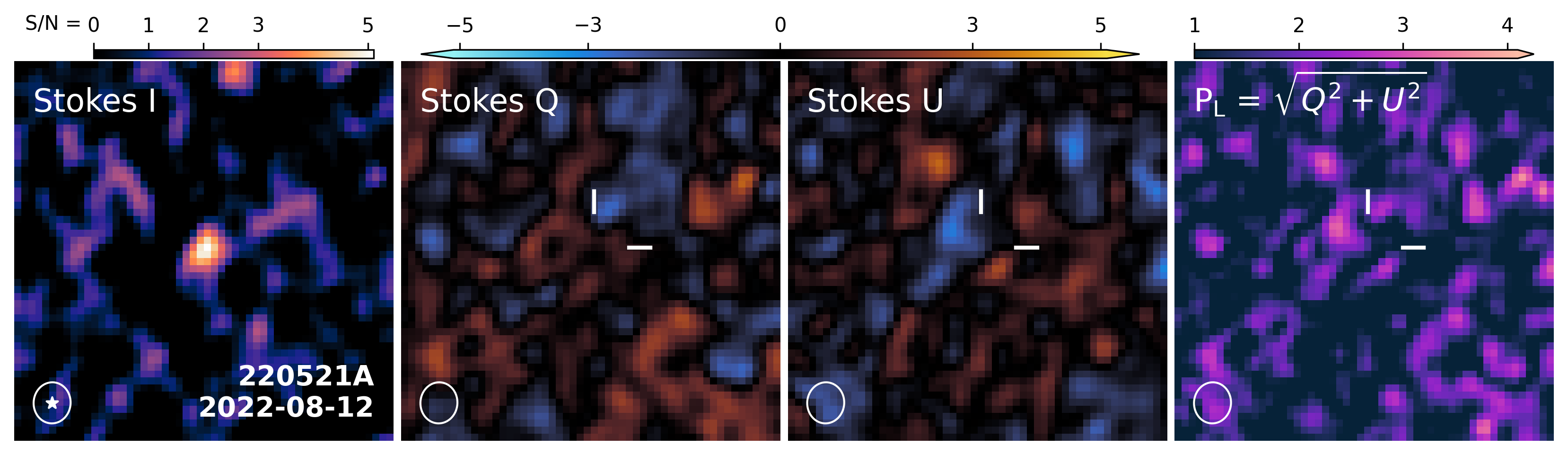}} \vfill
    {\includegraphics[width=\figwidth\textwidth]{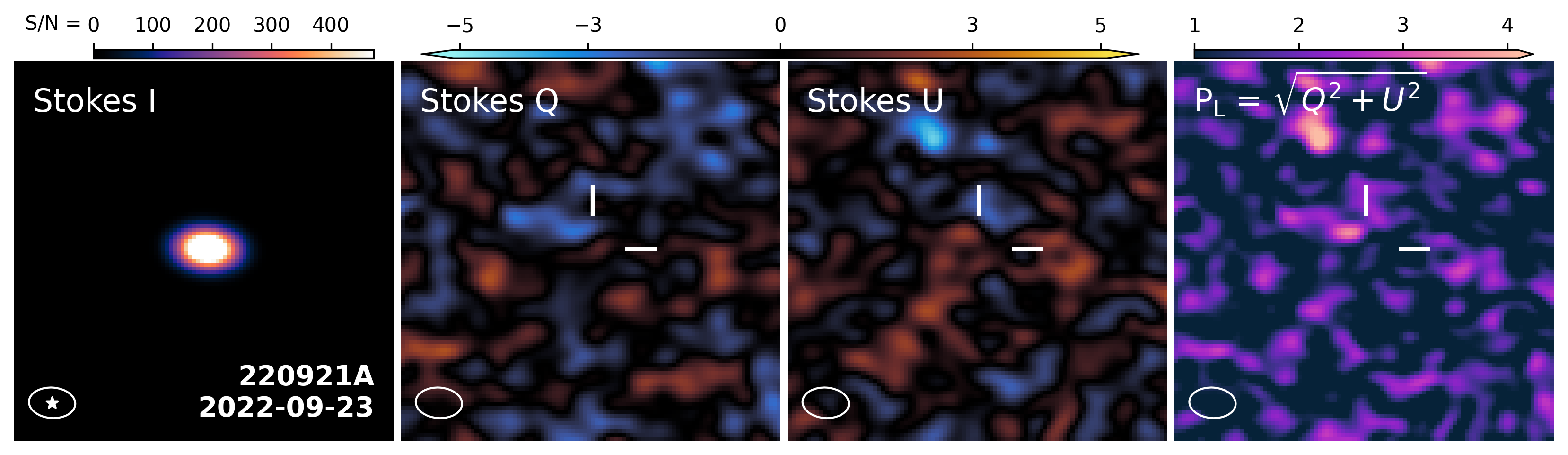}} \vfill
    {\includegraphics[width=\figwidth\textwidth]{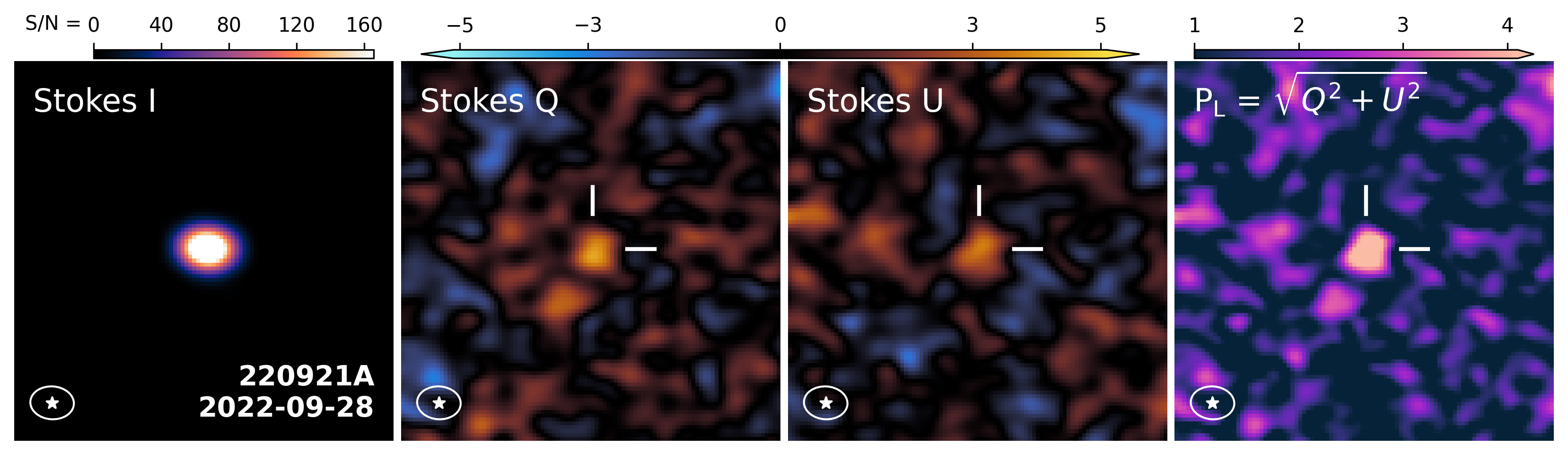}} \vfill  
    \caption[]{\emph{Continued.}}
\end{figure*}
\begin{figure*}[p!]
    \ContinuedFloat
    \centering
    {\includegraphics[width=\figwidth\textwidth]{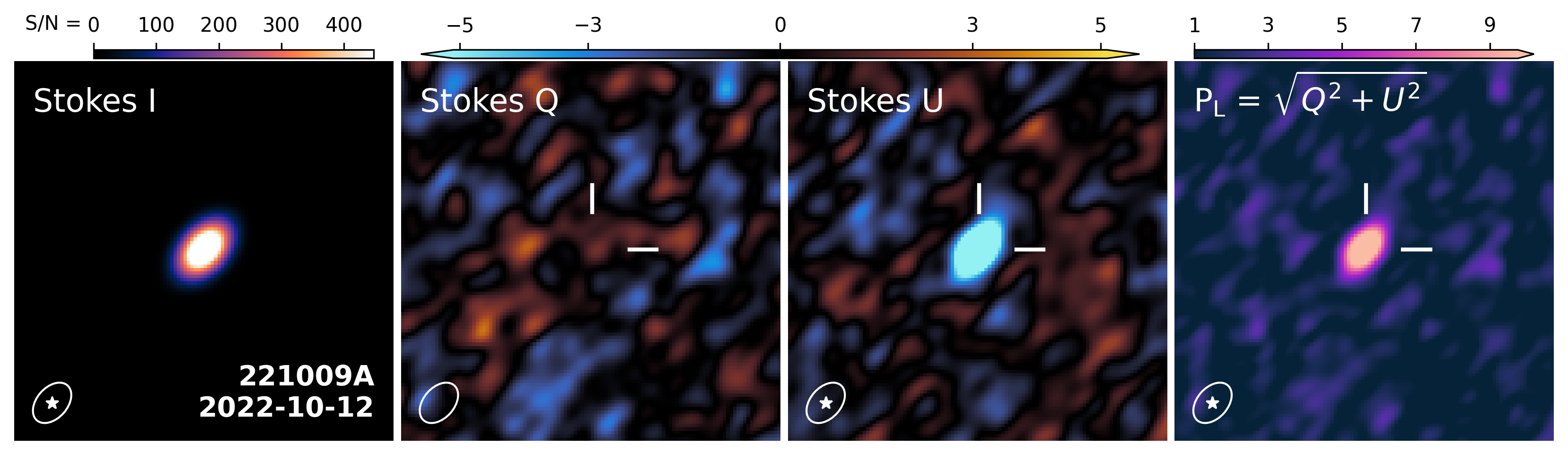}} \vfill
    {\includegraphics[width=\figwidth\textwidth]{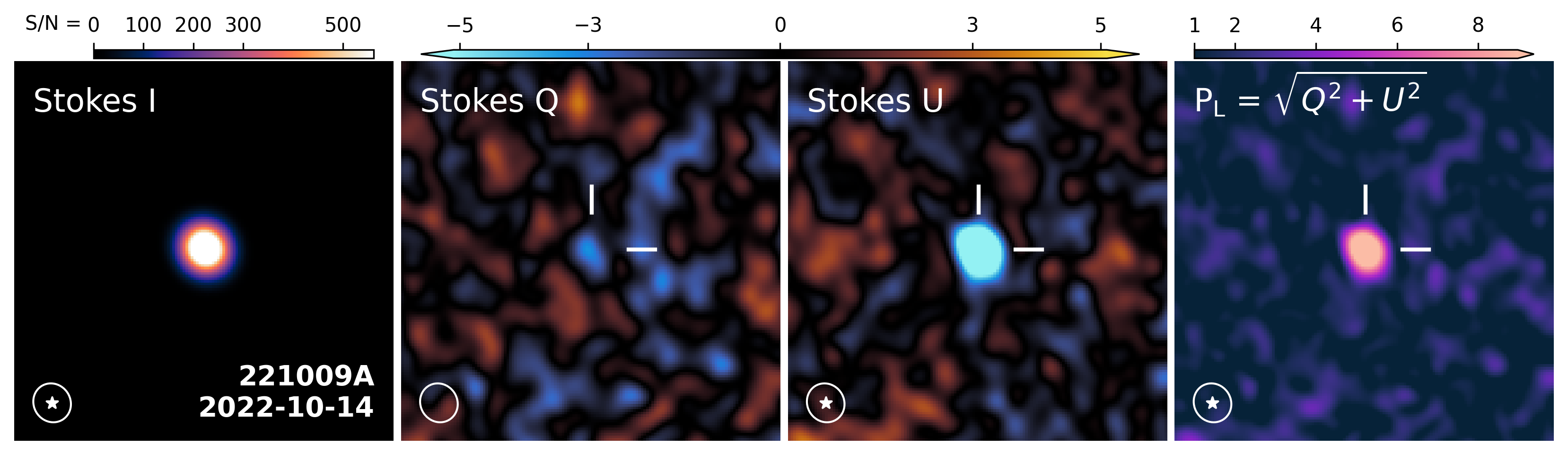}} \vfill
    {\includegraphics[width=\figwidth\textwidth]{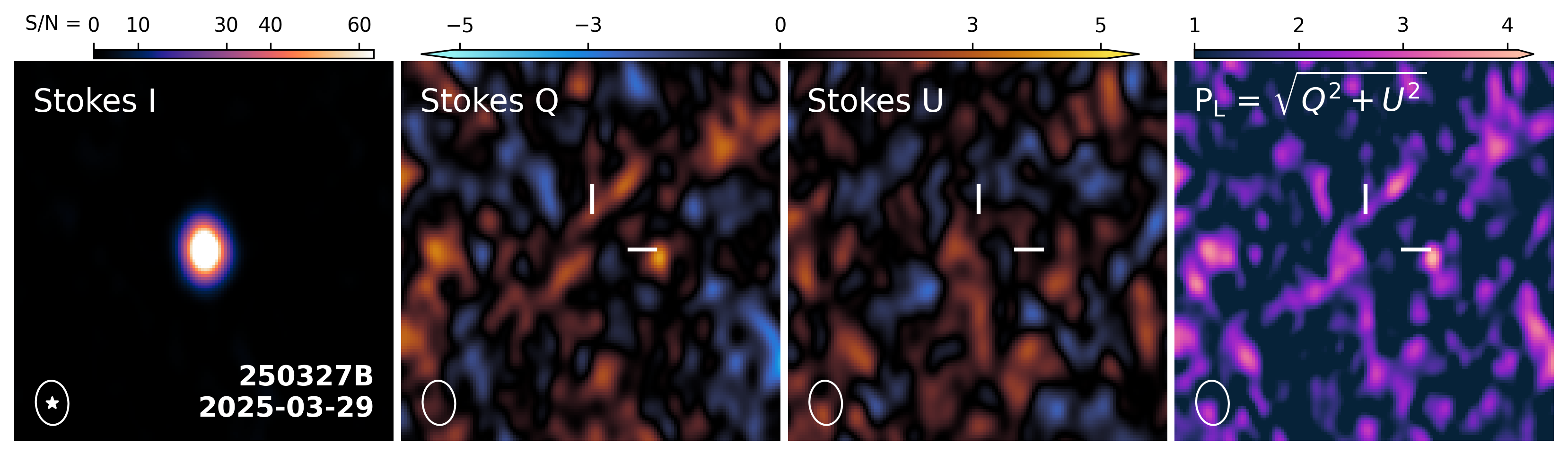}} \vfill
    {\includegraphics[width=\figwidth\textwidth]{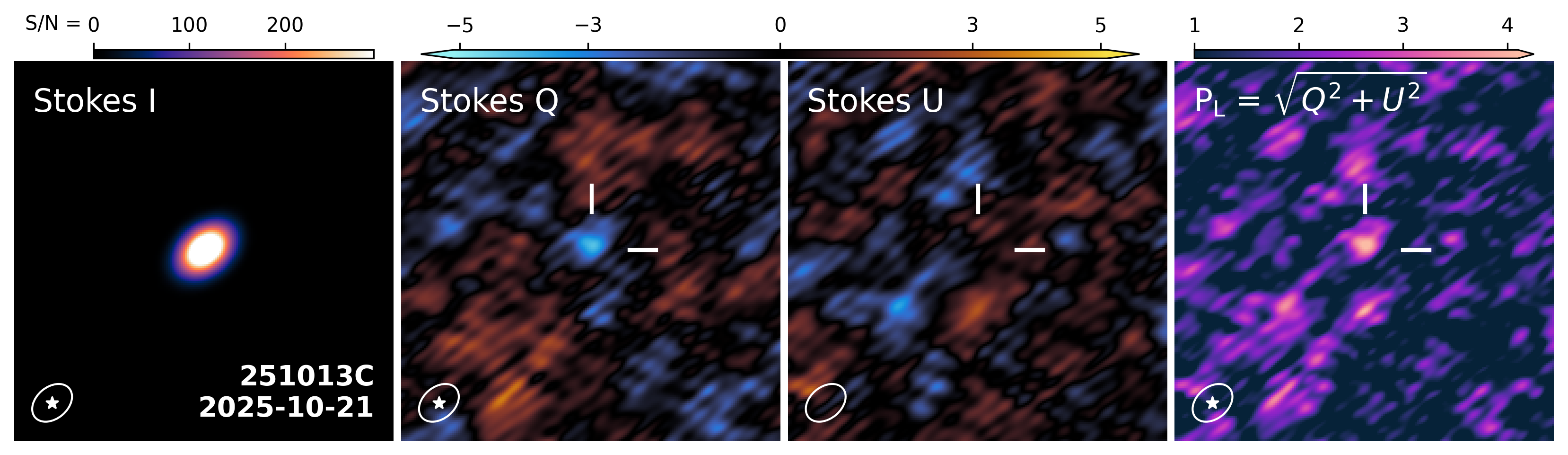}} \vfill   
    \caption[]{\emph{Continued.}}
\end{figure*}
%\clearafterfulltwocolumnpage

\section{Results} \label{sec:results}
Our 20 new ALMA polarimetric observations of 11 GRBs span $0.3$--87\,days post trigger (Table~\ref{tab:grb_pol}). In addition to the Stokes $I$  detections in each case, we detect significant ($\gtrsim3\sigma$) polarized emission in either Stokes $Q$ or $U$ in 8 observations across 6 events: GRBs\,190114C (1.12 and 2.06\,days), 190829A (2.5\,days), 191221B (0.48\,days), 220921A (6.8\,days), 221009A (3.4 and 5.3\,days), 251013C (8.2\,days). 
We test for the robustness of these linear polarization detections and upper limits against potential instrumental systematics \citep{lhc20}. The details of this framework are discussed in Appendix~\ref{sec:sys-checks}. In cases where the tests reveal potential data quality issues, our framework allows us to isolate these in the $uv$ data, flag the relevant issue (e.g., bad antennas or channels) and repeat the data reduction until all reductions pass all quality checks. For the observations with polarization detections, we also test for time- and frequency- dependent polarization systematics in both the target and gain-calibrator observations (see Figures~\ref{fig:polsys-timevar} in Appendix~\ref{sec:sys-checks}). 
All detections reported in this work pass all systematic tests. We note that several of our measurements yield a significant detection in either $Q$ or $U$ but not both; however, a Monte Carlo test confirms that this is expected for $QU$ measurements with large error bars drawn from a sample spanning a random distribution of polarization angles. 

Since $\PL$ is a positive-definite quantity, the statistical significance of polarized intensity measurements must be evaluated carefully \citep{wk74,vai06,hp15}. To assess the significance of a polarization detection, we treat the measured Stokes parameters $Q$ and $U$ as independent Gaussian random variables with uncertainties $\sigma_Q$ and $\sigma_U$. The test statistic,  
\begin{align}
T=\frac{Q^2}{\sigma_Q^2}+\frac{U^2}{\sigma_U^2},
\end{align}
follows a $\chi^2$ distribution with two degrees of freedom under the null hypothesis of zero intrinsic polarization. We quantify the significance of the polarization detection  using the tail probability of the $\chi^2$ distribution, 
\begin{align}
p_{\rm tail} = P\!\left(\chi^2_2 \ge T\right)=
\exp\left(-\frac{T}{2}\right), 
\end{align}
for two degrees of freedom and convert $p_{\rm tail}$ into a Gaussian-equivalent detection significance using the one-sided Gaussian significance, 
\begin{align}
Z = \Phi^{-1}(1-p_{\rm tail}),
\end{align}
where $\Phi^{-1}$ is the inverse cumulative distribution function of the standard normal distribution. 

\begin{figure}
    \centering
    \includegraphics[width=\linewidth]{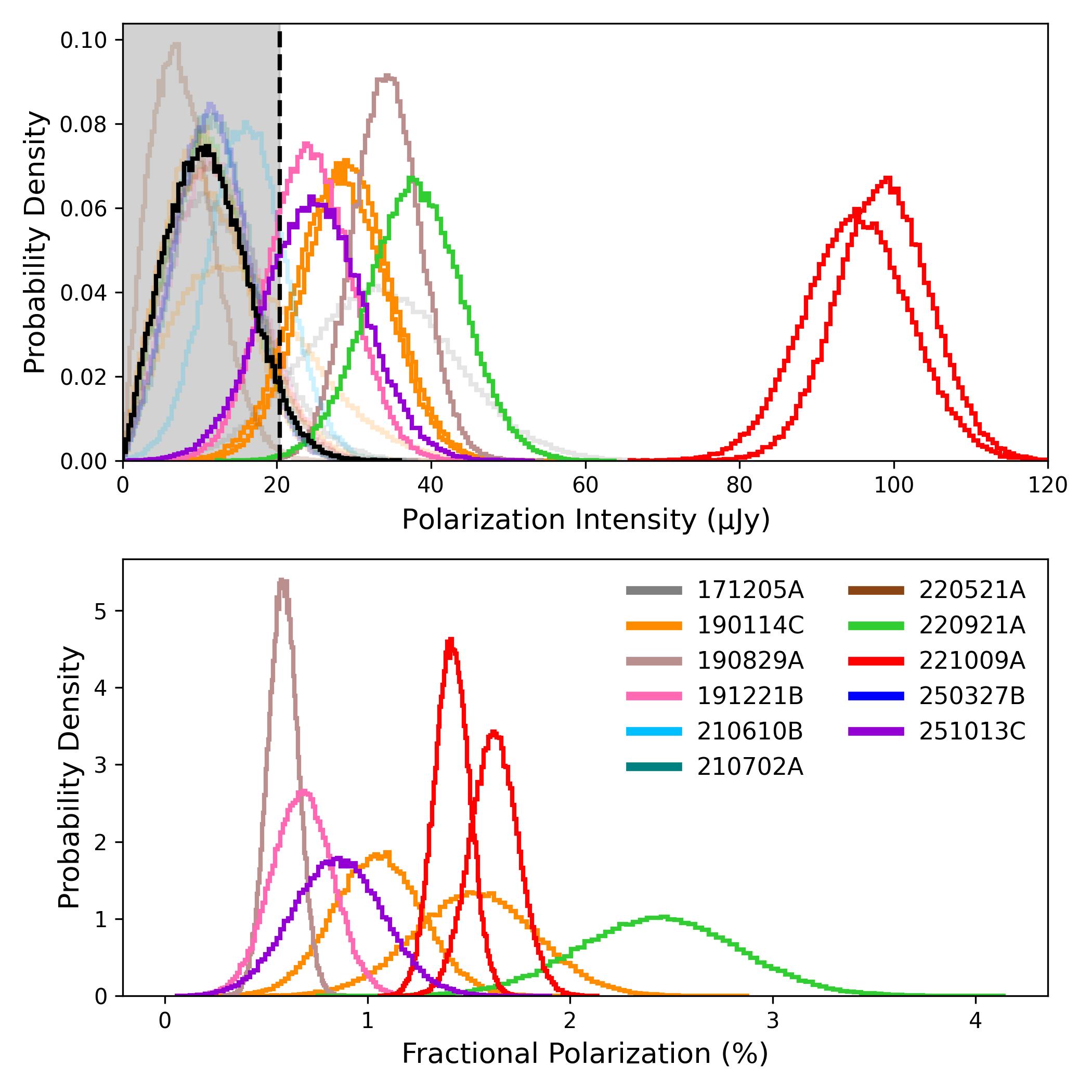}
    \caption{Distribution of polarized flux density (top panel) and polarization fraction (bottom panel) derived from Monte Carlo realizations of the measured Stokes $IQU$ parameters assuming Gaussian uncertainties. For polarization detections, we report the median and 68\% confidence intervals, while for non-detections we report 95\% upper limits based on the corresponding posterior. The black curve shows the distribution obtained from Monte Carlo realizations using the average noise level in the Stokes $Q$ and $U$ images, and the shaded region indicates the area below its 95\% quantile.}
    \label{fig:P_monte_carlo}
\end{figure}

\begin{figure}
    \centering    \includegraphics[width=\linewidth]{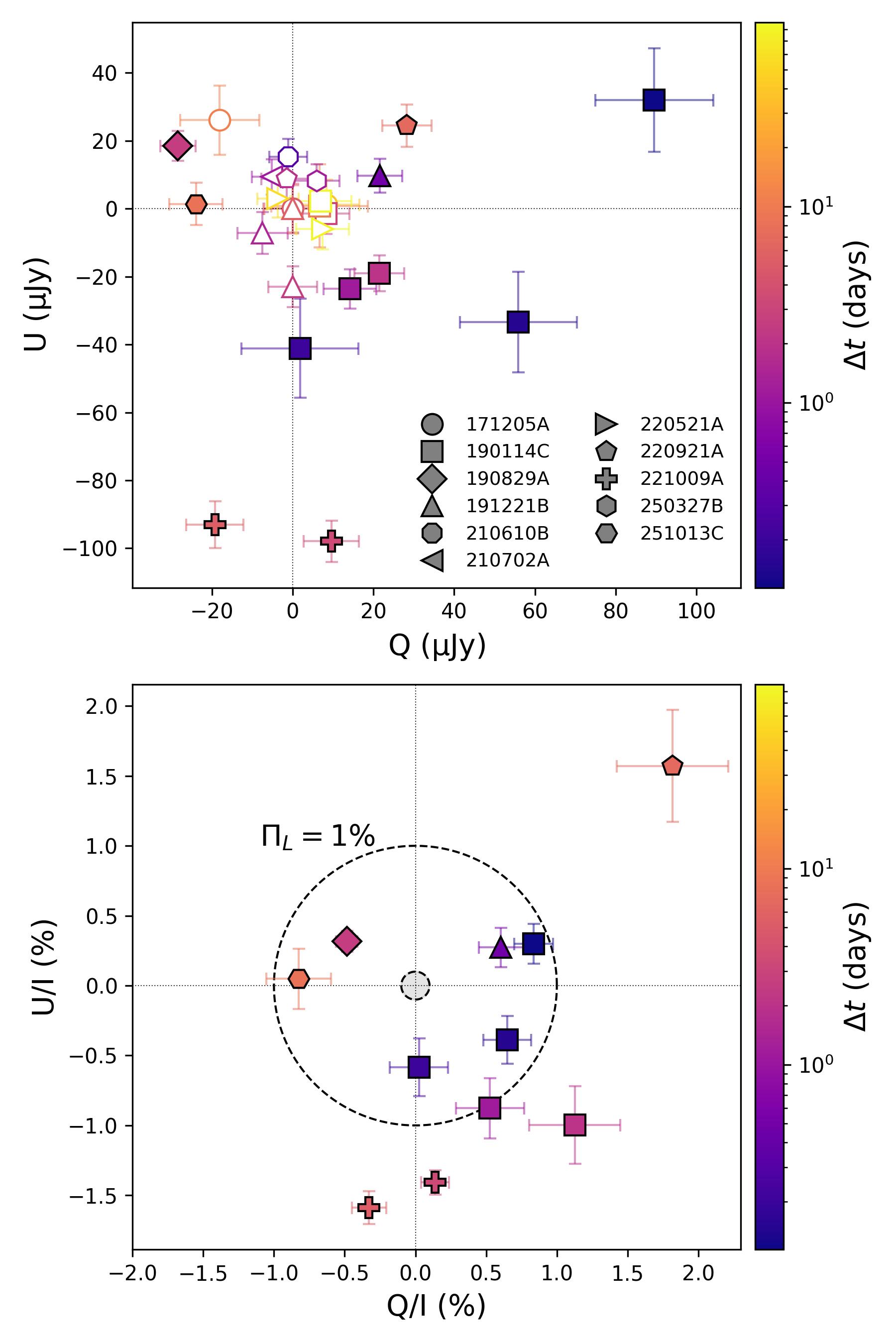}
    \caption{Stokes $U$ vs $Q$ measurements for all GRBs in our sample (top panel) and $U/I$ vs $Q/I$ for the detected sources (bottom panel). Filled markers in both panels denote polarization detections ($Z\gtrsim3$), open markers indicate non-detections,  data points are color-coded by time since burst, and error bars represent the $1\sigma$ uncertainties. Non-detections cluster near $(Q,U)=(0,0)$, consistent with noise. In the lower panel, the inner and outer dashed circles indicate $\Pi_L=0.1\%$ and $\Pi_L=1\%$, respectively.} 
    \label{fig:QU}
\end{figure}

\begin{figure}
    \centering
    \includegraphics[width=\linewidth]{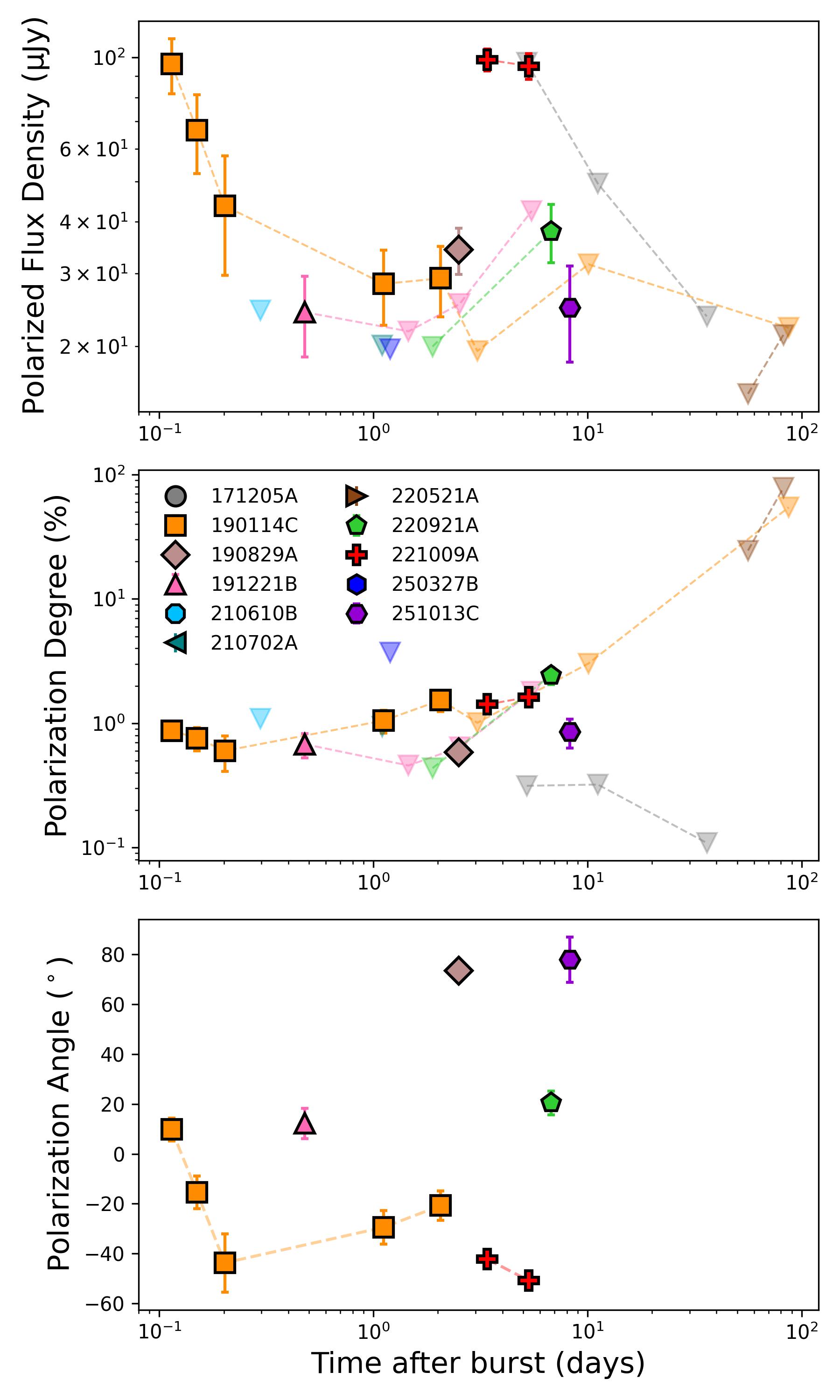}
    \caption{Linear polarized flux density ($\PL$, top panel), linear polarization degree ($\PiL$, middle panel), and polarization position angle ($\chi$, bottom panel) as a function of time since the burst in days. Non-detections are shown as downward-pointing triangles corresponding to 95\% upper limits. Multiple epochs for the same event are connected by dashed lines to guide the eye.\\
    }
    \label{fig:pol_lightcurves}
\end{figure}

We report the Gaussian-equivalent significance in each case in Table~\ref{tab:grb_pol}. This procedure indicates a linear polarized intensity significantly ($\gtrsim3\sigma$) different from zero in 7 of the above 8 observations; we find a marginal ($2.97\sigma$) detection for GRB\,251013C, which we consider together with the detections in the rest of this work, but note that this could be impacted, e.g., by non-Gaussian noise in the image. 

To derive an estimate for the linear polarized flux, $\PL$, and the linear polarization fraction, $\PiL=P/I$ for the polarization detections, we use Monte Carlo realizations of the measured $IQU$ Stokes parameters using their associated uncertainties and a Gaussian error model, and calculate $\PL$ and $\PiL$ in each case (Figure~\ref{fig:P_monte_carlo}). We also calculate the polarization position angle, $\chi=\frac12\tan^{-1}(U/Q)$ for each detection (see Figure~\ref{fig:QU} for a plot of $U$ vs $Q$). We report the median and 68\% credible interval for $\PL$, $\PiL$, and $\chi$ in Table~\ref{tab:grb_pol}. For the non-detections, we compute upper limits on $\PL$ and $\PiL$ by computing 95\% quantiles on their respective posteriors using the Monte Carlo realizations. 

Our analysis reveals a range of $\PiL$ spanning $\sim0.5$-$2.5\%$. For the non-detections, we constrain $\PiL\lesssim4\%$ in all cases except for both epochs of GRB\,220521A and the final epoch of GRB\,190114C, which have low ($\lesssim0.1$\,mJy) Stokes $I$ flux density. In the remaining seven cases, we obtain deep upper limits, $\PiL\lesssim1.1\%$ (most notably GRB\,171205A, with $\lesssim0.11\%$ at $\approx36$~days). We note that these uncertainties are statistical only, which dominate over the expected typical $1\sigma$ systematic uncertainty \citep{nnp+16}. For observations with strong ($Z>5$) detections, we also divide the calibrated target data in frequency and image the lower sideband (LSB) and upper sideband (USB) separately. For these observations, we report the Stokes parameters, polarized intensity, polarization fraction, and position angle for each frequency bin in Table~\ref{tab:sidebands}.

Six of the sources in our sample were observed at multiple epochs and, of these, the two observations of GRB\,221009A were also made at different frequencies (145\,GHz and 97.5\,GHz). This temporal sampling allows us to probe the time dependent behavior of the polarization in a few cases. For example, GRB\,190114C was previously found to exhibit significant polarization at $0.1$--0.2\,days; here, we extend that temporal baseline to $\approx87$\,days and find that statistically significant linear polarization continues to be detected in this event at the $\approx1\%$ level until at least $t\approx2.1$\,days. In the case of GRB\,220921A, we find strong evidence for late-rising polarized emission, with $\PiL<0.44\%$ at $t\sim2$\,days and $\PiL\sim2.4\%$ at $t\sim7$\,days. For GRB\,191221B, polarized emission is detected in the first epoch only at a level inconsistent with subsequent upper limits, indicating a decline in the polarization degree with time.

\begin{table*}
\centering
\caption{Polarimetric properties per sideband for four observations with strong ($Z>5$) polarization detections. Reported uncertainties correspond to the 68\% credible interval derived from Monte Carlo realizations of the measured Stokes parameters assuming Gaussian errors. The spectral indices ($F_\nu\propto \nu^\beta$) derived from the Stokes $I$ measurements suggest $\nu_{\rm obs}\lesssim\nu_{\rm m}$ for GRB\,190829A ($\beta=0.12\pm0.01$) and $\nu_{\rm m}\lesssim\nu_{\rm obs}$ for GRBs 220921A ($\beta=-0.48\pm0.06$) and 221009A ($\beta=-0.18\pm0.03$ at 145\,GHz at 3.39 days and $\beta=-0.28\pm0.02$ at 97.5\,GHz at 5.32 days).}
\begin{tabular}{l l l c c c c c c c c r}
\hline
GRB & $\nu$ & $t$ & $I$ & $Q$ & $U$ & $V$ & $P_L$ & $\PiL$ & $\chi$\\
 & (GHz) & (days) & (mJy) & ($\mu$Jy) & ($\mu$)Jy & ($\mu$Jy) & ($\mu$Jy) & (\%) & ($^\circ$)\\

\hline
190829A & 91.5  & 2.50 & $5.818\pm0.007$ & $-36.9\pm5.7$ & $ 27.1\pm6.2$ & $ -7.2\pm6.7$ & $40.3^{+6.0}_{-5.8}$ & $0.69^{+0.10}_{-0.10}$ & $68.2^{+3.6}_{-3.8}$ \\
190829A & 103.5 & 2.50 & $5.905\pm0.007$ & $-20.2\pm6.3$ & $  6.0\pm6.0$ & $ 10.0\pm6.2$ & $16.0^{+6.0}_{-6.1}$ & $0.27^{+0.10}_{-0.10}$ & $55.5^{+22.8}_{-7.3}$ \\
220921A & 91.5  & 6.75 & $1.608\pm0.010$ & $ 36.4\pm8.9$ & $ 29.0\pm9.6$ & $-14.7\pm9.0$ & $38.2^{+9.1}_{-9.3}$ & $2.38^{+0.56}_{-0.58}$ & $13.5^{+5.8}_{-5.7}$ \\
220921A & 103.5 & 6.75 & $1.516\pm0.007$ & $ 20.0\pm7.2$ & $ 19.4\pm6.9$ & $  5.7\pm7.2$ & $21.9^{+6.8}_{-6.8}$ & $1.44^{+0.45}_{-0.45}$ & $15.0^{+7.2}_{-7.4}$ \\
221009A & 139.0 & 3.39 & $6.996\pm0.012$ & $ 14.9\pm8.9$ & $-96.3\pm8.3$ & $ 12.5\pm8.7$ & $89.5^{+8.3}_{-8.3}$ & $1.28^{+0.12}_{-0.12}$ & $-43.2^{+2.6}_{-2.6}$ \\
221009A & 151.0 & 3.39 & $6.895\pm0.013$ & $ -5.1\pm8.5$ & $-95.7\pm8.5$ & $ 21.2\pm9.1$ & $87.6^{+8.5}_{-8.6}$ & $1.27^{+0.12}_{-0.12}$ & $-49.0^{+2.5}_{-2.5}$ \\
221009A & 91.5  & 5.32 & $5.959\pm0.009$ & $  2.7\pm8.8$ & $-98.5\pm8.7$ & $  7.1\pm9.1$ & $90.3^{+8.6}_{-8.8}$ & $1.52^{+0.14}_{-0.15}$ & $-46.8^{+2.5}_{-2.6}$ \\
221009A & 103.5 & 5.32 & $5.758\pm0.010$ & $-34.1\pm9.5$ & $-92.3\pm9.4$ & $ 14.5\pm8.9$ & $89.6^{+9.2}_{-9.3}$ & $1.56^{+0.16}_{-0.16}$ & $-57.9^{+2.7}_{-2.8}$\\
\hline
\end{tabular}
\label{tab:sidebands}
\end{table*}

In Figure~\ref{fig:pol_lightcurves}, we present the detections and upper limits on $\PL$ and $\PiL$ for our targets, along with the published linear polarization detections (GRB\,190114C; \citealt{lag+19}) and upper limits (GRB\,171205A; \citealt{uth+19,lhc20}) from the literature, as a function of time since the trigger in each case. Additionally, for the sources with significant $\PL$, we show $\chi$ as a function of time. Our observations increase the number of GRBs with mm-band polarization detections from one (GRB\,190114C) to six, now also including GRBs 190829A, 191221B, 220921A, 221009A, and 251013C.  

\section{Summary of Theoretical Models}
\label{sec:modelsummary}
We now describe the space of models that our data may be expected to help constrain. In general, a non-zero net polarization requires some degree of non-uniformity within the visible region (of angular size $\thetavis\sim1/\Gamma$ around our line of sight), arising either from the anisotropy and/or non-axisymmetry of the magnetic-field structure or from asymmetries in the outflow geometry and viewing configuration \citep{gl99,gw99,sar99,gk03,no04}. Existing afterglow polarization models differ primarily in the origin and structure of the magnetic field and in the mechanism by which the symmetry of the GRB image on the plane of the sky is broken.

The polarization degree and its temporal evolution have been calculated for a variety of magnetic-field configurations and viewing geometries and are sensitive to the hydrodynamic evolution of the emitting region via its Lorentz factor $\Gamma(R)$, which determines \thetavis$(t)$, the visible emitting region at a given observer time $t$, and the structure of the afterglow image on the sky \citep{gk03,rlsg04,lcg+04,gt05,bgb+24,bgb26}. The form of $\Gamma(R)$ is different for the post-shock fluid behind the FS (which is expected to follow the \citet[][BM]{bm76} self-similar solution with the local value of the isotropic equivalent kinetic energy, $E_{\rm k,iso}(\theta)$ (or $E_{\rm k,iso}(\theta,\phi)$ for a non-axisymmetric jet) and the shocked ejecta (which generally exhibit a more complex dynamical evolution; \citealt{kob00,ks00,kz03,gt05,gra12,gg23}).

\begin{table*}
\centering
\caption{Representative classes of afterglow polarization models discussed in the literature. The three classes differ primarily in the coherence scale of the magnetic field and in the mechanism responsible for breaking symmetry within the visible region.}
\begin{tabular}{lllll}
\hline
Model & Field origin & Field scale & Polarization source & Comparison to Data \\
\hline
Toroidal / ordered & Magnetized ejecta & $\theta_B\gtrsim\thetavis$ & Global field coherence & Figure~\ref{fig:polmodels-toroidalB}  \\
Random anisotropic & Shock-generated & Microscopic & Off-axis viewing & Figures~\ref{fig:polmodels-randomB} \& \ref{fig:polmodels-randomB-221009A} \\ 
Patchy field & Shock-generated or ejecta & $\theta_B\ll\thetavis$ & Finite number of patches & Section~\ref{sec:patchyfields} \\
\hline
\end{tabular}
\label{tab:modeltypes}
\end{table*}

For FS synchrotron radiation, magnetic fields may be generated locally at the shock front, producing fields that are effectively random on microscopic scales \citep{ml99} and in the simple idealized picture lie completely within the shock plane. Alternatively, they may arise from compression of a pre-existing upstream field \citep{gk03,rdsfkdgdp15}, in which case the field may be inherited from the ambient medium into which the blast wave propagates, e.g., a stellar wind in collapsar-driven GRBs \citep{mga09}. For the latter case of shock-compressed fields, the downstream magnetization parameter ($\epsilon_B$) follows the upstream magnetization ($\sigma$), which, for a stellar wind of velocity, $v_w$, can be up to $\epsilon_B\sim\sigma\lesssim(v_w/c)^2\sim10^{-5}(v_w/1000\,\textrm{km}\,\textrm{s}^{-1})^2$, while for the interstellar medium (ISM), $\epsilon_B\sim\sigma\sim10^{-10}-10^{-8}$. Large-scale ordered fields may also be generated through the mixing of shocked GRB ejecta into the shocked ambient medium via instabilities at the contact discontinuity \citep{lcg+04,lpg05,dm13}.

For RS emission, the same shock-generated\footnote{This requires weakly magnetized ejecta, with magnetization $\sigma\lesssim10^{-4}$, for the Weibel (filamentation) instability to be the dominant source of magnetic field generation and efficient electron acceleration \citep[e.g.][]{Jikei+26}. While this is naturally expected in the FS, for the RS it is challenging to have such a low magnetization within the GRB ejecta, as magnetic fields are expected to play a key dynamical role in the jet launching.} and shock-compressed field configurations may arise. In this case, the magnetic field may be inherited directly from the relativistic ejecta, potentially retaining information about the central engine and jet-launching process \citep{mga09}. Such ejecta-origin fields may contain a large-scale ordered component, with toroidal configurations providing the canonical example \citep{gra03,lpb03,lyu09,bt16}.

The resulting magnetic-field geometries can be broadly divided into three classes (Table~\ref{tab:modeltypes}). The first class consists of globally ordered magnetic fields, most naturally associated with magnetized ejecta and therefore primarily relevant for RS emission \citep{gra03,lpb03,gt05}. In these models the field remains coherent over angular scales larger than the visible region ($\theta_B\gtrsim\thetavis$), allowing polarization levels approaching the synchrotron maximum, $\Pimax$. 

The second class consists of shock-generated random magnetic fields that are locally axisymmetric about the shock normal \citep{gl99,sar99,gra03,gk03,rlsg04,gg18,gkg21,bgb+24,bgb26}. These fields are coherent only on microscopic scales and are commonly parameterized by \citep{gk03}
\begin{equation}
b \equiv \frac{2\langle B_\parallel^2\rangle}{\langle B_\perp^2\rangle}\;,
\end{equation}
where $B_\parallel$ and $B_\perp$ denote the magnetic-field components parallel and perpendicular to the local shock normal, respectively, and the angle brackets indicate volume average -- or, for models assuming emission from a 2D surface identified with the shock front -- a radial average of their energy densities over the width of the shocked region. 
An isotropic three-dimensional random field ($b=1$) produces no net polarization, while anisotropic configurations ($b\neq1$) can generate polarization when combined with a global asymmetry of the GRB image, which is usually attributed to an axisymmetric jet viewed at an angle $\thetaobs>0$ from its symmetry axis. Alternatively, the jet or outflow could have a non-axisymmetric energy distribution, $E_{\rm k,iso}(\theta,\phi)$ \citep[a ``patchy shell", e.g.][]{gk03,no04}, or a non-axisymmetric brightness distribution, e.g. arising from density clumps in the external medium or non-axisymmetric refreshed shocks \citep{gk03}. All of these create brighter and dimmer regions within the observed region that break its symmetry in a random way. Moreover, in a patchy shell different bright spots come into view as the jet decelerates, while different external density clumps or refreshed shocks similarly cause the non-axisymmetric pattern in the observed region to vary with time.
This in turn leads to random variation in $\Pi_L$ and $\chi$, as well as variability in the afterglow lightcurve, as was observed in GRB\;021004 \citep{rwf+03,npg03,no04} and GRB\;030329 \citep{gkr+03,gnp03}.

The third class consists of patchy-\emph{field} models, in which the magnetic field is ordered within coherent regions of angular scale $\theta_B\ll\thetavis$, but the field orientation varies randomly between patches \citep{gw99,gk03,no04,gt05,ktts24}. The observer receives emission from $N$ independent patches, causing the Stokes parameters to undergo a random walk in the $Q$--$U$ plane. The resulting polarization is reduced by approximately $N^{-1/2}$ relative to the polarization from a single coherent patch, which can itself approach $\Pimax$. Unlike locally axisymmetric random-field models, patchy-field models can produce detectable polarization even for an observer located close to the jet symmetry axis. In this model, the polarization angle $\chi$ is expected to undergo significant and random variations on the dynamical time (i.e. over $\Delta t\sim t$, as enough new patches come into view, or the individual turbulent eddies are replenished). For large scale magnetic magnetic fields, on the other hand, $\chi$ is expected to remain constant, and for random shock-generated field from an axisymmetric jet, $\chi$ may at most flip by 90$^\circ$ at the same time as $\PiL$ passes through zero, near the jet break time ($\tjet$).

The temporal evolution of the polarization in these models is governed by the angular size of the visible region relative to the characteristic angular scale of the jet ($\thetac$), as well as the viewing geometry, $q\equiv\thetaobs/\thetac$. Consequently, the polarization degree is commonly presented as a function of ($\tovertjet\equiv t/\tjet$). To facilitate model comparisons, we estimate $\tjet$ for each event in Appendix~\ref{sec:tjet}. In several cases only lower limits on \tjet\ are available, yielding corresponding upper limits on $\tovertjet$. Comparisons between the observations and model predictions should be regarded as qualitative for events with poorly constrained jet-break times.

\begin{figure*}
    \centering    
    \includegraphics[width=\linewidth]{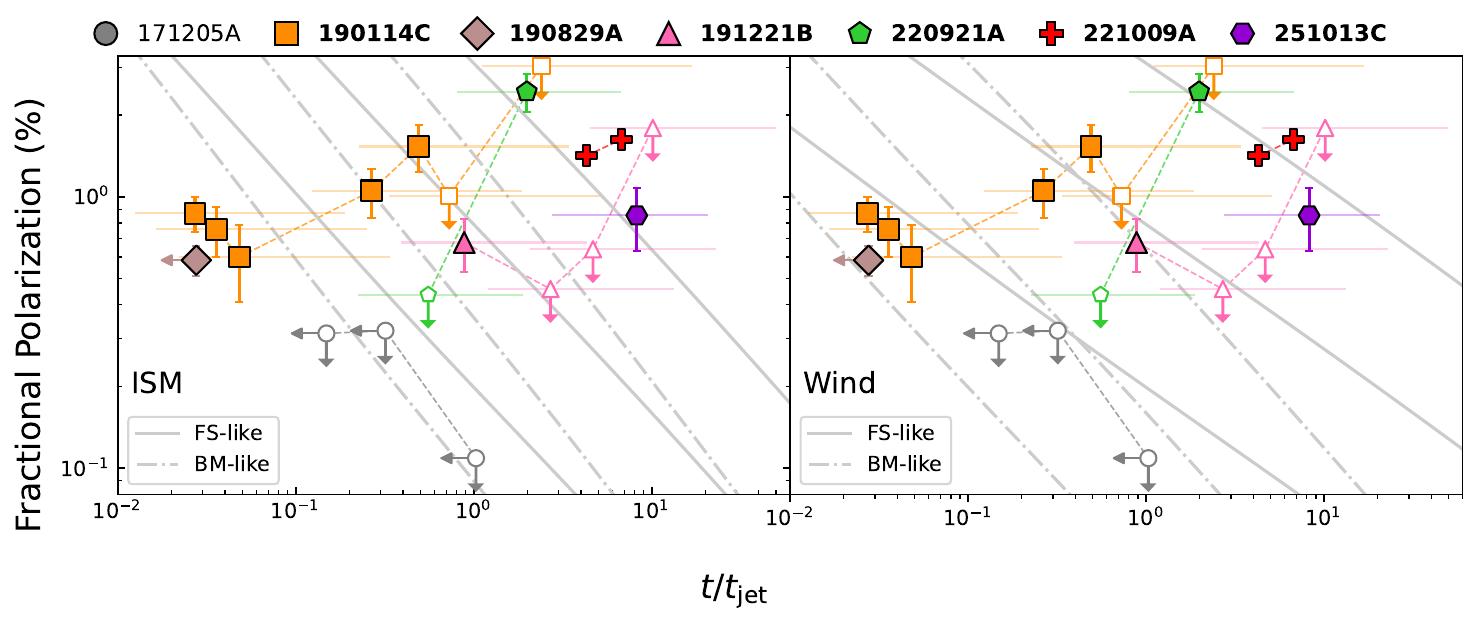}
    \caption{ALMA mm-band polarimetry for GRBs in our sample with constrained jet-break times and either deep ($\lesssim0.5\%$; 171205A) upper limits or at least one linear polarization detection\footnote{The final polarization observation of GRB\,190114C at $\approx87$\,days (corresponding to $\tovertjet\approx21$) yields only a weak upper limit ($\lesssim55\%$) and does not appear on this plot.}, along with theoretical model polarization light curves for a toroidal magnetic field in a top-hat (i.e., ``uniform'') jet for ISM-like and wind-like environments from \citet{gt05} as relevant for RS emission. The top-hat jet models are presented for a range of off-axis viewing angles, $q=\theta_{\rm obs}/\theta_{\rm jet}=[0.2, 0.1, 0.05]$ (from top to bottom in each panel). We assume $\Pi_{\rm 0,max}=(p+1)/(p+7/3) \approx 72\%$ (corresponding to $p=2.5$) in each case. For the solid lines, the Lorentz factor of the ejecta is assumed to remain equal to that of the freshly shocked fluid just behind the forward shock (``FS-like''), while for the dashed lines, it is assumed to follow the BM self-similar solution (``BM-like''). These two cases bracket the dynamics of the shocked ejecta in terms of its $\Gamma(R)\propto R^{-g}$ with $\frac{3-k}{2}\leq g\leq\frac{7-k}{2}$ \citep{ks00,gt05} where $\rho_{\rm ext}\propto R^{-k}$ with $k=0$ for an ISM and $k=2$ for a steady stellar wind. The horizontal bars for each data point correspond to the statistical uncertainty on $\tjet$ for that event (see Appendix~\ref{sec:tjet}) and the horizontal position of the points is sensitive to the assumed $\tjet$.}
    \label{fig:polmodels-toroidalB}
\end{figure*}

Given the large number of possible magnetic-field configurations and viewing geometries, a comprehensive interpretation of the observed polarization properties for any individual event in our sample would require detailed modeling of its multi-wavelength light curves, decomposition of the radio emission into FS and RS components along with a self-consistent calculation of  $\Gamma(R,\theta)$ for both, followed by calculation of bespoke polarization light curves for different jet structures and viewing geometries. Such an analysis is beyond the scope of the present work. Instead, we focus on two representative classes of models that bracket much of the parameter space explored in the literature: outflows containing a globally ordered toroidal magnetic field (Fig.~\ref{fig:polmodels-toroidalB}) and predominantly random magnetic field configurations viewed away from the jet symmetry axis (Figs.~\ref{fig:190114C}, \ref{fig:polmodels-randomB}, and \ref{fig:polmodels-randomB-221009A}). For the latter, we employ $b=0.66$, which yields the highest possible polarization level at fixed $q$ while remaining consistent with constraints from polarimetric observations of the radio afterglow of GRB\,170817A \citep{gg20}. We consider patchy-field models in Section~\ref{sec:patchyfields}. 
We discuss each event individually, assess the level of instrumental systematic uncertainty, and use context from published multi-wavelength analyses, where possible together with our polarization measurements, to place qualitative constraints on the magnetic-field structure and viewing geometry in each case. 

\section{Constraints on Theoretical Models}

\label{sec:individual_targets}
We now discuss our ALMA mm-band polarization observations, including both detections and upper limits, within the broader framework developed for synchrotron emission from relativistic outflows.
\subsection{GRB\,171205A}
There are three epochs of full Stokes observations of GRB\,171205A in the ALMA archive (PI: Urata). For the first epoch taken at 5.2\,days, \citet{uth+19} published a detection of $\PiL=(0.27\pm0.04)\%$. Subsequent re-analysis suggested that the observation was dominated by unremovable instrumental systematics, resulting in an upper limit on the polarization of $\PiL<0.30\%$ \citep{lhc20}. We include these results in Table~\ref{tab:grb_pol}. The second and third epochs for this event (taken at $\approx11.2$ and 36.0 days, respectively) are presented here for the first time and also result in upper limits. 

This event exhibits the brightest mm afterglow in our sample, with a 3\,mm Stokes $I$ flux decreasing from $(32.44\pm0.03)$\,mJy from $\approx5.2$\,days to $15.4\pm0.1$\,mJy between the first two epochs, corresponding to a decline rate $\alpha=-0.91\pm0.01$ and then increases from 11.2 to 36.0\,days with $\alpha=0.29\pm0.01$. The prompt $\gamma$-ray emission for this event exhibits a low isotropic-equivalent $\gamma$-ray energy ($\Egammaiso\approx2\times10^{49}$\,erg) given the observed $\gamma$-ray peak energy, $E_{\rm p}\approx125$\,keV, placing this event in the category of low-luminosity GRBs \citep{dcd+18}. Modeling of the multi-wavelength afterglow emission has led to suggestions that the afterglow was observed off-axis, with $\theta_{\rm jet}\approx34^\circ$ and $\theta_{\rm obs}\approx42^\circ$ (corresponding to $q\approx1.2$ \citep{mc21,kpg+22,lzhx24}. For this event, we use the lower limit on the jet break time of $\gtrsim71$\,days \citep{mc21}. 

We find that if this event is truly observed off-axis with $q\gtrsim1$ and the radio emission is dominated by the RS, then the toroidal magnetic field models are fully ruled out (Figure~\ref{fig:polmodels-toroidalB}). The last (and deepest) upper limit at $\approx36$\,days corresponds to $\tovertjet\lesssim0.5$; Therefore, the data do not constrain $b=0.66$ FS polarization models, highlighting the critical role of long-term polarimetric monitoring of these rare, bright afterglows.   

\subsection{GRB\,190114C}
\begin{figure*}
    \centering    
    \includegraphics[width=0.48\linewidth]{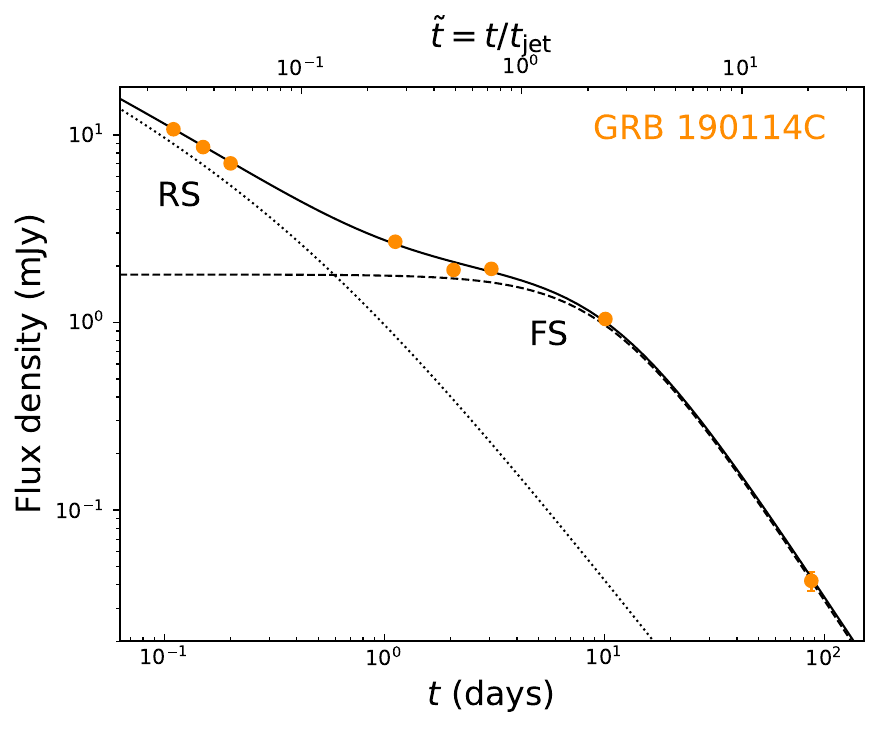}
    \includegraphics[width=0.48\linewidth]{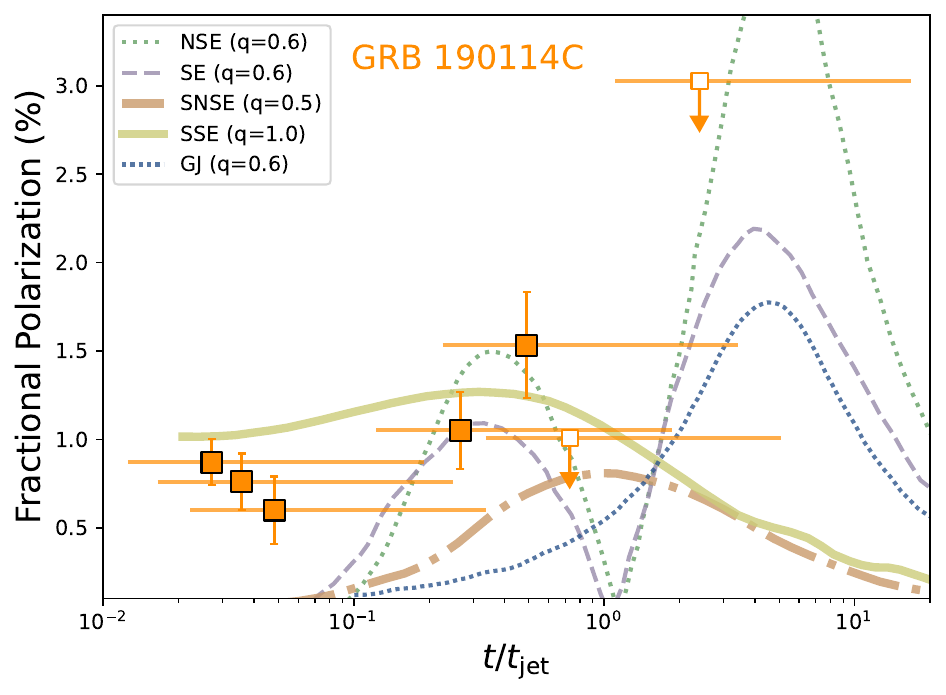}
    \caption{\textbf{GRB\,190114C.} Left panel: The ALMA 3mm Stokes $I$ light curve with time since burst in days (lower axis) and scaled to $\tjet=4.2$\,days (Appendix~\ref{sec:tjet}; upper axis) appears to comprise two emission components (broken power-law models presented to guide the eye). Right panel: $\PiL$ measurements and upper limits, compared with theoretical model polarization curves from \citet{rlsg04} for a random magnetic field scaled by $\Pi_{\rm rnd}/\Pimax\approx(b-1)/(b+1)\sim0.2$ corresponding to $b=0.66$ \citep{gg20}. Horizontal bars indicate the expected range of $\tovertjet$ for each observation (Appendix~\ref{sec:tjet}). The legend describes the off-axis viewing angle ($q\equiv\thetaobs/\thetajet$) and jet model: non-sideways expanding (NSE), sideways expanding (SE), (power-law) structured non-sideways expanding (SNSE), structured sideways expanding (SSE), and Gaussian jet (GJ). Our new polarization observations at $\tovertjet\gtrsim0.1$ are consistent with top-hat jet (green dotted and blue dashed) FS models. 
    }
    \label{fig:190114C}
\end{figure*}
We present five new full Stokes ALMA 3\,mm observations of GRB\,190114C spanning from 1.12 to 86.97 days post GRB trigger date. Combining with the observation presented in \citet{lag+19}, which we split further into three sub-epochs (which we label epochs 1.1, 1.2, and 1.3, respectively), yields  eight total measurements of $\PiL$ spaced roughly logarithmically in time spanning from 0.114 to 86.97 days post-burst ($\tovertjet\approx0.03$--21. In addition to the detections in epoch 1 presented in \cite{lag+19}, our analysis yields additional polarization detections in the observations on 2019-01-15 (epoch 2, 1.12\,days; Stokes $U$ only) and on 2019-01-16 (epoch 3, 2.06\,days; both Stokes $Q$ and $U$). The remaining observations yield non-detections in both Stokes $Q$ and $U$ (Table~\ref{tab:grb_pol}).    

Before discussing the interpretation of this polarization light curve, we make a few notes about the data and reduction quality. In both epochs 2 and 3, we find a statistically significant detection and variability in the Stokes $V$ flux density for the gain calibrator (from 0.3 to 2.3 mJy); however, these correspond only to $V/I\lesssim0.15\%$ (Stokes $I\sim 1600$\,mJy), which is  consistent with the expected level of instrumental systematics in Stokes $V$ \citep{nnp+16}. The data pass all remaining systematics checks. 

The Stokes $I$ light curve (Table~\ref{tab:grb_pol} and Fig.~\ref{fig:190114C}) declines as $\alpha\approx-0.6$ from $\approx0.1$--1.2\,days, is roughly flat from 1.2--4 days, continues to fade as $\alpha\approx-0.5$ from $\approx4$--10 days, and then declines rapidly ($\alpha\approx-1.8$) thereafter. Following \citet{lag+19} we consider the radio emission at $\lesssim1$\,day to be dominated by the RS \citep[although, see also][for an alternative interpretation]{mrk+21}. The shallower light curve observed at $\gtrsim1$\,day is unexpected in the RS model and is suggestive of an emerging, underlying FS component. We present a fiducial decomposition of the mm-band light curve into two components using the sum of two broken power-laws in Fig.~\ref{fig:190114C}, and note that a more complete decomposition requires more detailed multi-wavelength modeling. 

The polarization degree drops from $\approx0.9\%$ to $\approx0.6\%$ in the first day (corresponding to $\tovertjet\approx0.03$--0.05, while $\chi$  rotates from $\approx10^\circ$ to $-44^\circ$ \citep{lag+19}. We find that a $\chi^2$ test rules out the null hypothesis that $\chi$ (the polarization position angle) is constant during this time, consistent with \citet{lag+19}. The gradually changing $\chi$ rules out models of toroidal magnetic fields in the ejecta as well as a models with a dominant shock-generated magnetic field, since $\chi$ is expected to be constant in both (at $\tilde{t}\ll1$ for the latter, where even at $\tilde{t}\sim1$ only an abrupt 90$^\circ$ change in $\chi$ is possible). Instead, these observations are suggestive of patchy magnetic fields and this model is discussed further in Section~\ref{sec:patchyfields}. 

In our new polarimetric observations spanning $\tovertjet\approx0.3$--21, we find that $\PiL$ rises from $0.6\%$ to $\approx1.5\%$ over $\tovertjet\approx0.05$--0.5, then drops to $\lesssim1\%$ at $\tovertjet\gtrsim0.8$. During this period, a $\chi^2$ test cannot rule out the null hypothesis that $\chi$ is constant; and in fact, we do not find strong evidence for a change in $\chi$ at $\gtrsim0.15$~days (including and after epoch 1.2; corresponding to $\tovertjet\approx0.04$), indicating that the strongest evidence for changing $\chi$ arises in between epochs 1.1 and 1.2 ($\tovertjet\approx0.03$--0.04). 

If the 3\;mm emission at $\tovertjet\gtrsim0.3$ is indeed FS-dominated, our observations mark the first detection of polarized radio emission from a GRB FS. The relative invariance of $\chi$ at this time is also consistent with a dominant FS contribution. The increasing polarization degree at $\tovertjet\gtrsim0.3$ is once again inconsistent with toroidal $B$ fields in the post-shock region (Fig.~\ref{fig:polmodels-toroidalB}), which are also not expected in the FS. On comparing the observations with random $B$-field models (scaled to $b=0.66$; Fig.~\ref{fig:polmodels-randomB}), we find that (i) power-law structured jet models produce a broader peak in $\PiL(\tovertjet)$ than observed; (ii) the Gaussian jet model with $q\approx0.6$ (required to match the maximum value of $\PiL$) peaks at $\tovertjet\approx4$, contrary to the observations (which peak at $\tovertjet\approx0.5$), and (iii) top-hat jet models (in particular, the non-spreading jet model) with $q\approx0.6$ are roughly consistent with the $\PiL$ curve. While detailed investigation of these features is beyond the scope of the present work, we anticipate that these observations may provide additional constraints on the jet structure when directly modeled together with the multi-frequency Stokes $I$ light curves.  

\subsection{GRB\,190829A}
\label{sec:190829A}
There is one epoch of full-polarization observations of GRB\,190829A in the ALMA archive (PI: Urata). The ALMA pipeline products indicate a highly significant polarization detection (Table~\ref{tab:grb_pol_pipe}), which we confirm through our manual reduction. The Stokes $I$ images exhibit low-level ($\sim2\%$) residual gain calibration errors that are substantially reduced after three rounds of self-calibration (20 min phase-only, followed by 20 min and 10 min amplitude+phase), improving the image dynamic range by a factor of $\approx4$. We detect polarized emission in both Stokes $Q=-28\pm4\,\mu$Jy and $U=19\pm4\,\mu$Jy (Fig.~\ref{fig:imgs}; Table~\ref{tab:grb_pol}), corresponding to $\PiL=(0.58\pm0.07)\%$ and $\chi=(73.4\pm3.7)^\circ$. At $I\approx5.9$,mJy, this is the fourth-brightest afterglow in our sample and the second most significant polarization detection after GRB\,221009A (Section~\ref{sec:221009A}).

The data quality warrants some caution. The derived instrumental leakages are larger than for any other observation in our sample (Fig.~\ref{fig:dterm}), consistent with notes in the ALMA QA report, although no specific issue was identified and the observation passes all standard quality checks. Relative to the pipeline reduction, our self-calibrated images yield a slightly higher ($\sim2\sigma$) polarization significance. The two sidebands are consistent in Stokes $Q$ but differ in Stokes $U$ at the $3.4\sigma$ level, resulting in a lower polarized intensity in the USB by $\approx4\sigma$. The gain calibrator exhibits good polarization stability (at $\lesssim0.2\%$) and we cannot isolate a clear instrumental explanation for this frequency dependence. We therefore adopt the sideband-averaged polarization values reported in Table~\ref{tab:grb_pol}, while noting that residual instrumental systematics may contribute to systematic uncertainty.

GRB\,190829A exhibited VHE emission together with an early optical/X-ray re-brightening, and its broadband afterglow has been interpreted using a variety of models, including off-axis jets and combinations of FS and RS emission \citep{soy+21,rvdhf+20,dtl+22,sry+22,fvb+21,zrh+21,zmvm21}. Optical polarimetry yielded an upper limit of $Q\lesssim6\%$ \citep{dtl+22}. 

We adopt a conservative lower limit of $\tjet\gtrsim90$\,days from the absence of a clear X-ray jet break (Appendix~\ref{sec:tjet}), corresponding to $\tovertjet\lesssim3\times10^{-2}$ for our observation.
Comparing our measurement with the toroidal magnetic-field models (Fig.~\ref{fig:polmodels-toroidalB}), we find that the observed polarization is consistent only with a viewing geometry close to the jet axis ($q\lesssim0.05$). Because only a lower limit on $\tjet$ is available, the corresponding upper limit on $\tovertjet$ prevents a meaningful comparison with the random $B$-field models. Better constraints on the jet-break time would therefore be required before the radio polarimetry is used to discriminate among those models.

\subsection{GRB\,191221B}
\begin{figure*}
    \centering    
    \includegraphics[width=0.48\linewidth]{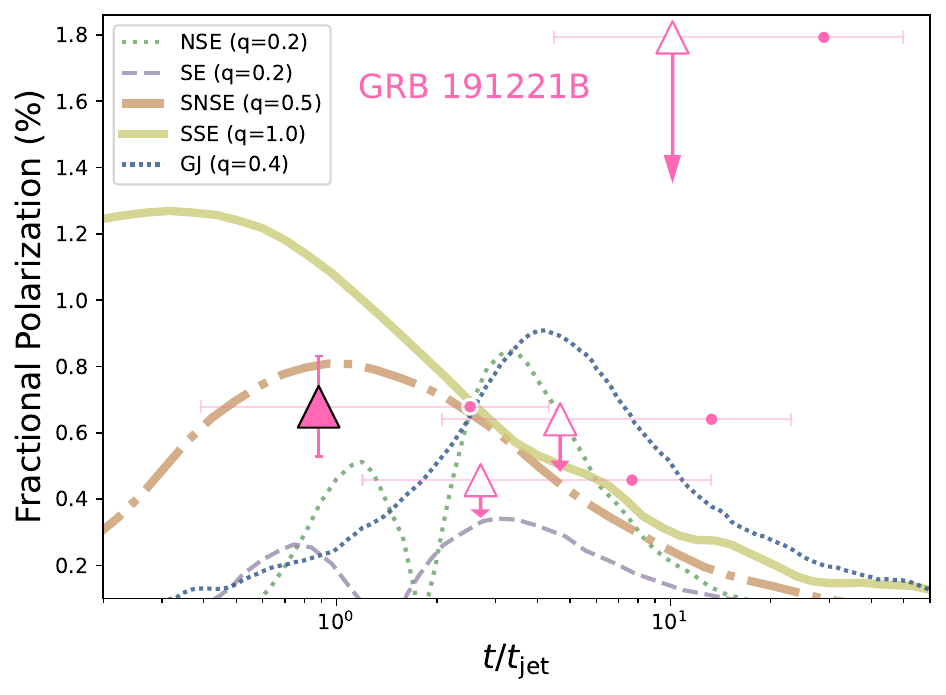}
    \includegraphics[width=0.48\linewidth]{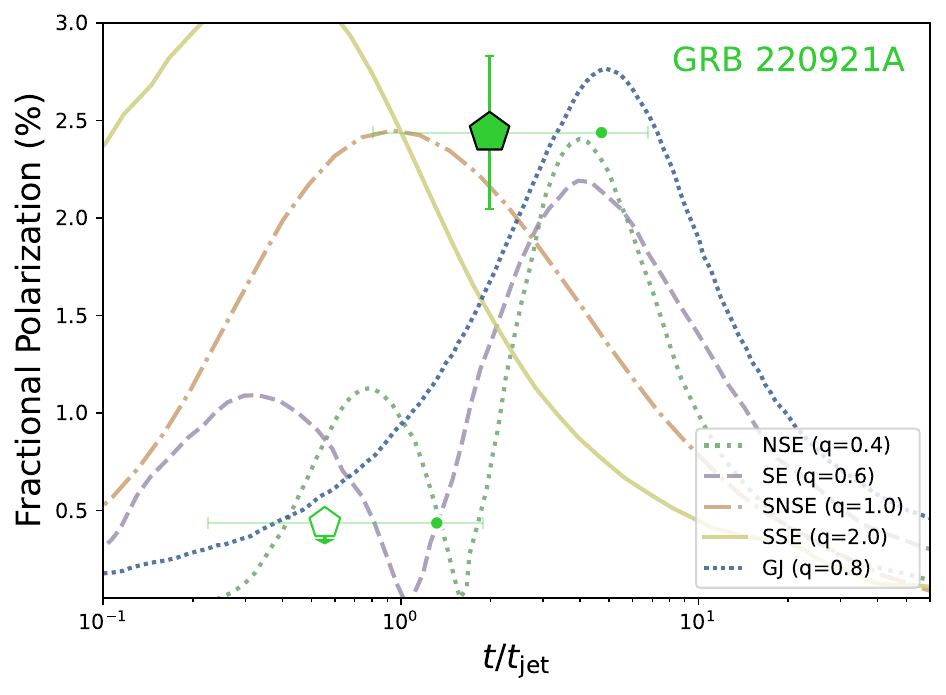}
    \caption{ALMA mm-band polarimetry (data points) for GRBs\,191221B and 220921A compared with theoretical model polarization light curves from \citet{rlsg04} for a random magnetic field scaled by $\Pi_{\rm rnd}/\Pimax=(b-1)/(b+1)\sim0.2$ corresponding to $b=0.66$ \citep{gg20}. Horizontal bars indicate the expected range of $\tovertjet$ for each observation (Appendix~\ref{sec:tjet}). The legend in each panel describes the off-axis viewing angle ($q\equiv\thetaobs/\thetajet$) and jet model: non-sideways expanding (NSE), sideways expanding (SE), (power-law) structured non-sideways expanding (SNSE), structured sideways expanding (SSE), and Gaussian jet (GJ). For GRB\,191221B, the observed polarization curves are roughly consistent with the NSE ($q=0.2$), SE ($q\approx0.2$--0.4), SNSE ($q=0.5$), and SSE ($q=1$) models (the dot corresponds to $\tjet = 0.19$\,days), while the GJ model is ruled out. For GRB\,220921A, the polarization measurements are consistent with NSE ($q=0.4$) and SE ($q=0.6$) models (the dot corresponds to $\tjet = 1.43$\,days), and broadly with the GJ models. 
    }
    \label{fig:polmodels-randomB}
\end{figure*}
Two ALMA polarimetric upper limits of GRB\,191221B at $t=2.52$\,days ($\PiL<0.64\%$) and $5.48$\,days ($\PiL<1.79\%$) were previously reported by \citet{utc+23}. We present two earlier observations obtained at 0.48 and 1.46\,days. Whereas the ALMA pipeline images indicate non-detections in both epochs, our re-reduction reveals a statistically significant detection in Stokes $Q$ during the first epoch ($Q\approx3.7\sigma_Q$), corresponding to $\PiL=(0.68\pm0.15)\%$, while the second epoch at 1.46\,days remains a non-detection. 

Although statistically significant, this detection lies close to the systematic floor of ALMA polarimetry. The measured polarization fraction is comparable to both the estimated instrumental polarization ($\sim0.3\%$ of Stokes $I$) and the observed variability in the target ($\sim0.6\%$ of Stokes $I$), suggesting that residual calibration systematics could plausibly account for the signal. We therefore regard this polarization detection as tentative and interpret it with caution.

We next discuss constraints on the magnetic field and viewing geometry for this event under the assumption that the detection is astrophysical. Comparing the observations with the RS polarization models (Fig.~\ref{fig:polmodels-toroidalB}), we find that the early polarization detection followed by deep upper limits is  consistent with toroidal $B$-field models for $q\approx0.1$. In the FS models (Fig.~\ref{fig:polmodels-randomB}), a range of models remain plausible given the uncertainty in the jet break time. We find that top-hat jet models with $q=0.2$ (green dotted curve, labeled ``NSE'' in Fig.~\ref{fig:polmodels-randomB}), structured jet models (``SNSE'' with $q=0.5$ and ``SSE'' with $q=1.0$) and Gaussian jet models (``GJ'' with $q=0.4$) are all roughly consistent with the data under the allowed values for the jet-break time. Therefore, the polarization curve remains consistent with both an FS or RS origin under several different models. 

\subsection{GRB\,210610B}
We observed GRB\,210610B at a single epoch at 0.30 days, the earliest observation in our sample post-burst. The observation yields a Stokes $I$ detection of $2.235\pm0.006$\,mJy. We find no significant signal in Stokes $Q$ ($-1\pm5\,\mu$Jy) and a marginal signal in Stokes $U$ ($15\pm5\,\mu$Jy), with a Stokes $V$ measurement consistent with noise ($-8\pm6\,\mu$Jy). The one-sided Gaussian significance of the polarized emission is $2.27\sigma$, below our detection threshold, and we therefore place a 95\% upper limit of $P<24\,\mu$Jy ($\PiL<1.09\%$).

From our X-ray light curve analysis (Appendix~\ref{sec:tjet}), we estimate a jet break time of $t_\text{jet}=0.53^{+0.80}_{-0.33}$\,days for this event. Our ALMA observation at $t=0.30$\,days therefore corresponds to $\tovertjet\approx0.57^{+0.55}_{-0.26}$, placing this measurement near or slightly before the jet break time. 

Optical polarimetry of this event was presented by \citet{afdupt+24}, who detected $\PiL=(4.27\pm1.45)\%$ at $t\approx0.12$\,days, followed by a decline to zero ($\lesssim0.5\%$ at 0.24--0.28\,days) and a subsequent rise to $(2.27\pm0.22)\%$ at $t\approx1.3$\,days, accompanied by a $(54\pm9)^\circ$ rotation in $\chi$. This behavior was interpreted as a transition from refreshed-shock-dominated emission at $\lesssim0.12$\,days to FS-dominated emission thereafter, with the late-time polarization arising from geometric effects or large-scale turbulent magnetic fields. Our ALMA upper limit of $\PiL\lesssim1.09\%$ at $t=0.30$\,days is contemporaneous and consistent with the optical polarization limit at 0.24--0.28\,days, and does not provide additional constraints on the magnetic field or geometry. 

\subsection{GRB\,210702A}
We observed GRB\,210702A at a single epoch at 1.10 days post-trigger. The observation yields a Stokes $I$ detection of $2.133\pm0.005$\,mJy. We find no significant signal in either Stokes $Q$ ($-5\pm5\,\mu$Jy) or Stokes $U$ ($9\pm5\,\mu$Jy), and place a 95\% upper limit on the polarized intensity of $P<20\,\mu$Jy ($\PiL<0.94\%$).

\citet{dwlg+24} interpret the 97.5\,GHz light curve of this event as dominated by FS emission. From our X-ray light curve analysis (Appendix~\ref{sec:tjet}), the X-ray afterglow of this event exhibits a single power-law decline with no evidence of a jet break until the last detection at $\approx9.4$\,days, yielding a lower limit of $t_\text{jet}\gtrsim9.4$\,days \citep[see also][]{dwlg+24}. Our ALMA observation at $t=1.10$\,days therefore corresponds to $\tovertjet\lesssim0.12$, firmly in the pre-jet-break regime. The upper limits on both $\tovertjet$ and $\PiL$ do not allow us to strongly constrain FS polarization models for this event. 

\subsection{GRB\,220521A}
We observed GRB\,220521A at 56.12 and 82.09 days post-trigger. Both observations resulted in polarization non-detections. Since Stokes $I$ was relatively faint (64 and 31\,$\mu$Jy, respectively), the limits on $\PiL$ are not constraining for the models. We therefore are unable to rule out any models for this event.

\subsection{GRB\,220921A}
We observed GRB\,220921A at $t=1.89$\,days and 6.75\,days. The first observation resulted in a non-detection, with $\PL\lesssim20\,\mu$Jy ($\PiL\lesssim0.44\%$), while the second yielded a significant polarization detection of $\PL\approx38\,\mu$Jy ($\PiL\approx2.43\%$), consistent with the ALMA pipeline products.

The first epoch exhibits an unusual frequency-dependent ``ramping'' in the D-term amplitudes that persists under recalibration with a different reference antenna, indicating residual instrumental systematics that could reduce our sensitivity to weak polarization signals. In the second epoch, we identified highly structured noise in the lower sideband of the first execution block, which produced anomalous variability in the gain calibrator. Flagging this portion of the data ($\sim1/6$ of the total observation) substantially improved the image quality and reduced the apparent variability of the calibrator. Both epochs used the same gain calibrator, whose polarization remained stable at the $\lesssim0.26\%$ level, well below the polarization detected from the GRB in the second epoch.

Since the origin of the mm-band emission is unknown, we compare the observations with both RS and FS polarization models. If the emission is RS-dominated, the rapid rise in polarization between the two epochs is inconsistent with toroidal magnetic-field models, which predict a monotonic decline in $\PiL$ (Fig.~\ref{fig:polmodels-toroidalB}). If instead the emission is FS-dominated, the observations are broadly consistent with random magnetic-field models for top-hat jets (NSE with $q\approx0.4$ and SE with $q\approx0.6$) and Gaussian jets ($q\approx0.8$), whereas power-law structured jets predict a polarization peak earlier than observed (Fig.~\ref{fig:polmodels-randomB}).

\subsection{GRB\,221009A}
\label{sec:221009A}
\begin{figure}
    \centering    
    \includegraphics[width=0.9\columnwidth]{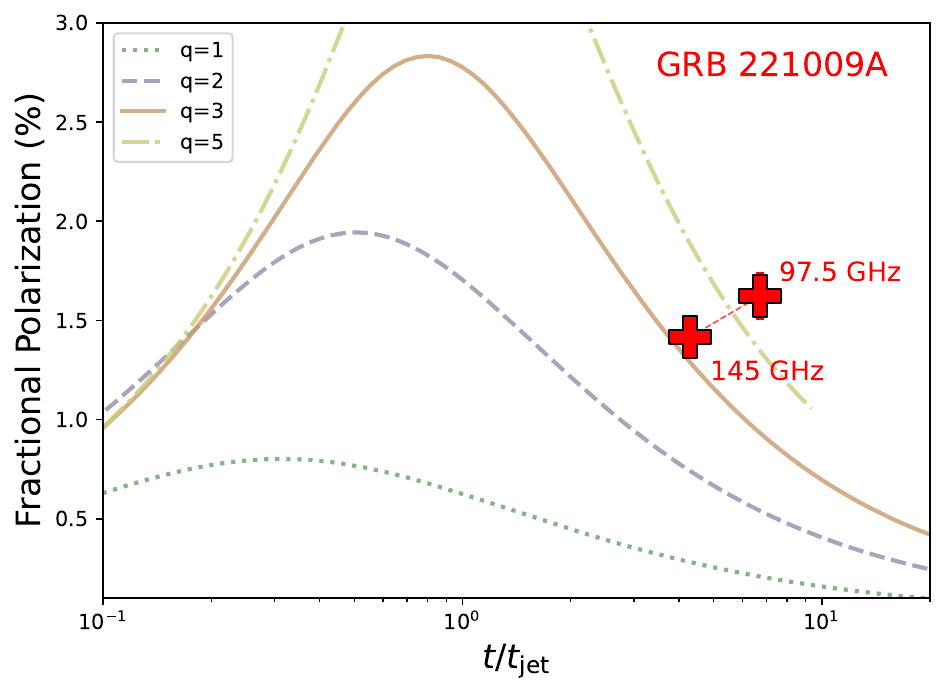}    
    \caption{ALMA mm-band polarimetry (data points) for GRB\,221009A assuming $\tjet\sim0.8$\,days \citep{otr+23,gg23}, compared with FS polarization models from \citet{bgb+24} with a random magnetic field and power-law energy profile index, $a=1$, scaled by $\Pi_{\rm rnd}/\Pimax=(b-1)/(b+1)\sim0.2$ for $b=0.66$ \citep{gg20}. These models are only consistent with the data for a highly off-axis viewing geometry $q\gtrsim3$, which is not consistent with the light curves \citep{gg23}. 
    }
    \label{fig:polmodels-randomB-221009A}
\end{figure}
There are two polarimetric observations of GRB\,221009A in the ALMA archive (DDT project 2022.A.00009.T, PI: Huang), obtained at 3.39 and 5.32\,days post-trigger at 145\,GHz (Band~4) and 97.5\,GHz (Band~3), respectively. We reduced both datasets following the procedures described in Section~\ref{sec:reduction}, including one round of phase-only self-calibration. Both observations yield bright Stokes $I$ detections, making GRB\,221009A the second-brightest afterglow in our sample after GRB\,171205A. We detect strong polarized emission in both epochs, primarily through Stokes $U$ (Fig.~\ref{fig:imgs}; Table~\ref{tab:grb_pol}), yielding the most significant polarization detections in our sample (16.3$\sigma$ and 13.3$\sigma$, respectively). Stokes $Q$ is consistent with noise in the first epoch and only marginally detected ($\approx2.7\sigma$) in the second.

The observations pass all systematic checks. The gain calibrator exhibits negligible variability with time and frequency in both epochs. The target likewise exhibits no significant sideband-dependent polarization variability, with the exception of a modest ($\PiL\approx1.5\%$) variation in the upper sideband of the second epoch.

The radio afterglow exhibits a long-lived spectral peak at cm wavelengths, whose evolution differs from that of the optical and X-ray afterglows, and has been interpreted as RS emission from a structured jet \citep{brf+23,lam+23,otr+23,gg23,zwz24}. At the time of the ALMA observations, the mm-band emission is therefore expected to be RS dominated. The optical and X-ray light curves exhibit evidence for a quasi-achromatic break at $\approx0.8$\,days \citep{otr+23,gg23}, commonly interpreted as the jet break of a narrow jet core. We adopt $\tjet\approx0.8$\,days for comparison with polarization models, noting that more detailed broadband modeling is required to establish this value robustly.

The measured polarization degree is consistent with being constant between the two epochs, with $\PiL=(1.42\pm0.09)\%$ at 145\,GHz and $(1.62\pm0.12)\%$ at 97.5\,GHz, corresponding to $\tovertjet\approx4.2$--6.7. We find only marginal evidence for evolution of the polarization position angle, $\Delta\chi=(8.6\pm2.9)^\circ$, driven entirely by the weak change in Stokes $Q$, while Stokes $U$ remains unchanged between the two observations. Overall, the data are consistent with a stable polarization position angle over this interval.

We first compare the observations with the available FS polarization models (Fig.~\ref{fig:polmodels-randomB-221009A}). The random magnetic-field models of \citet{bgb+24} require extreme viewing angles ($q\gtrsim3$) to reproduce the observed polarization, inconsistent with the broadband afterglow modeling \citep{gg23}. For RS models, a toroidal magnetic-field configuration require nearly on-axis viewing angles ($q\approx0.2$) and predict a declining polarization degree, contrary to the observed evolution (Fig.~\ref{fig:polmodels-toroidalB}). An alternative possibility is that the RS magnetic field is patchy rather than globally ordered (as previously inferred from ALMA polarimetry of the RS-dominated radio/mm emission of GRB\,190114C; \citealt{lag+19}). This model is discussed further in Section~\ref{sec:patchyfields}. 

\subsection{GRB\,250327B}
Our polarization observation of GRB\,250327B at $t=1.20$\,days yields a non-detection with $P_L\lesssim20\,\mu$Jy and $\PiL \lesssim 3.72\%$. The jet-break time is poorly constrained. A fit to the X-ray light curve, supplemented by sparse optical observations reported in GCN Circulars, suggests $\tjet\approx0.2$\,days, although substantially later jet breaks remain consistent with the available data.

If the mm-band emission is RS dominated, the polarization limit places only a weak constraint on the viewing geometry, requiring $q\lesssim0.4$ in the toroidal magnetic-field models. If instead the emission is FS dominated, the upper limit is consistent with all of the random magnetic-field models considered here (e.g., Fig.~\ref{fig:polmodels-randomB}). Consequently, the current observations do not provide strong constraints on the magnetic-field geometry or jet structure.

\subsection{GRB\,251013C}
We observed GRB\,251013C at a single epoch 8.24\,days post-trigger in ALMA Band~3 (97.5\,GHz). The observation yields a bright Stokes $I$ detection ($\approx2.9$\,mJy) together with a marginal Stokes $Q$ detection ($-(24\pm7)\,\mu$Jy; $\approx3.4\sigma$), while Stokes $U$ is consistent with zero. This corresponds to a polarized flux density of $\PL=(25\pm6)\,\mu$Jy and a linear polarization fraction of $\PiL\approx0.85\%$. Although the polarized intensity is detected at only $2.97\sigma$, the significant Stokes $Q$ signal persists after recalibration, including reductions using different reference antennas with distinct antenna architectures, suggesting that it is unlikely to arise from the calibration procedure.

Given the marginal detection, we examined the instrumental systematics carefully. The D-term amplitudes are well-behaved (Figure~\ref{fig:dterm}). This observation was taken in the C-8 extended configuration and employed a short cycle time of 83\,s with a large number of calibrator scans (Figure~\ref{fig:polsys-timevar}), providing dense sampling of the gain calibrator, which exhibits low-level drift in Stokes $I$, $Q$, and $U$ (peak-to-peak variations of $\sim0.3\%$, $\approx0.2\%$, and $0.5\%$ of Stokes $I$, respectively; Fig.~\ref{fig:polsys-timevar}), which may indicate slow variations in the instrumental polarization during the observation. We therefore regard the polarization detection as tentative.

From the X-ray light-curve analysis (Appendix~\ref{sec:tjet}), we estimate $\tjet=1.0^{+2.0}_{-0.6}$\,days, placing our observation at $\tovertjet\approx8.2$, well after the jet break. If the mm-band emission is RS dominated, the measured polarization is consistent with toroidal magnetic-field models viewed close to the jet axis ($q\approx0.2$; Fig.~\ref{fig:polmodels-toroidalB}). If instead the emission is FS dominated, the observed polarization is consistent with a broad range of random magnetic-field models. Consequently, this single, marginal detection does not provide strong constraints on the magnetic-field geometry or jet structure. 

\newcommand{\summarycolwidth}{13.5cm}
\begin{table*}
    \caption{Discussion Summary}
    \begin{tabularx}{\textwidth}{lrcl}
        \hline\hline
        GRB & $t$ & Pol. Det.? & Discussion Summary \\
        \hline
        \multirow{3}{*}{171205A} & 5.20 & \xmark & \multirow{3}{\summarycolwidth}{Deep polarization upper limits. If the radio emission is RS dominated, toroidal magnetic-field models are ruled out for $q\gtrsim0.05$. The uncertain jet-break time prevents meaningful constraints on FS models.}\\
         & 11.17 & \xmark & \\
         & 36.04 & \xmark & \\
        \hline
        \multirow{8}{*}{190114C} & 0.11 & \cmark & \multirow{8}{\summarycolwidth}{Best-sampled $\PiL(t)$ in the sample. Early evolution favors patchy magnetic fields in the RS, while the later (likely FS-dominated) polarization is broadly consistent with both\quad random-field top-hat jet models and patchy-field models.}\\
         & 0.15 & \cmark & \\
         & 0.20 & \cmark & \\
         & 1.12 & \cmark & \\
         & 2.06 & \cmark & \\
         & 3.06 & \xmark & \\
         & 10.09 & \xmark & \\
         & 86.97 & \xmark & \\
        \hline
        \multirow{3}{*}{190829A} &
        \multirow{3}{*}{2.50} &
        \multirow{3}{*}{\cmark} &
        \multirow{3}{\summarycolwidth}{%
        \raggedright
        Strong, single-epoch detection. Assuming RS-dominated emission, toroidal magnetic-field models require a nearly on-axis viewing geometry ($q\lesssim0.05$). The uncertain jet-break time precludes meaningful comparison with FS models.%
        } \\
        & & & \\
        & & & \\
        \hline
        \multirow{4}{*}{191221B} & 0.48 & \cmark & \multirow{4}{\summarycolwidth}
        {Tentative early polarization detection followed by deep upper limits. Residual instrumental systematics cannot be excluded, so only weak constraints on the magnetic-field geometry are obtained.}\\
         & 1.46 & \xmark & \\
         & 2.52 & \xmark & \\
         & 5.48 & \xmark & \\
        \hline
        210610B & 0.30 & \xmark & Consistent with contemporaneous optical limit.\\
        \hline
        210702A & 1.10 & \xmark & Upper limit does not strongly constrain FS polarization models.\\
        \hline
        \multirow{2}{*}{220521A} & 56.12 & \xmark & \multirow{2}{\summarycolwidth}{Two late-time polarization upper limits. The faint afterglow yields weak, non-constraining limits on $\PiL$.} \\
         & 82.09 & \xmark & \\
        \hline
        \multirow{2}{*}{220921A} & 1.89 & \xmark & \multirow{2}{\summarycolwidth}{Rapidly rising $\PiL$ is inconsistent with RS toroidal magnetic-field models but broadly consistent with several FS random-field models.} \\
         & 6.75 & \cmark & \\
        \hline
        \multirow{3}{*}{221009A} & 3.39 & \cmark & \multirow{3}{\summarycolwidth}{Highest-significance polarization detections in the sample. Most current RS polarization models do not reproduce both the observed polarization evolution and the viewing geometry inferred from broadband modeling; patchy RS magnetic fields are a plausible explanation.}\\
        & 5.32 & \cmark & \\
        &      &        & \\
        \hline
        250327B & 1.20 & \xmark & 
        Yields weak constraints on RS models; consistent with all FS random magnetic-field models.\\
        %This limit broadly constrains $q \lesssim 0.6$ consistent with a nearly on-axis jet.\\
        \hline
        251013C & 8.24 & \cmark%\tablenotemark{\dag} 
        & Marginal detection is consistent with both RS and FS polarization models.\\
        \hline
    \end{tabularx}
    \label{tab:summary-by-event}
\end{table*}

\begin{table*}
\caption{Polarimetric Constraints on Theoretical Models}
\begin{tabularx}{\textwidth}{llclll}
\hline\hline
$B$-type\tablenotemark{$*$} & Model & Location & $\PiL\lesssim3\%$ Allowed? & $0\ne|\Delta\chi|_{\rm max}\lesssim60^\circ$ & Possible for: \\
\hline
Large scale & Toroidal field (Fig.~\ref{fig:polmodels-toroidalB}) & RS & $q\ll1$ & \xmark & 191221B, 221009A, \\
  & & & & & 251013C\\
Large scale & Ordered in shock plane & RS, FS$^\dag$ & \xmark & \xmark & None \\
Large scale & Patches ($N$ within $\thetavis$) & RS, FS$^\ddag$ & $\thetaB\lesssim\theta_{\rm N}$ & \cmark &  All \\ 
Microscopic scale & $b=0$ & RS? & $q\ll1$ & \xmark\,$^\clubsuit$ & All \\
Microscopic scale & $b\sim1$ & FS, RS? & \cmark & \xmark\,$^\clubsuit$ & All \\
\multirow{2}{*}{Microscopic scale} &  
\multirow{2}{*}{$B_{\rm rnd}$+non-axisymmetry$^\diamondsuit$}
& \multirow{2}{*}{FS, RS?} & \multirow{2}{*}{$b\sim1$ or mild asymmetry} & \multirow{2}{*}{\cmark} & {? (Expect} \\
 & 
 &  &  
 &  & LC variability)\\
Combination & $B_{\rm ord}+B_{\rm rnd}$ & FS, RS? & $I_{\rm rnd}\gtrsim100\left(\frac{\Pi}{0.7\%}\right)^{-1}I_{\rm ord}$ & $P_{L,\textrm{ord}}\sim P_{L,\textrm{rnd}}$  & All \\
\multirow{2}{*}{Combination} & \multirow{2}{*}{$B_{\rm ord}$+patches/turbulence} & \multirow{2}{*}{RS} & 
$\thetaB\lesssim\theta_{\rm N}$ & \multirow{2}{*}{$P_{L,\textrm{ord}}\lesssim P_{L,\textrm{p/t}}$} & \multirow{2}{*}{All} \\
& & & $I_{\rm p/t}\gtrsim100\left(\frac{\Pi}{0.7\%}\right)^{-1}I_{\rm ord}$ & & \\
\hline
\end{tabularx}
\tablenotetext{*}{``Large-scale'' and ``microscopic-scale'' $B$-fields are sometimes referred to in the literature as ``hydrodynamic-scale'' and ``plasma-scale'', respectively.}
\tablenotetext{\dag}{Shock-compressed}
\tablenotetext{\ddag}{From turbulence or a rapidly growing coherence length of the shock-generated field}
\tablenotetext{\clubsuit}{\,\,Requires a sufficiently non-axisymmetric jet configuration.}
\tablenotetext{\diamondsuit\,}{\,\,This can arise, e.g., from a non-axisymmetric jet with $E_{\rm k,iso}(\theta,\phi)$, density clumps in the external medium or non-axisymmetric refreshed shocks, which all create brighter and dimmer regions within the observed region.}
\tablenotetext{}{In column 4, $\theta_{\rm N}=10^{-3}\left(\frac{\Gamma}{10}\right)^{-1}\left(\frac{\Pi_{\rm N}}{0.7\%}\right)^{2/d}$, where $\Gamma$ is the shock Lorentz factor, $d$ is the number of dimensions (i.e., 2D or 3D), and $\Pi_{\rm N}\equiv\Pimax/{\sqrt{N}}$ for $N$ visible patches of coherence length $\thetaB$. $I_{\rm ord}$, $I_{\rm rnd}$, and $I_{\rm p/t}$ refer to the Stokes $I$ (total) flux density from the ordered, random, and patchy/turbulent magnetic field, respectively.}
\label{tab:modelconstraints}
\end{table*}

\subsection{Patchy-Field Models}
\label{sec:patchyfields}
We now discuss our polarization observations in the context of the patchy-field model (third model in Table~\ref{tab:modeltypes}), in which the magnetic field is ordered within patches of angular size $\theta_B$ but varies randomly between patches. In this framework, the observed polarization degree constrains the magnetic-field coherence scale through
\begin{align}
\thetaB\sim\theta_{\rm N}\equiv10^{-3}\left(\frac{\Gamma}{10}\right)^{-1}\left(\frac{\PiL}{0.7\%}\right)^{2/d}, 
\end{align}
where $\Gamma$ is the Lorentz factor of the emitting region and $d$ is the dimensionality of the patch distribution (2D or 3D). The observed polarization degree is reduced by averaging, 
\begin{equation}
\Pi_L \approx \frac{\Pimax}{\sqrt{N}},
\end{equation}
where $N$ is the number of visible coherent patches. The polarization position angle is expected to vary randomly with time. 

Interpreting a given measurement in this framework requires knowledge of $\Gamma(R)$ at the time of observation. As discussed in Section~\ref{sec:modelsummary}, the evolution of $\Gamma$ depends on whether the emission is dominated by the FS or RS. For the FS, $\Gamma(R)$ follows the BM solution, whereas for the RS the evolution may be substantially different depending on whether the reverse shock is relativistic or non-relativistic \citep{ks00}. A robust determination of $\thetaB$ therefore requires decomposition of the observed emission into FS and RS components together with a normalization constraint\footnote{This can be achieved either by deriving the initial Lorentz factor of the ejecta, $\Gamma_0$, from the RS emission or by estimating the explosion energy and circumburst density and computing the FS Lorentz factor from the Blandford--McKee solution.} on $\Gamma(R)$. Such modeling is beyond the scope of the present work. We therefore restrict ourselves to order-of-magnitude estimates for four events: GRBs\,171205A, 190114C, 191221B, and 221009A. 

\paragraph{GRB\,171205A:}
Our deepest upper limit is obtained for GRB\,171205A, for which $\PiL<0.11\%$ at $t\sim36$\,days. Even under the conservative assumption that the radio emission arises from a non-relativistic outflow with $\Gamma\sim1$, the patchy-field model implies a coherence scale of $\thetaB\lesssim2\times10^{-3}$\,rad.

\paragraph{GRB\,190114C:}
ALMA polarimetry of this event at $0.11$--0.20\,days has previously been interpreted in the context of the RS emission with a patchy field, yielding a coherence scale of $\thetaB\approx10^{-3}$\,rad in the ejecta \citep{lag+19}. If the observed mm emission at the time of the subsequent polarization detections $1.1$--3.1\,days ($\tovertjet\approx0.3$--0.7) is now dominated by the FS, a similar argument, using the FS parameters of \citet{lag+19} (for which $\Gamma_{\rm FS}\approx15$ at 1--3\,days) yields an identical constraint\footnote{This is because \citet{lag+19} used $\Gamma_{\rm ej}\approx15$ at the time of the RS polarization measurement.} of $\thetaB\approx10^{-3}$ for an intrinsic maximum polarization degree, $\Pi_0\sim0.6$. 

\paragraph{GRB\,191221B:}
We measure $\PiL=(0.68\pm0.15)\%$ at 97.5\,GHz at $t\sim0.48$\,days. \citet{cww+24} model the multi-wavelength observations of this event as FS emission from a two-component jet and find that the radio emission at the time of our polarization detection is dominated by a jet component with $E_{\rm K,iso}\approx5\times10^{53}$\,erg expanding into a uniform-density ($k=0$) medium with $n_0\approx20\,{\rm cm}^{-3}$. These parameters imply $\Gamma_{\rm FS}\approx7$ from the BM solution at the time of the polarization measurement, yielding $\thetaB\sim10^{-3}$\,rad.

In the patchy-field model, the polarization is expected to decrease with time as an increasing number of coherent patches become visible. At our second epoch ($t\sim1.46$\,days), the BM solution predicts $\Gamma_{\rm FS}\sim4.7$, corresponding to an expected polarization of $\PiL\sim0.39\%$. This is consistent with the observed upper limit of $\PiL<0.46\%$.

\paragraph{GRB\,221009A:} At both ALMA epochs (145\,GHz at $t\sim3.39$\,days and 97.5\,GHz at $t\sim5.32$\,days), the mm-band emission is dominated by the reverse shock \citep{gg23}. The measured polarization fractions are $\Pi_L=(1.42\pm0.09)\%$ and $(1.62\pm0.12)\%$, respectively, with spectral indices $\beta=-0.18\pm0.03$ and $-0.28\pm0.02$. For both the relativistic and non-relativistic reverse-shock models considered by \citet{gg23}, the Lorentz factor of the emitting material has already declined to $\Gamma_{\rm RS}\sim1$ by these epochs. Adopting this value yields an inferred coherence scale of order
$\thetaB \sim 2\times10^{-2}\ {\rm rad}$, with only a weak dependence on the assumed dimensionality of the patch distribution. The measured spectral indices imply $\Pimax=(1-\beta)/(5/3-\beta)\approx65\%$, so the observed polarization corresponds to averaging over roughly $N\sim 2\times10^3$
coherent patches. Unlike the FS-dominated case discussed above, little evolution is expected once the emitting material becomes only mildly relativistic, as the visible region no longer grows rapidly with time. The approximately constant $\PiL$ and minor variation in $\chi$ observed between the two epochs is therefore not inconsistent with the patchy-field interpretation.

In summary, the patchy-field model provides a plausible explanation for the observed polarization properties of these four events. The inferred coherence scales span $\thetaB\sim10^{-3}$\,rad for the FS-dominated event GRB\,191221B (and possibly the later epochs of GRB\,190114C) to $\thetaB\sim2\times10^{-2}$\,rad for the RS-dominated event GRB\,221009A. The deep upper limit on GRB\,171205A is also consistent with this framework, implying $\thetaB\lesssim2\times10^{-2}$\,rad, under the conservative assumption that the emitting region is non-relativistic.

\subsection{Discussion Summary}
We summarize the observational results and their interpretation in Tables~\ref{tab:summary-by-event} and \ref{tab:modelconstraints}. Table~\ref{tab:summary-by-event} provides an event-by-event summary of the polarization detections, upper limits, and their implications for the magnetic-field structure and viewing geometry. Table~\ref{tab:modelconstraints} summarizes the classes of theoretical models considered in this work together with the constraints imposed by the ALMA polarimetric observations. Overall, the data indicate low levels of linear polarization ($\PiL\lesssim3\%$) and generally disfavor large-scale ordered magnetic-field models, while several models involving random, patchy, or mixed magnetic-field configurations remain broadly consistent with the observations.

\section{Conclusions} \label{sec:conclusion}
We present the first radio/mm polarimetric sample of GRB afterglows comprising 24 ALMA polarimetric observations of 11 long-duration GRB afterglows. We detect linear polarization in several events, with significant detections spanning $\PiL\approx0.6\%$--$2.4\%$, plus one tentative and one marginal detection. For non-detections, we obtain deep limits, with a median upper limit of $\PiL\lesssim1\%$. We interpret these observations in the context of synchrotron emission from both shocked ejecta and shocked circumstellar material. A summary of our main findings is as follows:

\begin{enumerate}
    \item GRB\,190114C provides the best-sampled polarization our sample. The early-time evolution of $\PiL$ and $\chi$ favors patchy magnetic fields in the RS, while the later epochs are broadly consistent with FS emission and a combination of random-fields in top-hat jets or patchy-field models.
    \item GRB\,190829A yields a strong single-epoch polarization detection. If the mm emission is RS dominated, toroidal magnetic-field models require a nearly on-axis viewing geometry ($q\lesssim0.05$). The uncertain jet-break time prevents a meaningful comparison with FS models.
    \item GRB\,220921A exhibits a rapid increase in $\PiL$, which is inconsistent with RS toroidal-field models, but is broadly consistent with several FS random-field models.
    \item GRB\,221009A represents the highest-significance polarization detections in the sample. The mm-band emission is RS dominated, but neither FS random-field models nor RS toroidal-field models reproduce both the observed polarization evolution and the viewing geometry inferred from broadband afterglow modeling. Patchy RS magnetic fields remain a plausible explanation.
    \item GRB\,171205A shows very deep polarization limits. These limits strongly disfavor RS toroidal-field models, while the lower limit on the jet-break time prevents meaningful constraints on FS models. Several additional single-epoch events provide only limited leverage on models: GRB\,210610B and GRB\,210702A yield upper limits, GRB\,220521A and GRB\,250327B are too faint to constrain the models strongly, GRB\,191221B remains tentative because of possible residual systematics, and GRB\,251013C is a marginal detection that is broadly consistent with both RS and FS models.
\end{enumerate}
Interpreting selected events in the patchy-field framework suggests magnetic-field coherence scales of order $\theta_B\sim10^{-3}$--$10^{-2}$\,rad, with the smallest values inferred for FS-dominated events and the largest values for the RS-dominated GRB\,221009A. 

Detailed modeling of the light curves and polarization observations presented here will place stronger constraints on the magnetic-field structure and viewing geometry of GRB outflows. Future radio/mm polarimetric studies of larger GRB samples, together with analogous observations of other relativistic jet sources such as tidal disruption events and AGN, will reveal how common different magnetic-field configurations are across relativistic outflows and provide new insight into the formation of jets and the physics of relativistic shocks.

\section{Acknowledgments}
This paper makes use of the following ALMA data: \url{ADS/JAO.ALMA\#2017.1.00801.T}, 
\url{ADS/JAO.ALMA\#2018.1.01405.T}, 
\url{ADS/JAO.ALMA\#2018.1.00579.T}, 
\url{ADS/JAO.ALMA\#2019.1.01482.T}, 
\url{ADS/JAO.ALMA\#2021.1.00657.T},
\url{ADS/JAO.ALMA\#2022.1.00009.T},
\url{ADS/JAO.ALMA\#2024.1.01174.T},
and  \url{ADS/JAO.ALMA\#2025.A.00003.T}. 
ALMA is a partnership of ESO (representing its member states), NSF (USA) and NINS (Japan), together with NRC (Canada), NSTC and ASIAA (Taiwan), and KASI (Republic of Korea), in cooperation with the Republic of Chile. The Joint ALMA Observatory is operated by ESO, AUI/NRAO and NAOJ. The National Radio Astronomy Observatory and Green Bank Observatory are facilities of the U.S. National Science Foundation operated under cooperative agreement by Associated Universities, Inc. This work made use of data supplied by the UK Swift Science Data Centre at the University of Leicester. N.F. acknowledges support from the National Science Foundation Graduate Research Fellowship Program under Grant No. DGE-2137419.

This work benefited from the use of generative AI (GAI) to improve visualizations \citep{chatgpt, claude}. All GAI outputs were carefully verified under human supervision.

\facilities{ALMA, ADS}

\software{
\texttt{astropy} \citep{astropy:2013,astropy:2018,astropy:2022}, \texttt{Jupyter} \citep{2007CSE.....9c..21P}, \texttt{matplotlib} \citep{Hunter:2007}, \texttt{numpy} \citep{numpy}, \texttt{pandas} \citep{mckinney_data_2010}, \texttt{python} \citep{python}, \texttt{scipy} \citep{2020SciPy-NMeth}, CASA \citep{mcmullin_casa_2007, casa_team_casa_2022}, {\tt ChatGPT} \citep{chatgpt}, and {\tt Claude} \citep{claude}.
}

\bibliography{grb_alpha,software,gcn,otherbib}{}
\bibliographystyle{aasjournalv7}

%%%%%%%%%%%%%%%%%%%%%%%
% APPENDICES
%%%%%%%%%%%%%%%%%%%%%%%

\appendix

\section{Custom ALMA Polarization Calibration Pipeline}
\label{sec:obspipe}

We build a custom calibration pipeline for reduction of ALMA full-Stokes observations, with the following steps\footnote{Most of these steps follow the NRAO 3C286 Polarization Calibration tutorial: \url{https://casaguides.nrao.edu/index.php/3C286_Polarization}}: 
\begin{enumerate}
    % Script 1: alma_precal.py
    \item Pre-calibration
    \begin{enumerate}
        \item If necessary, convert the raw ASDM files into MS file using CASA {\tt importasdm}. Each MS file corresponds to an individual observation and is treated separately for pre-calibration.
        \item Perform {\it a priori} flagging on the dataset including (i) flagging autocorrelations of non-radiometer data, (ii) flagging pointing and sideband ratio calibration scans, and (iii) flagging data for antenna shadowing.
        \item Run CASA {\tt fixsyscaltimes}
        \item Create and apply a water vapor radiometer gain calibration table.
        \item Create and apply a system temperature gain calibration table.
        \item Concatenate the MS files into a single file and remove non-science spectral windows.
    \end{enumerate}

    % Script 2: calibrate.py
    \item Parallel Hand Calibration
    \begin{enumerate}
        \item Set the flux-density scale. Since the flux-density calibrators are variable, we query the ALMA flux calibrator service page\footnote{\url{https://almascience.nrao.edu/sc/}} for the flux at a time closest to that of the science observation.  
        \item Calibrate the bandpass. We use {\tt refantmode=`strict'} to enforce\footnote{This causes all antennas to be flagged if the reference antenna is flagged at any time, thus requiring a reference antenna that is not only near the array center but one that is present for most or all of the observations.} a single reference antenna, required for polarization calibration.
        \item Derive the complex gain calibration tables for the bandpass, flux, and phase calibrators. The phase and amplitude solutions are derived separately. 
        \item Bootstrap the flux density scale from the flux calibrator to the gain calibrator.
        \item Apply the parallel hand bandpass, complex gain, and flux calibration to all (except leakage) calibrators. 
    \end{enumerate}

    \item Polarization Calibration
    \begin{enumerate}
        \item Solve for the complex gains on the leakage calibrator.
        \item Calculate the linear polarization of the leakage calibrator and plot the gain amplitude ratio of the two parallel hands per scan. Visually inspect this plot and locate the scan where the gain ratio is closest to its mean value (corresponding to a time when the source polarization dominates the cross-hand phase relative to the instrumental polarization). Use this scan for the cross hand delay calibration.
        \item Derive the cross-hand delay calibration. Confirm that applying the cross-hand delays to the leakage calibrator removes any frequency-dependent phase ramps previously present within each spectral window. 
        \item Derive the residual cross-hand phase of the reference antenna and the intrinsic polarization of the gain and leakage calibrators. 
        \item Revise the complex gains using the full polarization model of the gain calibrator. Running CASA {\tt polfromgain} on the leakage calibrator after this step should reveal a gain amplitude ratio consistent with $\approx 1$.
        \item Solve for the leakage (D-terms) using CASA {\tt polcal}. Check that the D-terms are small ($\lesssim 5$\%) and relatively stable with frequency.
        \item Derive the cross-hand gain amplitude calibration table from the leakage calibrator.
        \item Apply all calibration tables to the target field.
    \end{enumerate}
\end{enumerate}

\section{Tests of Instrumental Systematics}\label{sec:sys-checks}
Radio polarization measurements are susceptible to a variety of instrumental systematics that can introduce spurious polarization signals \citep{lhc20}. Assessing these effects is therefore essential when interpreting polarization detections. Below, we summarize the systematic checks used in this work and briefly describe how each is performed.

\subsection{Checks for Instrumental Systematics}\label{app:systematic-steps}
\begin{enumerate}
    \item Check that the residual polarization of the leakage calibrator is consistent with noise after polarization calibration. To perform this check we use CASA {\tt polfromgain} to calculate the residual Stokes $QUV$ of the leakage calibrator and verify that $\left|{\rm QUV}\right|\lesssim 0.1\%$ of Stokes I. 
    \item Check that the ratio of the parallel hand gains (e.g. XX/YY) is uniform and low for all antennas after polarization calibration. This is done by plotting the ratio of the parallel hand gains as a function of frequency and checking that this ratio is within $\approx 1.0 \pm 0.1$ after calibration.
    \item Check that the rms gain ratio is uniform across antennas after polarization calibration. This is done by calculating the mean and RMS of the X/Y gain amplitude polarization ratio for each antenna and averaged over time. We verify that the polarization ratio is close to unity with an RMS of $\lesssim 0.05$ for each antenna.
    \item Check the magnitude of the leakage (D-terms) is consistent with known values for the array. We verify this by plotting the D-term amplitude as a function of frequency. For ALMA, we typically expect D-term amplitudes $\lesssim 5\%$ \citep{nnp+16,lhc20}. 
    \item Check the net (averaged over baselines) instrumental polarization is consistent with expectation for the array. These values are reported in the CASA log during our calibration process.  
    \item When self-calibrating, ensure that solutions are calculated only at intervals larger than the minimum theoretical solution interval. The minimum theoretical solution interval can be found using \citep{ALMA_memo-620}
    \begin{align} \label{eq:solint_min}
        t_{\rm solint} > 9(N-3)t_{\rm int}\left(\frac{I_{\rm pk}}{\sigma_I}\right)^{-2},
    \end{align}
    where $N$ is the number of antennas, $t_{\rm int}$ is the on source integration time, $I_{\rm pk}$ is the peak Stokes I flux density, and $\sigma_I$ pre-self-calibration off-source RMS. Assuming typical values for ALMA observations of GRB afterglows gives a minimum solution interval of
    \begin{align}
        t_{\rm solint} \gtrsim 4.3~{\rm min}~\left(\frac{N-3}{43}\right)\left(\frac{t_{\rm int}}{2{\rm hr}}\right)\left(\frac{I_{\rm pk}}{10 {\rm mJy}}\right)^{-2}\left(\frac{\sigma_I}{100 {\rm \mu Jy}}\right)^2
    \end{align}
    \item When self-calibrating, keep the phase offset between the two polarizations fixed. This is done by setting \texttt{gaintype=`T'} in \texttt{gaincal}. 
    \item Test the source for time-variable polarization by dividing the data up into time bins. Any fluctuations in the polarization signal as a function of time should be consistent within statistical uncertainty. 
    \item Test the source for frequency-dependent polarization by dividing the data up into frequency bins (e.g. by spectral window). Any fluctuations in the polarization signal as a function of frequency should be consistent within statistical uncertainty. 
    \item Image the gain calibrator, check that it appears as a point source, and verify that any secondary peaks in both Stokes $Q$ and $U$ images are consistent with noise. To check whether any secondary peaks in both Stokes $Q$ and $U$ are consistent with noise we compare the peak flux density of the noise peak with the flux density of the Q and U detections in the gain calibrator. Any secondary noise peaks should be $\lesssim3\sigma$ relative to the $Q$ and $U$ image rms.  
    \item Test the gain calibrator for time and frequency-dependent polarization by dividing the data into time and frequency bins.
    \item Check Stokes V for spurious polarization signal in both the target and the gain calibrator. We perform this test by fitting a point source at the Stokes $I$ position  and then comparing the flux density to the Stokes $V$ image rms and to the Stokes $I$ flux density. Note that $V/I\lesssim0.1$\% is consistent with known systematics for ALMA polarization images \citep{nnp+16}. 
    \item Compare the results against the ALMA pipeline to check for other calibration and imaging artifacts that could be influencing the polarization detection.
\end{enumerate}
Our publicly available polarimetric calibration pipeline can generate diagnostic plots required for some of these tests during run time. In addition, we also provide scripts for many of the other tests (e.g., a script to image the source in full polarization as a function of time). We next highlight the application of some of these tests on the GRB sample presented in this work. 

\subsection{Checks 1 \& 5: Quantifying Instrumental Systematics}
After applying the calibrator polarization model to the leakage calibrator, regenerating a leakage solution should return a vector consistent with noise (Check 1). We obtain ${Q,U}/I$ measurements by running {\tt polfromgain} on the calibrated leakage calibrated data and present the results for the GRBs with polarization detections in Table \ref{tab:systematics}. We find residual polarization $\lesssim 0.15\%$ in all cases. 

During calibration, the CASA log reports the (real part of the) instrumental polarization averaged across all baselines per spectral window and these values are expected to be small. For our polarization detections, we report the minimum and maximum of these values in each case in Table \ref{tab:systematics}. 

\begin{table}
\centering
\caption{Instrumental Systematics}
\begin{tabular}{llrrrr}
\toprule
 GRB & Date & Q Res. Pol.\tablenotemark{*} & U Res. Pol.\tablenotemark{*} & Min. Instr. Pol.\tablenotemark{\dag} & Max. Instr. Pol.\tablenotemark{\dag} \\
 & & (\% of I) & (\% of I) & (\% of I) & (\% of I) \\
\hline
190114C & 2019-01-15 & -0.009& -0.010 & 0.08 & 0.28  \\
190114C & 2019-01-16 & 0.033 & -0.032 & 0.14 & 0.31  \\
190829A & 2019-09-01 & -0.004 & 0.022 & 0.07 & 0.20 \\
191221B & 2019-12-22 & -0.009 & -0.010 & 0.11 & 0.31 \\
220921A & 2022-09-28 & 0.139 & -0.079 & 0.11 & 0.35 \\
221009A & 2022-10-12 & -0.019 & -0.002 & 0.05 & 0.06 \\
221009A & 2022-10-14 & -0.001 & -0.031 & 0.22 & 0.34 \\
251013C & 2025-10-22 & 0.127 & -0.050 & 0.15 & 0.27 \\
\hline
\end{tabular}
\tablenotetext{*}{Residual polarization of the leakage calibrator from {\tt polfromgain} reported in \% of Stokes I of the leakage calibrator. The result of systematic check 1 in Appendix \ref{app:systematic-steps}.}
\tablenotetext{\dag}{The minimum and maximum instrumental polarization for the observation reported in \% of Stokes I of the target. The result of systematic check 5 in Appendix \ref{app:systematic-steps}.}
\label{tab:systematics}
\end{table}

\subsection{Check 4: Leakage ($D$-term) Amplitudes}

\begin{figure*}
    \centering
    \includegraphics[width=\linewidth]{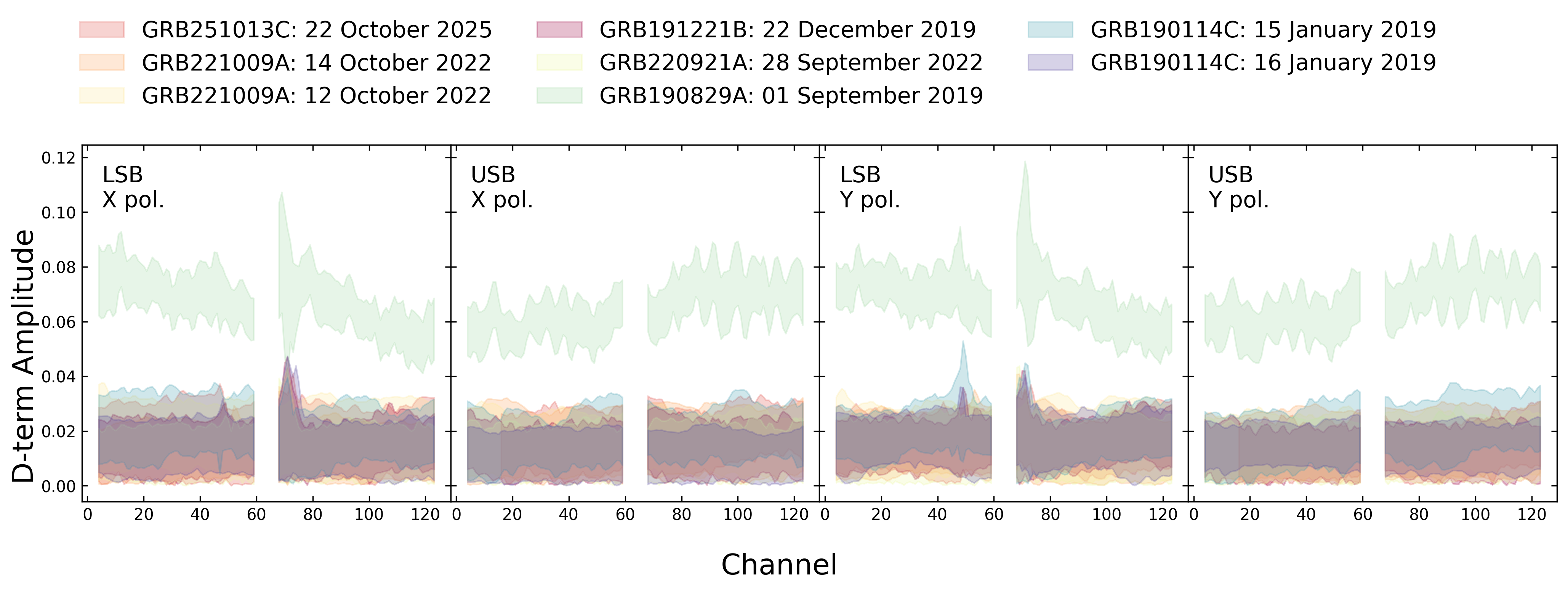}
    \caption{Leakage ($D$-term) amplitudes as a function of frequency for each observation (colors) with statistically significant detections (Table~\ref{tab:grb_pol}). GRB\,190829A exhibits larger $D$-terms than the others, the cause for which remains unknown; those data otherwise pass our quality checks.}
    \label{fig:dterm}
\end{figure*}

We plot the $D$-term amplitudes for all GRBs with a $3\sigma$ detection in either Stokes $Q$ or $U$ in Figure~\ref{fig:dterm}, split into the lower and upper sidebands and by polarization. Occasional spikes in D-terms are known to occur; for observations where these extend beyond $\gtrsim5\%$ on certain antennas, we flag the corresponding channels for those antennas. The significant outlier in these plots is GRB\,190829A with high $D$-terms that are also noted in the ALMA QA report. No obvious cause was identified for the high leakage in this case. 

\subsection{Checks 8, 9, \& 11: Variability Checks}

\begin{figure*}
    \includegraphics[width=0.33\linewidth]{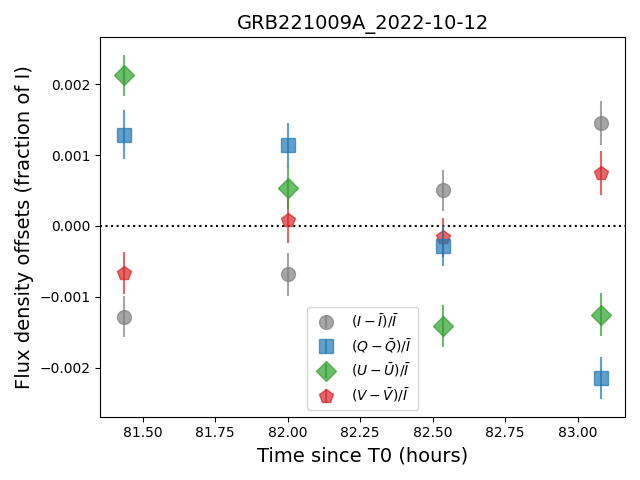}
    \includegraphics[width=0.33\linewidth]{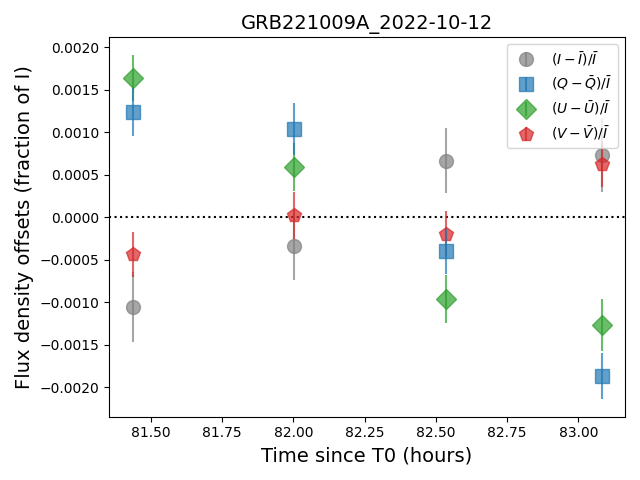}
    \includegraphics[width=0.33\linewidth]{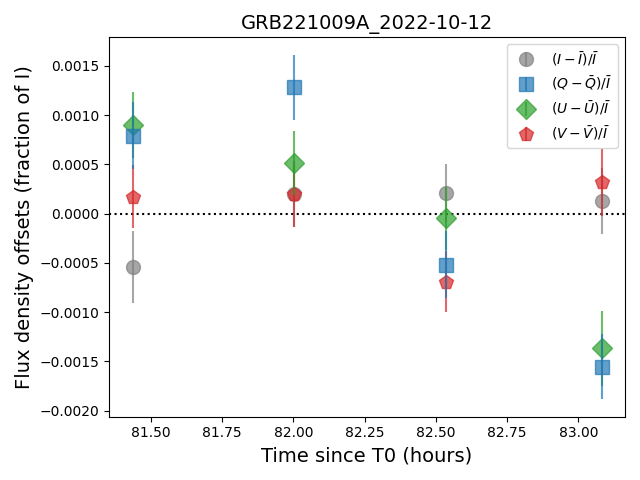}\\
    \includegraphics[width=0.33\linewidth]{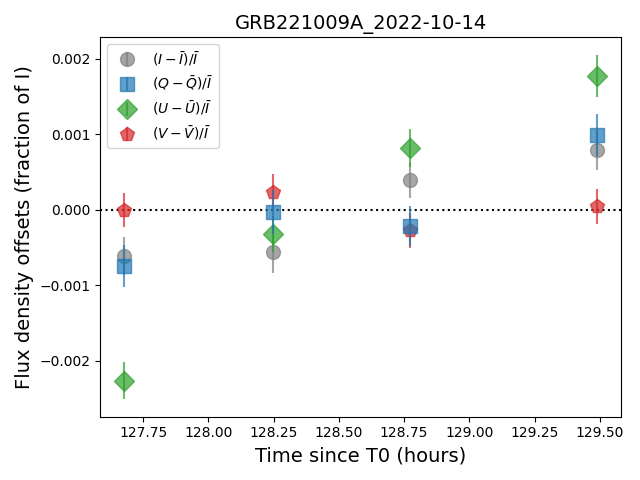}
    \includegraphics[width=0.33\linewidth]{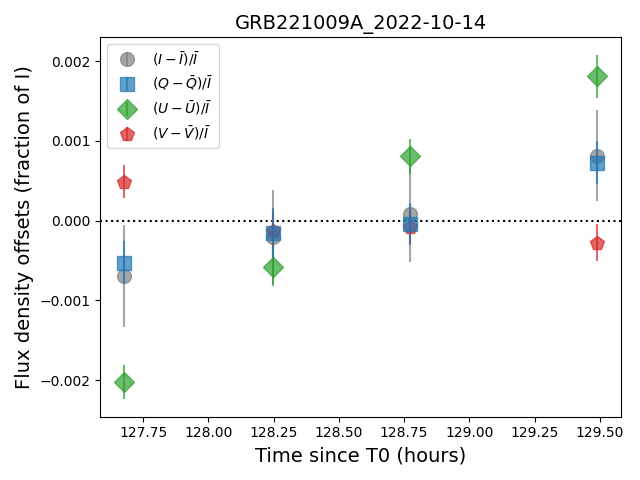}
    \includegraphics[width=0.33\linewidth]{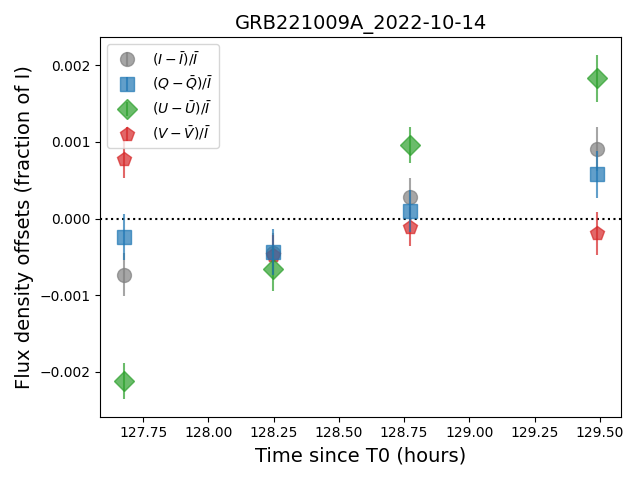}\\
    \includegraphics[width=0.33\linewidth]{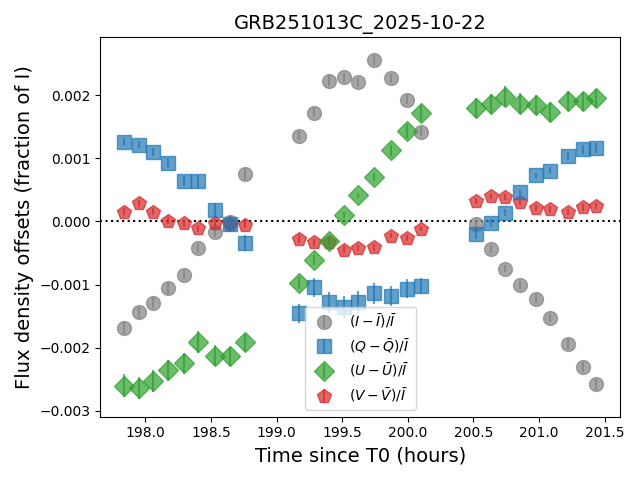}
    \includegraphics[width=0.33\linewidth]{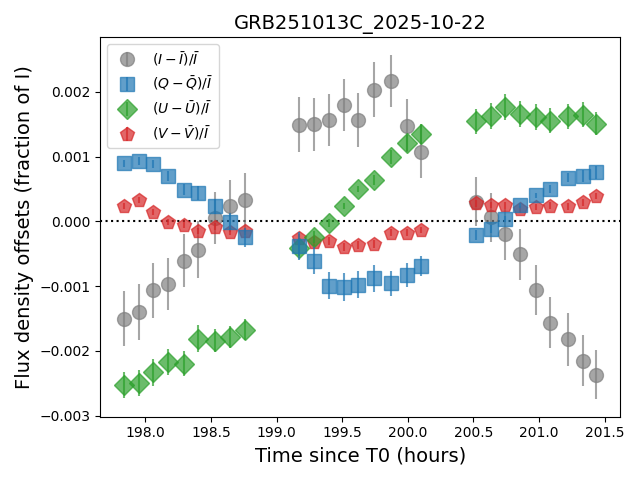}
    \includegraphics[width=0.33\linewidth]{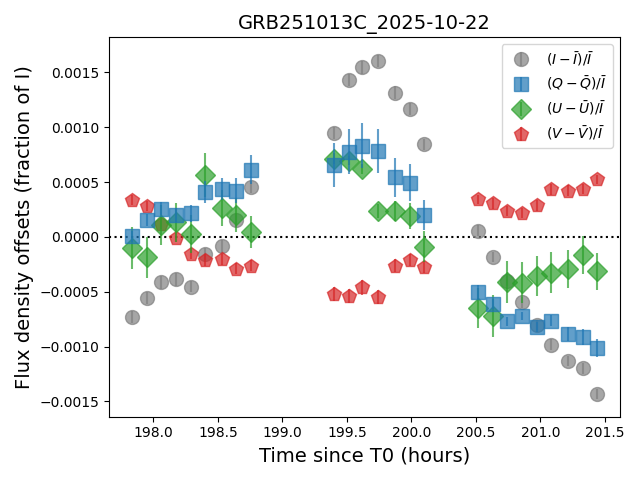}    
    \caption{Fractional variation (relative to Stokes $I$, gray; itself plotted relative to its mean value in each plot) in Stokes $Q$ (blue), Stokes $U$ (green), and Stokes $V$ (red) for the observed \textbf{gain calibrator} as a function of time within each epoch of observations of GRB\,221009A on 2022-10-12 in ALMA Band 3 (top row), GRB\,221009A on 2022-10-14 in ALMA Band 4 (middle row) and of GRB\,251013C in ALMA Band 3 (bottom row), split into lower and upper side bands (left and right columns, respectively) compared with the full band (center column). Observations of GRB\,251013C were taken in C-8 configuration and employed a short cycle time of 83\,s, yielding with a large number of calibrator scans.}
    \label{fig:polsys-timevar}
\end{figure*}

For most of the GRB afterglows and gain calibrators observed in our projects, we do not expect significant variability in the polarization properties as a function of time and frequency during the course of the observations. We therefore split the data in these cases into three frequency bins (upper sideband, lower sideband, and full band), with each frequency bin being further divided in time into multiple scan blocks. No significant variability was found in the target data. We show some examples of gain calibrator time variability in Fig.~\ref{fig:polsys-timevar}, which may be instrumental or intrinsic to the source. Given the limited literature on short-timescale polarization variability for these calibrators, we are unable to interpret these further at this time; however, we note that this variability may indicate previously unquantified instrumental systematics in ALMA polarization observations. 

\subsection{Checks 6 \& 7: Verifying Self-Calibration}
\begin{figure*}
    \centering
    \includegraphics[width=0.45\linewidth]{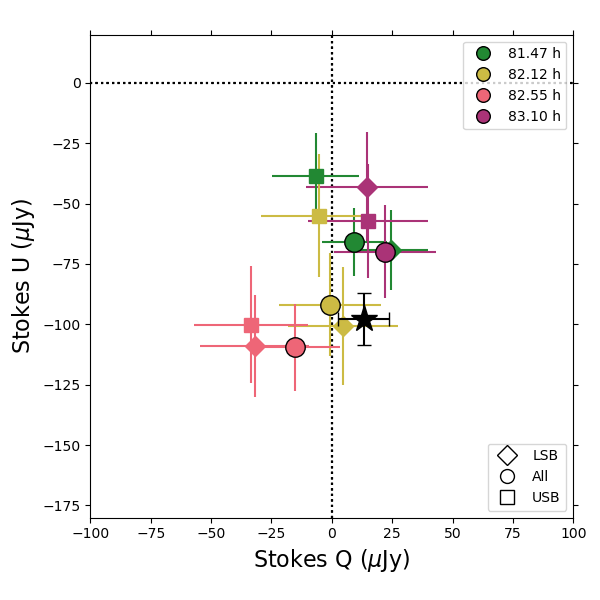}
    \includegraphics[width=0.45\linewidth]{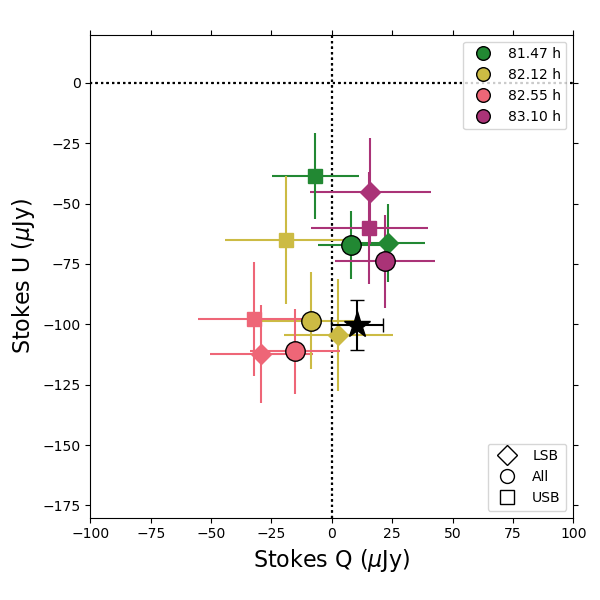}\\
    \includegraphics[width=0.45\linewidth]{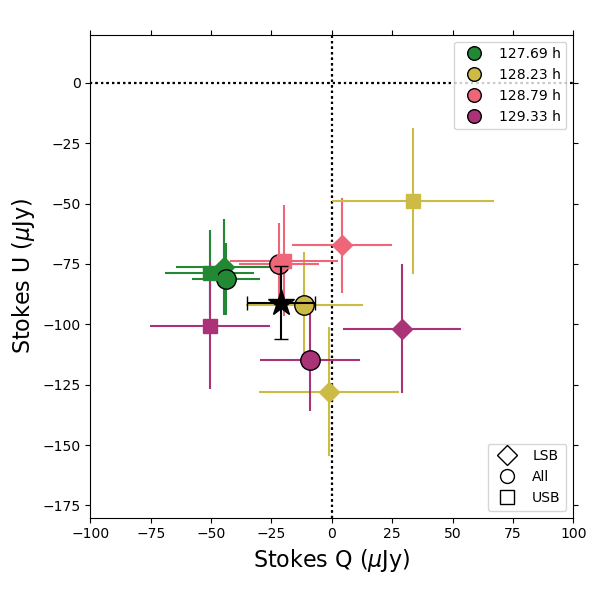}
    \includegraphics[width=0.45\linewidth]{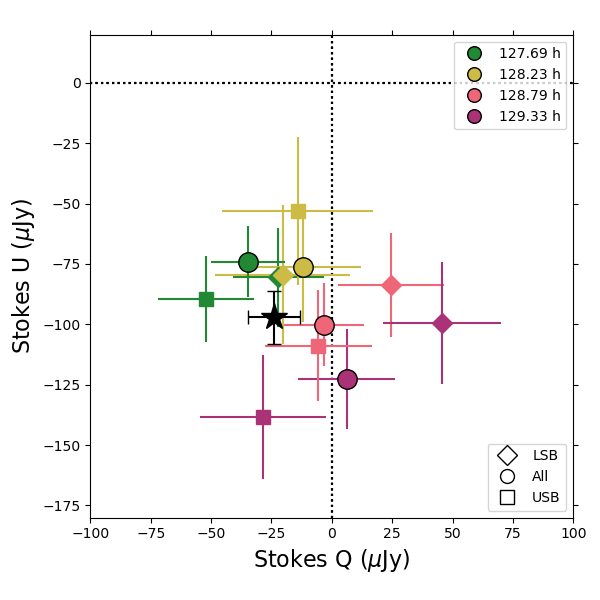}    
    \caption{Polarimetric observations of GRB\,221009A at 145 GHz (ALMA Band 4; top row) and 97.5 GHz (ALMA Band 3; bottom row) in the Stokes $QU$ plane before (left column) and after (right column) self-calibration, divided by time (colored symbols) and by frequency into lower sideband (diamond), upper sideband (square) and full (circles). The observations are consistent with negligible Stokes $Q$ but a significant Stokes $U$ signal both before and after self-calibration, indicating that these results are robust to this process.}
    \label{fig:selfcal-check}
\end{figure*}

The brightest targets benefit from self-calibration in reducing residual gain errors. However, because self-calibration is a non-linear process, inappropriate solution intervals or models can bias the recovered images, amplify noise, and produce spurious polarized emission and source structure \citep{mvm08,lhc20}. We validate that the observed polarized flux density for our detections is robust to any self-calibration applied, and present one example for GRB\,221009A in Fig.~\ref{fig:selfcal-check}. 

\section{ALMA Pipeline Polarization Results}

\begin{table*}
\centering
\caption{Radio polarization measurements derived from the ALMA pipeline images. Reported uncertainties correspond to the 68\% confidence interval derived from Monte Carlo realizations of the measured Stokes parameters assuming Gaussian errors. Upper limits are quoted at the 95\% confidence level for non-detections.}
\begin{tabular}{l l l c c c c c c c c c}
\hline
GRB & Date & $\nu$ & $t$ & $I$ & $Q$ & $U$ & $V$ & $P_L$ & $\PiL$ & $\chi$ & Z \\
 & UTC & GHz & d & $\rm mJy$ & $\rm \mu Jy$ & $\rm \mu Jy$ & $\rm \mu Jy$ & $\rm \mu Jy$ & \% & ($^\circ$) &  \\
\hline
171205A & 2017 Dec 16 & 97.6 & 11.17 & $15.409_{-0.142}^{+0.142}$ & $-18_{-10}^{+10}$ & $26_{-10}^{+10}$ & - & $<49$ & $<0.32$ & - & $2.47$ \\
171205A & 2018 Jan 10 & 97.5 & 36.04 & $21.690_{-0.174}^{+0.174}$ & $8_{-8}^{+8}$ & $1_{-7}^{+7}$ & - & $<24$ & $<0.11$ & - & $-0.21$ \\
190114C & 2019 Jan 15 & 97.5 & 1.12 & $2.647_{-0.013}^{+0.013}$ & $11_{-8}^{+8}$ & $-15_{-8}^{+8}$ & - & $<32$ & $<1.21$ & - & $1.54$ \\
190114C & 2019 Jan 16 & 97.5 & 2.06 & $1.994_{-0.020}^{+0.020}$ & $25_{-8}^{+8}$ & $-11_{-8}^{+8}$ & - & $<41$ & $<2.06$ & - & $2.77$ \\
190114C & 2019 Jan 17 & 97.5 & 3.06 & $1.952_{-0.012}^{+0.012}$ & $17_{-8}^{+8}$ & $1_{-7}^{+7}$ & - & $<32$ & $<1.65$ & - & $1.20$ \\
190114C & 2019 Jan 24 & 97.5 & 10.09 & $1.041_{-0.011}^{+0.011}$ & $9_{-8}^{+8}$ & $3_{-9}^{+9}$ & - & $<26$ & $<2.51$ & - & $0.07$ \\
190114C & 2019 Apr 11 & 97.5 & 86.97 & $0.042_{-0.005}^{+0.005}$ & $7_{-8}^{+8}$ & $2_{-8}^{+8}$ & - & $<22$ & $<54.39$ & - & $-0.33$ \\
190829A & 2019 Sep 01 & 97.5 & 2.50 & $5.697_{-0.025}^{+0.025}$ & $-23_{-4}^{+4}$ & $11_{-4}^{+4}$ & - & $26_{-4}^{+4}$ & $0.45_{-0.07}^{+0.07}$ & $77.6_{-4.6}^{+4.7}$ & $6.31$ \\
191221B & 2019 Dec 22 & 97.5 & 0.48 & $3.573_{-0.007}^{+0.007}$ & $8_{-6}^{+6}$ & $11_{-5}^{+5}$ & - & $<23$ & $<0.64$ & - & $1.75$ \\
191221B & 2019 Dec 23 & 97.5 & 1.46 & $4.827_{-0.009}^{+0.009}$ & $-7_{-6}^{+6}$ & $-8_{-7}^{+7}$ & - & $<23$ & $<0.48$ & - & $0.64$ \\
191221B & 2019 Dec 24 & 97.5 & 2.46 & $3.895_{-0.008}^{+0.008}$ & $1_{-11}^{+11}$ & $-17_{-10}^{+10}$ & $-7_{-11}^{+11}$ & $<37$ & $<0.95$ & - & $0.71$ \\
191221B & 2019 Dec 27 & 97.5 & 5.41 & $2.402_{-0.009}^{+0.009}$ & $-13_{-13}^{+13}$ & $-2_{-12}^{+12}$ & $10_{-12}^{+12}$ & $<37$ & $<1.54$ & - & $-0.22$ \\
210610B & 2021 Jun 11 & 97.5 & 0.30 & $2.148_{-0.012}^{+0.012}$ & $-2_{-5}^{+5}$ & $3_{-5}^{+5}$ & - & $<14$ & $<0.66$ & - & $-0.83$ \\
210702A & 2021 Jul 03 & 97.5 & 1.10 & $2.190_{-0.007}^{+0.007}$ & $-1_{-4}^{+4}$ & $9_{-5}^{+5}$ & - & $<18$ & $<0.81$ & - & $0.81$ \\
220521A & 2022 Jul 17 & 97.5 & 56.12 & $0.064_{-0.006}^{+0.006}$ & $-4_{-5}^{+5}$ & $3_{-6}^{+6}$ & - & $<15$ & $<24.48$ & - & $-0.45$ \\
220521A & 2022 Aug 12 & 97.5 & 82.09 & $0.031_{-0.007}^{+0.007}$ & $7_{-6}^{+6}$ & $-6_{-6}^{+6}$ & - & $<21$ & $<77.67$ & - & $0.45$ \\
220921A & 2022 Sep 23 & 97.5 & 1.89 & $4.633_{-0.007}^{+0.007}$ & $8_{-5}^{+5}$ & $16_{-6}^{+6}$ & - & $<28$ & $<0.61$ & - & $2.24$ \\
220921A & 2022 Sep 28 & 97.5 & 6.75 & $1.566_{-0.007}^{+0.007}$ & $33_{-6}^{+6}$ & $22_{-5}^{+5}$ & - & $40_{-6}^{+6}$ & $2.54_{-0.36}^{+0.36}$ & $17.0_{-4.0}^{+4.1}$ & $6.46$ \\
221009A & 2022 Oct 12 & 145.0 & 3.39 & $6.926_{-0.033}^{+0.033}$ & $-10_{-7}^{+7}$ & $-89_{-7}^{+7}$ & - & $90_{-7}^{+7}$ & $1.30_{-0.10}^{+0.10}$ & $-48.1_{-2.2}^{+2.2}$ & $12.47$ \\
221009A & 2022 Oct 14 & 97.5 & 5.32 & $5.822_{-0.014}^{+0.014}$ & $-7_{-9}^{+9}$ & $-98_{-8}^{+8}$ & - & $99_{-8}^{+8}$ & $1.70_{-0.14}^{+0.14}$ & $-47.2_{-2.7}^{+2.7}$ & $12.02$ \\
250327B & 2025 Mar 29 & 97.5 & 1.20 & $0.528_{-0.005}^{+0.005}$ & $3_{-6}^{+6}$ & $2_{-6}^{+6}$ & - & $<16$ & $<3.05$ & - & $-0.89$ \\
251013C & 2025 Oct 21 & 97.5 & 8.24 & $2.862_{-0.009}^{+0.009}$ & $-26_{-7}^{+7}$ & $-9_{-7}^{+7}$ & - & $29_{-7}^{+7}$ & $1.01_{-0.24}^{+0.24}$ & $-78.2_{-6.8}^{+10.1}$ & $3.32$ \\
\hline
\end{tabular}
\label{tab:grb_pol_pipe}
\end{table*}

For comparison with our manual analysis, we also present the results of fitting a point source to $IQUVP$ images obtained from the ALMA archive for each of our targets in Table \ref{tab:grb_pol_pipe}. With few exceptions (see Section~\ref{sec:individual_targets} and Table~\ref{tab:grb_pol}), our manual calibration results in polarization properties consistent with those derived from images obtained from the ALMA archive. 

\section{Estimates of Jet-break Times}
\label{sec:tjet}

\begin{figure*}
    \includegraphics[width=0.33\textwidth]{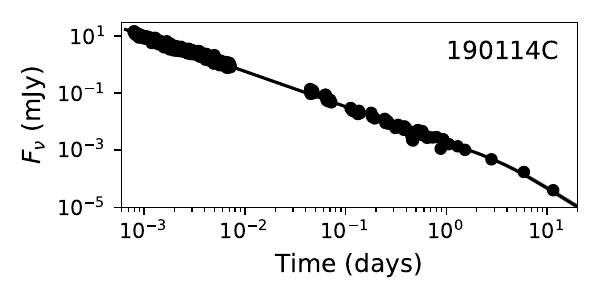} \includegraphics[width=0.33\textwidth]{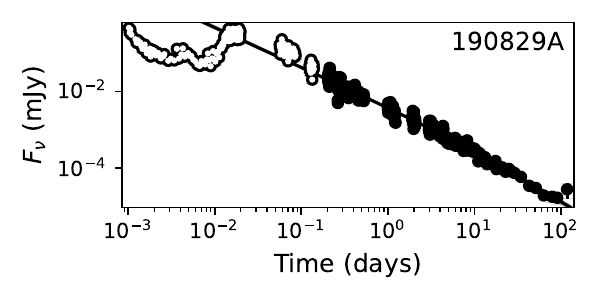}   
    \includegraphics[width=0.33\textwidth]{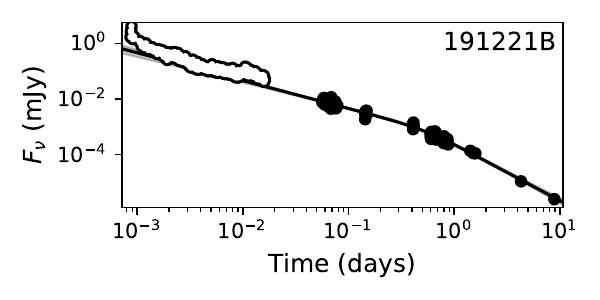}
    
    \includegraphics[width=0.33\textwidth]{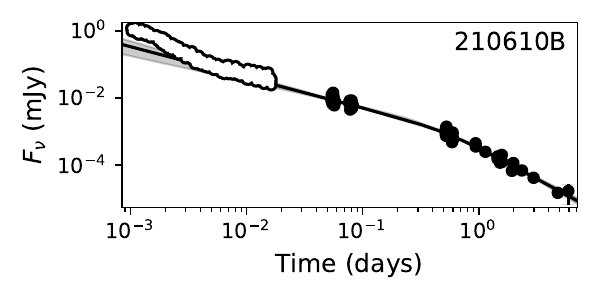}    
    \includegraphics[width=0.33\textwidth]{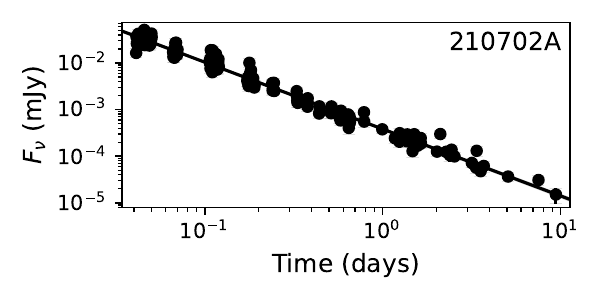}
    \includegraphics[width=0.33\textwidth]{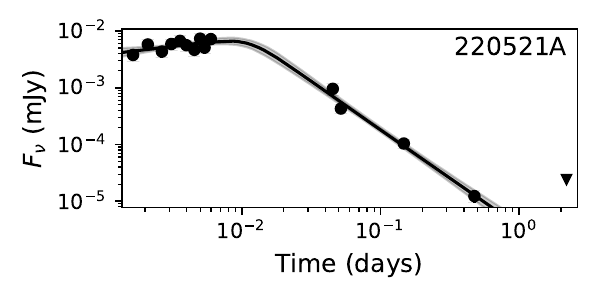}

    \includegraphics[width=0.33\textwidth]{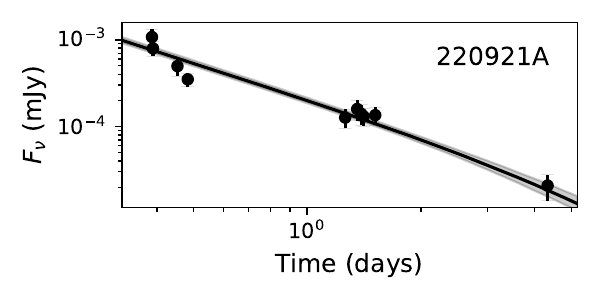}    
    \includegraphics[width=0.33\textwidth]{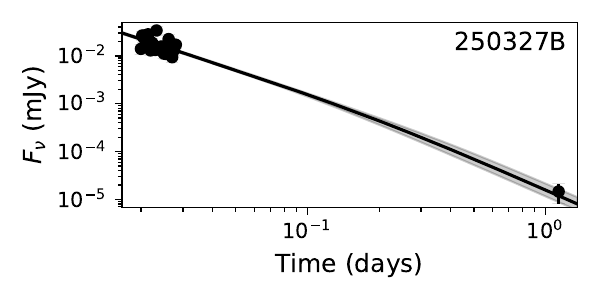}
    \includegraphics[width=0.33\textwidth]{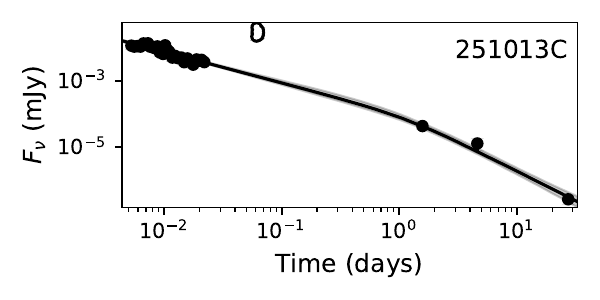}
    \caption{X-ray light curves of GRBs in our sample with broken power-law fits for jet break estimation. The jet-break time for GRB\,250327B is highly uncertain, while only lower limits on $\tjet$ can be derived for GRBs\,190829A and 220521A (Appendix~\ref{sec:tjet}).}
    \label{fig:jetbreaks}
\end{figure*}

In this section, we discuss constraints on jet break times for the events in our sample. We acknowledge that it can be difficult to determine a jet break using the light curve in a single band (e.g., X-ray), since additional features (e.g., plateaus) can result in breaks unrelated to the jet break. Thus, where possible, we use constraints from multi-wavelength modeling. In the remaining cases, we extract this information from the X-ray light curves, with the caveat that these may be subject to a larger uncertainty.

For GRB\,171205A, we use the lower limit of $\tjet\gtrsim71$\,days reported in \citep{mc21}. There is some debate about the jet-break time for GRB\,221009A in the literature spanning from $7.5\times10^{-3}$\,days to $\gtrsim475$\,days \citep{ndlo+23,lls+23,lca+23,lam+23,rvdhb+24, scb+24}. The multiwavelength light curves of this event exhibit evidence for a shallow jet structure \citep[e.g.][]{otr+23,gg23,bgb+24}, with a very narrow core possibly producing an early shallow jet break, while a larger outer angle of the shallow structure could produce a much later jet break. Here, we adopt the value of $\tjet=0.79\pm0.07$\,days from the achromatic (optical and X-ray) break found by \citet{otr+23}, but note that other values are possible.

For the remaining 9 events, we download the X-ray count-rate light curves from the \textit{Swift}/XRT website \citep{ebp+07,ebp+09}. Our workflow converts these to flux density at 1\,keV using the spectral parameters on the Swift website\footnote{The uncertainty from the flux calibration is typically much smaller than the uncertainties on the break time estimates inferred from the broken power law fits.}, after which we fit the light curve with a single or broken power-law of the form 
\begin{align*}
    F_\nu=F_0\left[
    (t/t_0)^{-s\alpha_1} +
    (t/t_0)^{-s\alpha_2}
    \right]^{-1/s},
\end{align*}
where $t_0$ is the break time, $s$ is the break smoothness, and $\alpha_1$, and $\alpha_2$ are the pre- and post-break decline rates, respectively. We discuss each case individually below, but in summary, we find evidence for breaks in the X-ray light curves consistent with a steepening due to a jet break in four cases (GRBs 190114C, 191221B, 210610B, 251013C) and a weak constraint under strong assumptions in one case (GRB 220921A). Three cases yield lower limits on \tjet\ (GRBs 190829A, 210702A, 220521A), and one case is inconclusive (GRB 250327B). 

\subsection{GRB\,190114C}
The last two XRT detections at $\approx5.9$ and $11.5$\,days reveal a rapidly fading afterglow, $\alpha_2=2.2\pm0.4$; fixing the post-break decay to this index and the smoothness to $s=3$, a broken power-law fit to the entire light curve yields a pre-break index of $\alpha_1 = -1.236\pm0.003$ and break time of $\tjet=4.2^{+4.9}_{-3.6}$\,days (Figure~\ref{fig:jetbreaks}). 

\subsection{GRB\,190829A}
The X-ray afterglow of GRB\,190829A exhibits a large flare peaking at $\approx(2\times10^{-2})$\,days. Masking the light curve at $\lesssim0.2$\,days and fitting with a broken power law yields  $\alpha_1 = -1.04\pm0.03$, $\alpha_2 = -1.37^{+0.07}_{-0.08}$ and a break time, $t_{\rm break}=7.1^{+12.3}_{-3.7}$\,days. Given the shallow post-break decline, we do not consider this to be the jet break; on the other hand, the last X-ray detection deviates by $\approx2\sigma$ from the best-fit broken power-law above, and hence we place a conservative lower limit of $\tjet\gtrsim90.6$\,days (corresponding to the mean time of the penultimate XRT observation) for this event. 

\subsection{GRB\,191221B}
The WT and PC-mode light curves of this event appear to exhibit an offset (or multiple break) and hence we exclude the WT-mode data from our analysis. Fitting the PC-mode data with a broken power-law yields a pre-break decline, $\alpha_1=-1.00\pm0.07$, post-break decline, $\alpha_2=-2.0\pm0.1$ (consistent with a jet break) and a break time, $\tjet=0.54^{+0.68}_{-0.43}$\,days.

\subsection{GRB\,210610B}
The WT-mode data for this event exhibits an early steep decline at $\lesssim4\times10^{-3}$\,days\footnote{\url{https://www.swift.ac.uk/xrt\_live\_cat/00945521/}} and hence we mask the WT-mode data in our fit. The (PC-mode) X-ray light curve (available at $\gtrsim0.05$\,days) can be fit with a broken power-law with a pre-break decline of $\alpha_1=-0.92^{+0.16}_{-0.09}$, a post-break decline of $\alpha_2=-1.9^{+0.1}_{-0.2}$ (consistent with a jet break) and a break time of $\tjet=0.53^{+0.80}_{-0.33}$ (Figure~\ref{fig:jetbreaks}; although see also \citealt{wwl+24} for an alternate interpretation of this break). 

\subsection{GRB\,210702A}
The X-ray light curve of this event exhibits a single power-law decline at $\gtrsim0.04$\,days until the last detection at $\approx9.4$\,days with a decline rate of $\alpha_{\rm X}=1.43\pm0.01$. Thus, we place a lower limit on the jet break time, $\tjet\gtrsim 9.4$\,days (see also, \citealt{dwlg+24}). 

\subsection{GRB\,220521A}
The X-ray light curve of this event exhibits a rising phase at $\lesssim6\times10^{-3}$\,days, followed by a decline at $\alpha_2=-1.7\pm0.1$. This decline appears shallow for a jet break and hence we set a limit of $\tjet\gtrsim2.2$\,days (corresponding to the last XRT datapoint) for this event, with the caveat that a low value of $p\lesssim2$ might be consistent with a lower $\tjet$. 

\subsection{GRB\,220921A}
MeerLICHT observations of the afterglow of this event at $\approx0.6$ and 2.0\,days after the burst yield a decline rate of $\alpha_{\rm opt}=-1.4\pm0.1$ \citep{gcn32572, gcn32582}. Fitting the X-ray light curve with a broken power-law with the pre-break decay fixed to $\alpha_{\rm opt}$, the post-break decay fixed to $\alpha_2=-2.2$, and the break smoothness fixed to $s=3$ yields a modest constraint on the break time of $\tjet = 3.4^{+5.0}_{-2.4}$\,days, which we use in our analysis. 

\subsection{GRB\,250327B}
The X-ray light curve of GRB\,250327B is poorly sampled\footnote{\url{https://www.swift.ac.uk/xrt\_curves/00019652/}}, with only two sets of data points at $\approx0.02$ and 1.1\,days. A power-law fit to the light curve yields a decline index of $\alpha_X=-1.8\pm0.1$. On the other hand, a fit to the optical light curve using data collected from the GCN circulars \citep{gcn39893, gcn39896, gcn39902, gcn39905, gcn39912} suggests $\alpha_{\rm opt}=-1.60\pm0.04$ from $\approx0.06$ to 0.24\,days. Assuming the X-ray light curve at this time follows a similar decline, fixing the pre-break decline to this value, and assuming a post-break decline of $t^{-2.2}$ yields a break time of $\tjet=0.20^{+0.34}_{-0.11}$\,days. This is the value of $\tjet$ used in Figures~\ref{fig:polmodels-toroidalB} and \ref{fig:polmodels-randomB}. However, we note that this value is extremely uncertain; for example, the observed X-ray and optical decline rates are equally consistent with the spectral ordering $\nu_c<\nu_{\rm X}$ with $p\approx3$ or $\nu_{\rm X}<\nu_c$ in a wind-like environment with $p\approx2.7$. Thus, we are unable to discuss the polarimetric observation of this burst relative to $\tovertjet$. 

\subsection{GRB\,251013C}
We combine the \textit{Swift}/XRT light curve of this event with the \textit{Chandra} measurement reported by \cite{gcn42642}. We mask the X-ray flare at $\approx0.06$\,days and fit the X-ray light curve with a broken power-law model with smoothness fixed to $s=3$. The break time is not strongly sensitive to the break smoothness, but is somewhat degenerate with $\alpha_2$. We find a pre-break index, $\alpha_1=-0.94^{+0.07}_{-0.6}$, post-break index, $\alpha_2=-1.8\pm0.2$, and break time, $\tjet=1.0^{+2.0}_{-0.6}$ (Figure~\ref{fig:jetbreaks}).

\newpage

\end{document}